\documentclass[12pt]{article}

\usepackage[utf8]{inputenc}
\usepackage[T1]{fontenc}

\usepackage{amsmath,amssymb,amsfonts,amsthm,mathtools,mathrsfs,dsfont}
\usepackage{bm,bbm}
\usepackage{scalerel}
\usepackage{booktabs}
\usepackage{color}
\usepackage[dvips,letterpaper]{geometry}
\usepackage{epsfig}
\usepackage{fullpage}
\usepackage{graphicx,psfrag,epsf}
\graphicspath{{./figures/}}
\usepackage{indentfirst}
\usepackage[authoryear]{natbib}
\usepackage{setspace}
\usepackage{subfigure}
\usepackage{times}
\usepackage{url}
\usepackage{verbatim}
\usepackage[symbol]{footmisc}

\usepackage{algorithm,algcompatible,algpseudocode}

\usepackage{adjustbox}
\usepackage{caption}
\usepackage{multicol,multirow}
\usepackage{float}

\usepackage{newtxtext,newtxmath} 
\usepackage{mhchem} 

\input cyracc.def

\def \R {\mathbb{R}}

\def \E    {\textrm{E}}

\def \p {\partial}

\newtheorem{proposition}{Proposition}

\title{\bf Scale-mixture Birnbaum-Saunders quantile regression models applied to personal accident insurance data}

\author{
{Alan Dasilva}$^{1}$, {Helton Saulo}$^{2,3}$, {Roberto Vila}$^{2,4}$\footnote{Corresponding author, rovig161@gmail.com.} \, and {Suvra Pal}$^{3}$\\[0.05cm]
{\normalsize $^{1}$Institute of Mathematics and Statistics, University of Sao Paulo, Sao Paulo, Brazil}\\[-0.05cm]
{\normalsize $^{2}$Department of Statistics, University of Brasilia, Brasilia, Brazil}\\[-0.05cm]
{\small $^{3}$Department of Mathematics, University of Texas at Arlington, Arlington, TX, USA}\\[-0.05cm]
{\normalsize $^{4}$
	Department of Mathematics and Statistics, McMaster University, Hamilton, Ontario, Canada
}\\[-0.05cm]
}

\date{}

\begin{document}

\maketitle


\noindent{\bf Abstract.} {The modeling of personal accident insurance data has been a topic of extreme relevance in the insurance literature. This kind of data often exhibits positive skewness and heavy tails. In this work, we propose a new quantile regression model based on the scale-mixture Birnbaum-Saunders distribution for modeling personal accident insurance data. The maximum likelihood estimates of the model parameters are obtained via the EM algorithm. Two Monte Carlo simulation studies are performed using the \texttt{R} software. The first study aims to analyze the performances of the EM algorithm to obtain the maximum likelihood estimates, and the randomized quantile and generalized Cox-Snell residuals. In the second simulation study, the size and power of the the Wald, likelihood ratio, score and gradient tests are evaluated. The two simulation studies are conducted considering different quantiles of interest and sample sizes. Finally, a real insurance data set is analyzed to illustrate the proposed approach.}\\



\section{Introduction}\label{section:01}

Birnbaum-Saunders (BS) regression models have proven to be good alternatives in insurance data modeling; see \cite{l:16} and \cite{NADERI2020125109}. Usually, this kind of data shows positive skewness and heavy tails. \cite{paula2012}, for example, used the BS-Student-$t$-BS regression model to study how some explanatory variables (covariates) influence the amount of paid money by an insurance policy. The Student-$t$-BS distribution is a special case of the class of scale-mixture Birnbaum-Saunders (SBS) distributions, which was proposed by \citet{balakrishnan2009} and is based on the relationship between the class of scale mixtures of normal (SMN) and BS distributions. The main advantage of the SBS distributions over the classical BS distribution \citep{l:16,balakundu:19} is their capability to make robust estimation of parameters in a way similar to that of SMN models, in addition to facilitating the implementation of an expectation maximization (EM) algorithm; see \citet{balakrishnan2009}. Recently, \citet{lachos2017} proposed a Bayesian regression model based on the SBS distributions for censored data.

The class of SMN distributions comprises a family of symmetric distributions studied initially by \citet{andrews1974}. These distributions have been gaining considerable attention with the publication of some works, as can be seen in \citet{efron1978}, \citet{west1987}, \citet{lange1993}, \citet{gneiting1997}, \citet{taylor2004}, \citet{walker2007} and \citet{lachos2007}. This family of distributions has as special cases flexible distributions with heavy-tails, which are often employed for robust estimation; see \citet{lange1989} and \citet{lucas1997}.

Quantile regression models emerged as a broader proposal in the analysis of the relationship between the response variable and the respective covariates of a regression model. These models were initially studied by \citet{koenker1978} and have since been increasingly used in regression analysis; see  \citet{koenker2001}, \citet{yu2001} and \citet{koenker2004}. Quantile regression models are robust alternatives in relation to the regression models for mean because the process of estimating parameters of the quantile regression model is not subject to the influence of outliers as in the case of linear regression. Moreover, these models are capable of showing different effects of covariates on the response along the quantiles of response variable.

In this work, we propose a new parametric quantile regression model for strictly positive data based on a reparameterization of the SBS distributions. We first introduce a reparameterization of the SBS model by inserting a quantile parameter, and then develop the new SBS quantile regression model. In addition to the quantile approach, another advantage in relation to the regression approach developed in \citet{lachos2017} is the modeling of dependent variable without the need of applying the logarithm transformation, which can cause problems with regard to interpretation and loss in power of the study; see \cite{huang:06} for an explanation. We illustrate the proposed methodology by using the insurance data set studied by \cite{paula2012}. The results show that a quantile approach provides a richer characterization of the effects of covariates on the dependent variable.

The rest of this paper is organized as follows. In Section \ref{section:02}, we describe the usual SBS distribution and propose a reparameterization of this distribution in terms of a quantile parameter. In this section, we also discuss some mathematical properties of the proposed reparameterized model. In Section \ref{section:03}, we introduce the SBS quantile regression model. In this section, we also describe the steps o the EM algorithm \citep{dempster1977} for maximum likelihood (ML) estimation of the model parameters, the standard error calculation, hypothesis tests, model selection criteria and residuals. In Section~\ref{section:04}, we carry out two Monte Carlo simulation studies. In the first simulation study, we evaluate the performance of the EM algorithm through the calculated biases, mean square errors (MSEs) and coverage probabilities of the asymptotic confidence intervals. We also evaluate the performances of residuals, such as the generalized Cox-Snell --GCS-- \citep{cs:68} and randomized quantile --RQ-- residuals \citep{ds:96}, for assessing the goodness-of-fit and identifying any departure from model assumptions. In the second simulation study, we assess the performances of formal test procedures such as the likelihood ratio test, score test, Wald test and gradient test. In Section \ref{section:05}, we apply the SBS quantile regression model to a real insurance data set. Finally, in Section \ref{section:06}, we make some concluding remarks and discuss potential future research problems.

\newpage

\section{Preliminaries}\label{section:02}

In this section, we first present the SBS distribution. Then, we present the proposed quantile-based SBS distribution, which we call as QSBS distribution. Some properties and special cases of the QSBS distribution are also discussed.

\subsection{SBS distribution}

Let $Y$ be a random variable (RV) whose stochastic representation can be written as follows:
\begin{equation}\label{eq:smn}
Y = \mu + \sqrt{g(U)} X,
\end{equation} 
where $\mu$ is a location parameter, $X \sim \text{N}(0,\sigma^2)$ with $\sigma>0$, $U$ is a RV distributed independent of $X$ and having cumulative distribution function (CDF) $H(\cdot)$, which is indexed by an extra parameter ${\nu}$ (or extra parameter vector $\bm{\nu}$), and $g(\cdot)$ is a {\color{black}  strictly} positive function. Note that when $g(U) = 1/U$, the distribution of $Y$ in \eqref{eq:smn} reduces to the normal/independent distribution, presented in \cite{lange1993}. Similarly, when $g(U) = U$ in \eqref{eq:smn}, the distribution of $Y$ reduces to the SMN distribution, studied by \cite{fernandez1999}.

The probability density function (PDF) of $Y$, a RV that follows a SMN distribution with location and scale parameters $\mu \in \R$ and $\sigma^2 > 0$, respectively, is given {\color{black}by the following Lebesgue-Stieltjes integral:}
\begin{equation}\label{eq:smn-fdp}
\phi_{\text{SMN}}(y) 
= 
\int_{\color{black}(0,\infty)\cap {\rm Supp}(U)} \phi(y; \mu, g(u) \sigma^2) dH(u), {\color{black}\quad y\in\mathbb{R},}
\end{equation}
where {\color{black}${\rm Supp}(U)$ is the support of $U$,}
$\phi({\color{black}\cdot}; \mu, g(u) \sigma^2)$ is the normal PDF with mean $\mu$ and variance $g(u) \sigma^2$, and $H(u)$ is the CDF of $U$. Let us denote $Y \sim \text{SMN}(\mu,\sigma^2;H)$. When $\mu = 0$ and $\sigma^2 = 1$, we denote $Y \sim \text{SMN}(H)$.

According to \cite{balakrishnan2009}, a RV $T$ follows a SBS distribution if it has the following stochastic representation:
\begin{equation}\label{eq:sbs}
T = \frac{\beta}{4} \left[ \alpha {\color{black} Y} + \sqrt{(\alpha {\color{black} Y})^2 + 4} \right]^2,
\end{equation}
where $Y = \sqrt{g(U)}Z \sim \text{SMN}(H)$, such that $Z \sim \text{N}(0,1)$ and $\alpha>0$ and $\beta>0$ are the shape and scale parameters, respectively, of the BS distribution. In this case, the notation $T \sim \text{SBS}(\alpha,\beta;H)$ is used. When $Y \sim \text{N}(0,1)$, then $T$ follows a classical BS distribution. The PDF of $T \sim \text{SBS}(\alpha,\beta;H)$ can be expressed as
\begin{equation}\label{eq:sbs-fdp}
f_T(t) = \phi_{\text{SMN}}(a(t)) a'(t), \quad t > 0,
\end{equation}
where $\phi_{\text{SMN}}(\cdot)$ is the PDF given in \eqref{eq:smn-fdp} with $\mu = 0$ and $\sigma^2 = 1$, $a(t) = \bigl(\sqrt{t/\beta} - \sqrt{\beta/t}\,\bigr)/\alpha$, and $a'(t) = t^{-3/2}(t + \beta)/(2 \alpha \beta^{1/2})$ is the derivative of $a(t)$ with respect to $t$. The CDF of $T$ is given by
\begin{equation}\label{eq:sbs-fda}
F_T(t) = \Phi_{\text{SMN}}(a(t)),  \quad t > 0,
\end{equation}
where $\Phi_{\text{SMN}}(\cdot)$ is the CDF of the $\text{SMN}(H)$  distribution. 
The $100q-$th quantile of $T \sim \text{SBS}(\alpha,\beta;H)$ is given by
\begin{equation}\label{eq:sbs-q}
Q = t_q = \frac{\beta}{4} \left[ \alpha y_q + \sqrt{(\alpha y_q)^2 + 4} \right]^2, 
\end{equation}
where $y_q$ is the $q\times 100$-th quantile of $Y \sim \text{SMN}(H)$.

\subsection{Quantile-based SBS distributions}

Consider a fixed number $q \in (0,1)$ and the one-to-one transformation 
$
(\alpha,\beta;H) \longmapsto (\alpha,Q;H) \nonumber,
$
where $Q$ is the $100q$-th quantile of $T \sim \text{SBS}(\alpha,\beta;H)$ defined in \eqref{eq:sbs-q}. Then, we obtain the following stochastic representation based on a quantile parameter:
\begin{equation}\label{eq:qsbs}
	T = \frac{Q}{\gamma_\alpha^2} \left[ \alpha \sqrt{g(U)} Z + \sqrt{\big(\alpha \sqrt{g(U)} Z\big)^2 + 4} \right]^2, 
\end{equation}
where $\gamma_\alpha = \alpha y_q + \sqrt{(\alpha y_q)^2 + 4}$, $Z \sim \text{N}(0,1)$, and $g(U) = 1/U$
is such that the distribution of $g(U)$ has a known PDF. The stochastic representation \eqref{eq:qsbs} can be used to generate random numbers, to obtain the moments, and also in the implementation of the EM algorithm.

From the stochastic representation \eqref{eq:qsbs}, we obtain a quantile-based reparameterization of the SBS distribution with CDF and PDF given, respectively, by
\begin{equation}\label{eq:qsbs-fda}
\begin{array}{lllll}
\displaystyle
F_T(t) = \Phi_\text{SMN}(a(t))\,
{\color{black}
	=
	\int_{(0,\infty)\cap {\rm Supp}(U)} \Phi\big(a_{\scaleto{\sqrt{g(u)}}{9pt}}(t)\big) dH(u)
},  \quad t > 0, 
\\[0,6cm]
\displaystyle
f_T(t) = \phi_\text{SMN}(a(t))a'(t)
\,
{\color{black}
	=
	\int_{(0,\infty)\cap {\rm Supp}(U)} \phi\big(a_{\scaleto{\sqrt{g(u)}}{9pt}}(t)\big) a'_{\scaleto{\sqrt{g(u)}}{9pt}}(t) dH(u)
}
, \quad t > 0, 
\end{array}
\end{equation}
where 
{\color{black} $\Phi(\cdot)$ and $\phi(\cdot)$ are the CDF and PDF of the standard normal distribution, respectively,
$a_x(t) 
= \bigl(\sqrt{\gamma_{\alpha x}^2 t/4Q} - \sqrt{4Q/\gamma_{\alpha x}^2 t}\,\bigr)/{(\alpha x)}
%
$, $a'_x(t) = (\gamma_{\alpha x}^2/2 + 2Q/t)/(\alpha x \gamma_{\alpha x} \sqrt{4Qt})$  and} $\gamma_\alpha = \alpha y_q + \sqrt{(\alpha y_q)^2 + 4}$. In this case, the notation $T \sim \text{QSBS}(\alpha,Q;H)$ is used. Some properties of the QSBS distribution are presented below.

{\color{black}
\begin{proposition}\label{limits-pdf}
	If $T \sim \text{QSBS}(\alpha,Q;H)$ then
	$
	\lim_{t\to 0^+}f_T(t)=\lim_{t\to \infty}f_T(t)=0.
	$
\end{proposition}
\begin{proof}
Since $f_T(\cdot)$ is a PDF, it is clear that $\lim_{t\to \infty}f_T(t)=0$. On the other hand, by \eqref{eq:qsbs-fda} we observe that
$f_T(t)=	\int_{(0,\infty)\cap {\rm Supp}(U)} f_{S_u}(t) dH(u)$, where $f_{S_u}(\cdot)$ is the corresponding PDF of a RV $S_u$  having a quantile-based Birnbaum-Saunders (QBS) distribution with parameters $\alpha\sqrt{u}$ and $Q$, denoted by $S_u\sim \text{QBS}\big(\alpha\sqrt{g(u)},Q\big)$. It is well-known that the QBS distribution is unimodal. Then there is a unique $t_0=t_0\big(\alpha\sqrt{g(u)},Q\big)>0$ such that $f_{S_u}(t)\leqslant \max\{{f_{S_u}(t)}: t>0\}=f_{S_u}(t_0)\leqslant \max\{f_{S_u}(t_0):u>0\}$. Applying the bounded convergence theorem, we have $\lim_{t\to 0^+}f_T(t)=\int_{(0,\infty)\cap {\rm Supp}(U)} \lim_{t\to 0^+}f_{S_u}(t) dH(u)=0$ because $\lim_{t\to 0^+}f_{S_u}(t)=0$. This completes the proof.
\end{proof}

\begin{proposition}
	If  $g(\cdot)$ in \eqref{eq:smn} is chosen such that the function 
	\begin{align*}
	G(t)=
	{\phi'_\text{SMN}(a(t))\over \phi_\text{SMN}(a(t))}+{a''(t)\over [a'(t)]^2}, \quad t>0
	\end{align*}
	has a unique zero $t=t_0$, then the QSBS PDF  \eqref{eq:qsbs-fda} is unimodal.
\end{proposition}
\begin{proof}
Suppose that $g(\cdot)$ is such that $G(t_0)=0$.
A simple calculus shows that $f_T'(t)=f_T(t)a'(t) G(t)$. Then $f_T'(t_0)=0$ because $G(t_0)=0$, $f_T(t)>0$ and $a'(t)>0$. That is, $t_0$ is the unique critical point of $f_T(\cdot)$. Furthermore, by Proposition \ref{limits-pdf}, $\lim_{t\to 0^+}f_T(t)=\lim_{t\to \infty}f_T(t)=0$. Consequently, the QSBS PDF is increasing on $(0,t_0)$ and is decreasing on $(t_0,\infty)$. This proves the unimodality.
\end{proof}

		In the proofs of the following propositions, we adopt the following notations:
		$a_x(t)=a_x(t;\alpha,Q) 
		$ and $a(t)=a(t;\alpha,Q)=a_1(t)$.
\begin{proposition}\label{prop-TU-conditional}
Let $T \sim \text{QSBS}(\alpha,Q;H)$. Then, $T$ allows the following conditional stochastic representation:
\begin{align*}
T\vert U=u\sim \text{QBS}\big(\alpha\sqrt{g(u)},Q\big).
\end{align*}
\end{proposition}
\begin{proof}
If $U=u$ then $T=	a^{-1}\big(\sqrt{g(u)} Z\big)$.
Thus, the conditional distribution of $T$ given $U=u$ is the same as the distribution of the elements presented as
$
	a^{-1}\big(\sqrt{g(u)} Z\big) \vert U=u.
$
This implies that $F_T(t\vert U=u)=\Phi\big(a(t)/\sqrt{g(u)}\big)=\Phi\big(a_{\scaleto{\sqrt{g(u)}}{9pt}}(t)\big)$ because $U$ is independent of $Z$. Then, the proof follows.
\end{proof}

Let $T \sim \text{QSBS}(\alpha,Q;H)$ and $f$ be a Borel measurable function. Then, by using the law of total expectation, we have
\begin{align}\label{formula-law-total-exp}
\mathbb{E}[f(T)]
=
\mathbb{E}[\mathbb{E}[f(T)\vert U]]
=
\int_{(0,\infty)\cap {\rm Supp}(U)} 
\mathbb{E}[f(T)\vert U=u] dH(u).
\end{align}

By taking $f(t)=\exp(ist)$, $s\in\mathbb{R}$, in \eqref{formula-law-total-exp}, where $i=\sqrt{-1}$ is the unit imaginary number, from
Proposition \ref{prop-TU-conditional} and Formula (2.13) of \cite{l:16}, we get the following result:
\begin{proposition}
Let $T \sim \text{QSBS}(\alpha,Q;H)$. Then, the characteristic function of $T$, denoted by $\phi_T(s)=\mathbb{E}(\exp(isT))$, can be written as
\begin{align*}
\phi_T(s)
=
{1\over 2}\,
\mathbb{E}\left[
\left(1+\frac{1}{\sqrt{1-2i({4Q}/{\gamma_\alpha^2})s^2 g(U)}}\right)
\exp\left(\frac{1-\sqrt{1-2i({4Q}/{\gamma_\alpha^2})s^2 g(U)}}{\alpha^2 g(U)}\right)
\right].
\end{align*}
\end{proposition}

Taking $k$ consecutive partial derivatives of $\phi_T(s)$ with respect to $s$, assuming that we can interchange the derivative with the expectation, and then evaluating at $s=0$ and dividing by $i^k$, we get the following result:
}
\begin{proposition}\label{prop:moments}
	Let $T \sim \text{QSBS}(\alpha,Q;H)$ and $g(U)$ be a RV as in \eqref{eq:smn} with finite moments of all order. Then, the $k$-th moment of $T$ has the following form:
	\begin{equation}
		\mathbb{E}[T^k] 
		= 
		\left(  \frac{4Q}{\gamma_\alpha^2}\right)^k \sum_{l = 0}^{k} \binom{2k}{2l} \sum_{j = 0}^{l} \binom{l}{j} \omega_{k + j - l} \left( \frac{\alpha}{2} \right) ^ {2(k + j - l)}, \nonumber
	\end{equation}
	where $\omega_r = \mathbb{E}[g^r(U)]$ denotes the $r$-th moment of $g(U)$.
\end{proposition}
{\color{black}
Another way to obtain the expression of the above proposition is to consider $f(t)=t^k$ in \eqref{formula-law-total-exp} and by using
Proposition \ref{prop-TU-conditional} and Formula (2.18) of \cite{l:16}.

The following two results prove that the QSBS distribution belongs to the family of scale and reciprocal invariant distributions:
}

\begin{proposition}
	Let $T \sim \text{QSBS}(\alpha,Q;H)$. Then, $c T \sim \text{QSBS}(\alpha, c Q; H)$ with $c > 0$.
\end{proposition}
	\begin{proof}
		{\color{black} It is simple to see that $a(t/c;\alpha,Q)=a(t;\alpha,cQ)$. Then, by using \eqref{eq:qsbs-fda}, we have }
$
			\mathbb{P}(cT \leqslant t) 
{\color{black}
			=
			\Phi_{\mathrm{SMN}}(a(t/c;\alpha,Q)) 
			=
			\Phi_{\mathrm{SMN}}(a(t;\alpha,cQ))
		}
$
		{\color{black} because $c>0$. This guarantees the result.}
	\end{proof}

\begin{proposition}\label{reciprocal-prop}
	Let $T \sim \text{QSBS}(\alpha,Q;H)$. Then, $1/T \sim \text{QSBS}(\alpha, {\color{black}\gamma_\alpha^4/16Q}; H)$.
\end{proposition}
	\begin{proof}
	{\color{black} A simple observation shows that $a(1/t;\alpha,Q)=-a(t;\alpha,\gamma_\alpha^4/16Q)$. Then, by using \eqref{eq:qsbs-fda} together with the identity $\phi_{\text{SMN}}(-y) =1-\phi_{\text{SMN}}(y)$, we get}
$
			\mathbb{P}(1/T \leqslant t) 
			= 
			{\color{black}
1-		\Phi_{\mathrm{SMN}}(a(1/t;\alpha,Q)) 
=
1-		\Phi_{\mathrm{SMN}}(-a(t;\alpha,\gamma_\alpha^4/16Q))
= 
\Phi_{\mathrm{SMN}}(a(t;\alpha,\gamma_\alpha^4/16Q)).
}
			%
$
{\color{black} Hence, the proof follows.}
	\end{proof}

{\color{black}
From Proposition \ref{reciprocal-prop}, we get the following formula for the negative moments of the QSBS distribution:
\begin{proposition}\label{prop:negative-moments}
	Let $T \sim \text{QSBS}(\alpha,Q;H)$ and $g(U)$ be a RV as in \eqref{eq:smn} with finite moments of all order. Then, the $k$-th negative  moment of $T$ has the following form:
	\begin{equation}
	\mathbb{E}[T^{-k}] 
	= 
	\left(  \frac{\gamma_\alpha^2}{4Q}\right)^k \sum_{l = 0}^{k} \binom{2k}{2l} \sum_{j = 0}^{l} \binom{l}{j} \omega_{k + j - l} \left( \frac{\alpha}{2} \right) ^ {2(k + j - l)}, \nonumber
	\end{equation}
	where $\omega_r = \mathbb{E}[g^r(U)]$ denotes the $r$-th moment of $g(U)$.
\end{proposition}
}

By using Proposition \ref{prop:moments}, if $T \sim \text{QSBS}(\alpha,Q;H)$, then the expectation, variance and coefficients of variation (CV), skewness (CS) and kurtosis (CK) are given, respectively, by
	
	\begin{eqnarray}
		\mathbb{E}[T] &=& \frac{2Q}{\gamma_\alpha^2}\big( 2 + \omega_1 \alpha^2\big), \quad
		\text{Var}[T] = \frac{4 Q^2 \alpha^2}{\gamma_\alpha^4}\left[ \omega_1 + (2\omega_2 - \omega_1^2)\alpha^2\right], \nonumber\\
		\text{CV}[T] &=& \frac{\alpha \sqrt{4 \omega_1 + (2 \omega_2  - \omega_1^2)\alpha^2}}{2 + \omega_1 \alpha^2}, \nonumber\\
		\text{CS}[T] &=& \frac{4 \alpha[3(\omega_2 - \omega_1^2) + (2 \omega_3 - 3\omega_1\omega_2 + \omega_1^3)\alpha^2 / 2]}{[4 \omega_1 + (2 \omega_2 - \omega_1^2)\alpha^2]^{3/2}}, \nonumber\\
		\text{CK}[T] &=& \frac{16 \omega_2 + (32\omega_3 - 48\omega_1\omega_2 + 24\omega_1^3)\alpha^2 + (8\omega_4 - 16\omega_1\omega_3 + 12\omega_1^2\omega_2 - \omega_1^4)\alpha^4}{[4\omega_1 + (2\omega_2 - \omega_1^2)\alpha^2]^2}, \nonumber
	\end{eqnarray}
	where $\omega_r = \mathbb{E}[g^r(U)]$ denotes the $r$-th moment of $g(U)$.

\subsection{Special cases of the QSBS distribution}

In this section, we present some particular cases of the QSBS family of distributions, i.e., QSBS models based on the contaminated-normal, slash, and Student-$t$ distributions. 

{\color{black}
	Let $T \sim \text{QSBS}(\alpha,Q;H)$, where $H$ is the CDF of the variable $U$, whose PDF (or probability mass function) is defined as $h_U$. By using \eqref{eq:smn-fdp} and by considering $g(u)=1/u$, for determined functions $h_U$, the following formulas (Table \ref{table:1}) for the PDF $\phi_{\text{SMN}}$ are obtained.
	\begin{table}[H]
		\caption{Examples of functions $h_U$ and $\phi_{\text{SMN}}$ associated with some QSBS distributions with $g(u)=1/u$.}
		\vspace*{0.15cm}
		\centering 
		\resizebox{\linewidth}{!}{
			\begin{tabular}{llll} 
				\hline
				Distribution & $h_U(u)$ & $\phi_{\text{SMN}}(y)$ &  Parameter 
				\\ [0.5ex] 
				\noalign{\hrule}
				Quantile contaminated-normal BS
				& $[\nu\updelta_{u \delta} + (1 - \nu)\updelta_{u 1}]\mathds{1}_{(0,1)}(u)$ & $\bigl[\nu \sqrt{\delta} \phi(\sqrt{\delta}y) + (1 - \nu) \phi(y)\bigr]\mathds{1}_{(-\infty,\infty)}(y)$ & $0 < \nu < 1, \; 0 < \delta < 1$ 
				\\ [1ex] 
				Quantile slash BS
				& $\nu u ^{\nu - 1} \mathds{1}_{(0,1)}(u)$  & $\bigl[\nu \int_{0}^{1} u^{\nu - 1} \phi\bigl(y; 0 , {1\over u}\bigr) du\bigr] \mathds{1}_{(-\infty,\infty)}(y)$ & $ \nu > 0 $   
				\\ [1ex]
				Quantile Student-$t$ BS
				& $\frac{({\nu\over 2})^{\nu/2}}{\Gamma({\nu\over 2})} u^{\nu/2 - 1} \exp\left( - \frac{\nu u}{2} \right) \mathds{1}_{(0,\infty)}(u)$ &  $\frac{\Gamma(\frac{\nu + 1}{2})}{\sqrt{\pi} \sqrt{\nu} \Gamma(\frac{\nu}{2})} 
				\left( 
				1 + \frac{y^2}{\nu}  
				\right)^{-\frac{\nu + 1}{2}} \mathds{1}_{(-\infty,\infty)}(y)$ & $\nu > 0$   
				\\ [1ex]
				\hline	
			\end{tabular}
		}
		\label{table:1} 
	\end{table}
\noindent
In Table \ref{table:1}, $\updelta_{i j}$ denotes the delta de Kronecker, $\mathds{1}_{A}$ is the indicator function of an event $A$, $\phi(\cdot)$ is the standard normal PDF and $\Gamma(\cdot)$ is the complete gamma function.
For the rest of this paper, we consider $g(u)=1/u$.

}

\subsubsection{Quantile contaminated-normal BS distribution}

Let $T \sim \text{QSBS}(\alpha,Q;H)$ {\color{black} and $h_U$ be as given in Table \ref{table:1} }.
By using 
\eqref{eq:qsbs-fda} {\color{black} and $\phi_{\text{SMN}}$ in Table \ref{table:1}}, it follows that the PDF of $T$ can be expressed as {\color{black} (for $t>0$)}
\begin{equation}
f_T(t) 
= 
\left[ 
\nu \sqrt{\delta} \phi\left({\sqrt{\delta}\over \alpha \gamma_\alpha}\sqrt{4Q\over t} \left({\gamma_\alpha^2t\over 4Q} - 1\right)\right) + (1 - \nu) \phi\left({1\over \alpha \gamma_\alpha}\sqrt{4Q\over t} \left({\gamma_\alpha^2t\over 4Q} - 1\right)\right)
\right] 
\frac{1}{\alpha \gamma_\alpha \sqrt{4Qt}} \left( \frac{\gamma_\alpha^2}{2} + \frac{2Q}{t} \right). \nonumber
\end{equation}
Let us denote $T \sim \text{CN-BS}(\alpha,Q,\bm{\nu})$ with $\bm{\nu} = (\nu,\delta)$. The PDF of the conditional distribution of $U|T = t$ is given by
\begin{equation}
	h_{U|T}(u|t) = \nu p(t,u) {\color{black} \updelta_{u \delta}} + (1 - \nu) p(t,u)  {\color{black} \updelta_{u 1}}, \nonumber
\end{equation}
such that we obtain $p(t,u)$ by calculating
\begin{equation}
	p(t,u) 
	=
	 \frac{\sqrt{u} 
		\exp\left(- u\, \frac{2Q}{t \alpha^2 \gamma_\alpha^2}   \left( \frac{\gamma_\alpha^2t}{4Q} - 1 \right)^2\right) }{\nu \sqrt{\delta} 
		\exp\left( 
		-\delta\,  \frac{2Q}{t \alpha^2 \gamma_\alpha^2}   \left( \frac{\gamma_\alpha^2t}{4Q} - 1 \right)^2
		\right) + (1-\nu) 
		\exp\left(
		 -\frac{2Q}{t \alpha^2 \gamma_\alpha^2}   
		 \left( \frac{\gamma_\alpha^2t}{4Q} - 1 \right)^2 
		 \right)  }. \nonumber
\end{equation}
Then, the conditional expectation of $U| T = t$ can be written as
\begin{equation} \label{eq:meancnbs}
\mathbb{E} [U| T = t] 
= 
\frac{1 - \nu + \nu \delta^{3/2} 
	\exp \left(
	 (1 - \delta)\, \frac{2Q}{t \alpha^2 \gamma_\alpha^2}   \left( \frac{\gamma_\alpha^2t}{4Q} - 1 \right)^2
	 \right)  }{1 - \nu + \nu \sqrt{\delta} 
	 \exp\left( 
	 (1-\delta)\, \frac{2Q}{t \alpha^2 \gamma_\alpha^2}   \left( \frac{\gamma_\alpha^2t}{4Q} - 1 \right)^2
	 \right) }. 
\end{equation}

\subsubsection{Quantile slash-BS distribution }

Let $T \sim \text{QSBS}(\alpha,Q;H)$
{\color{black} and $h_U$ be as given in Table \ref{table:1}, meaning that $U \sim \text{Beta}(\nu,1)$.}
From 
\eqref{eq:qsbs-fda} {\color{black} and $\phi_{\text{SMN}}$ in Table \ref{table:1}}, we can write the PDF of $T$ as
\begin{equation}
f_T(t) = 
\left[ 
\nu \int_{0}^{1} u^{\nu - 1} \phi\left({1\over \alpha \gamma_\alpha}\sqrt{4Q\over t} \left({\gamma_\alpha^2t\over 4Q} - 1\right); 0 , {1\over u}\right) du 
\right] \frac{1}{\alpha \gamma_\alpha \sqrt{4Qt}} \left( \frac{\gamma_\alpha^2}{2} + \frac{2Q}{t} \right), \quad t > 0. \nonumber
\end{equation}
In this case, the notation $T \sim \text{SL-BS}(\alpha,Q,\nu)$ is used. Note that 
\begin{align*}
U|T = t \sim \text{Gama}\left({1\over 2} + \nu, \frac{2Q}{t \alpha^2 \gamma_\alpha^2}   \left( \frac{\gamma_\alpha^2t}{4Q} - 1 \right)^2\right)
\end{align*}
truncated to the interval $[0,1]$. We can then obtain the conditional expectation as
\begin{equation} \label{eq:meanslbs}
\mathbb{E}[U|T = t] = 
\left[  \frac{1 + 2\nu}{\frac{4Q}{t \alpha^2 \gamma_\alpha^2}   \left( \frac{\gamma_\alpha^2t}{4Q} - 1 \right)^2 } \right]  
\frac{P_1\left(\frac{3}{2} + \nu, \frac{2Q}{t \alpha^2 \gamma_\alpha^2}   \left( \frac{\gamma_\alpha^2t}{4Q} - 1 \right)^2  \right)}{P_1\left(\frac{1}{2} + \nu, \frac{2Q}{t \alpha^2 \gamma_\alpha^2}   \left( \frac{\gamma_\alpha^2t}{4Q} - 1 \right)^2 \right)}, 
\end{equation}
where $P_x(a,b)$ is the gamma CDF evaluated at $x$ and with parameters $a$ and $b$. 

\subsubsection{Quantile Student-$t$ BS distribution}

Let $T \sim \text{QSBS}(\alpha,Q;H)$
{\color{black} and $h_U$ be as given in Table \ref{table:1}, meaning that $U \sim \text{Gamma}(\nu/2,\nu/2)$.}
Then, based on 
\eqref{eq:qsbs-fda} {\color{black} and $\phi_{\text{SMN}}$ in Table \ref{table:1}}, we obtain the PDF of $T$ as
\begin{equation}\label{eq:tbs}
f_T(t) 
= 
\frac{\Gamma(\frac{\nu + 1}{2})}{\sqrt{\pi} \sqrt{\nu} \Gamma(\frac{\nu}{2})} 
\left[ 
1 + \frac{4Q}{\nu t \alpha^2 \gamma_\alpha^2}   \left( \frac{\gamma_\alpha^2t}{4Q} - 1 \right) ^2 
\right]^{-\frac{\nu + 1}{2}} 
\frac{1}{\alpha \gamma_\alpha \sqrt{4Qt}} \left( \frac{\gamma_\alpha^2}{2} + \frac{2Q}{t} \right), \quad t > 0,
\end{equation}
with notation $T \sim t_\nu \text{- BS}(\alpha,Q,\nu)$. In this case, we have 
\begin{align*}
U|T = t \sim \text{Gamma}\left({\nu + 1\over 2},{\nu\over 2} + \frac{2Q}{t \alpha^2 \gamma_\alpha^2} \left( \frac{\gamma_\alpha^2t}{4Q} - 1\right)^2\right)
\end{align*}
and
\begin{equation} \label{eq:meantbs}
\mathbb{E}[U|T = t] = \frac{\nu + 1}{\nu + \frac{4Q}{t \alpha^2 \gamma_\alpha^2}   \left( \frac{\gamma_\alpha^2t}{4Q} - 1 \right)^2 }.
\end{equation}

\begin{figure}[H]
	\centering
	\subfigure[$\text{CN-BS}(\alpha,1,(0.2,0.2))$, $q=0.25$]{\includegraphics[scale=0.23]{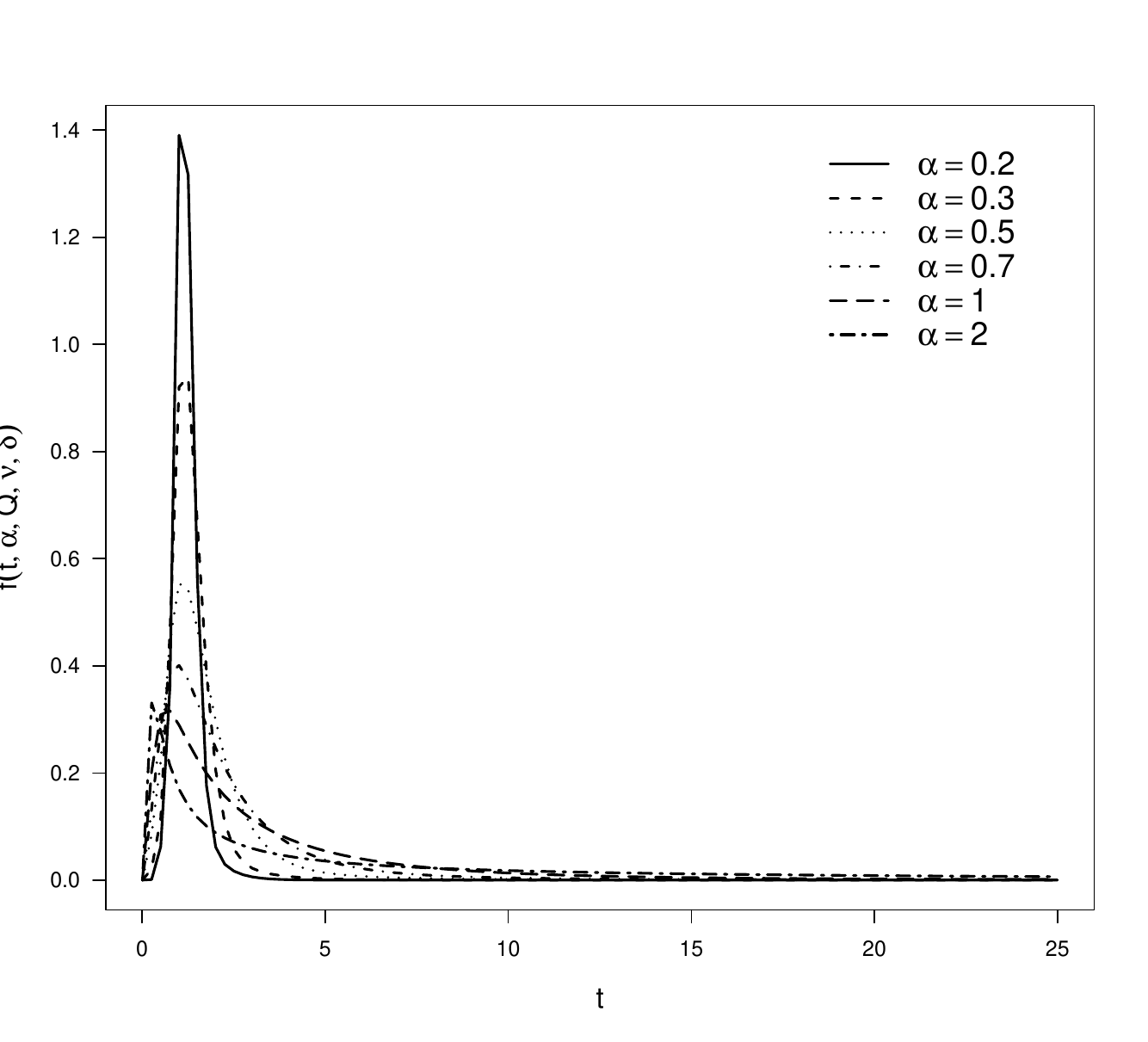}}
	\subfigure[$\text{CN-BS}(1,Q,(0.2,0.2))$, $q=0.50$]{\includegraphics[scale=0.23]{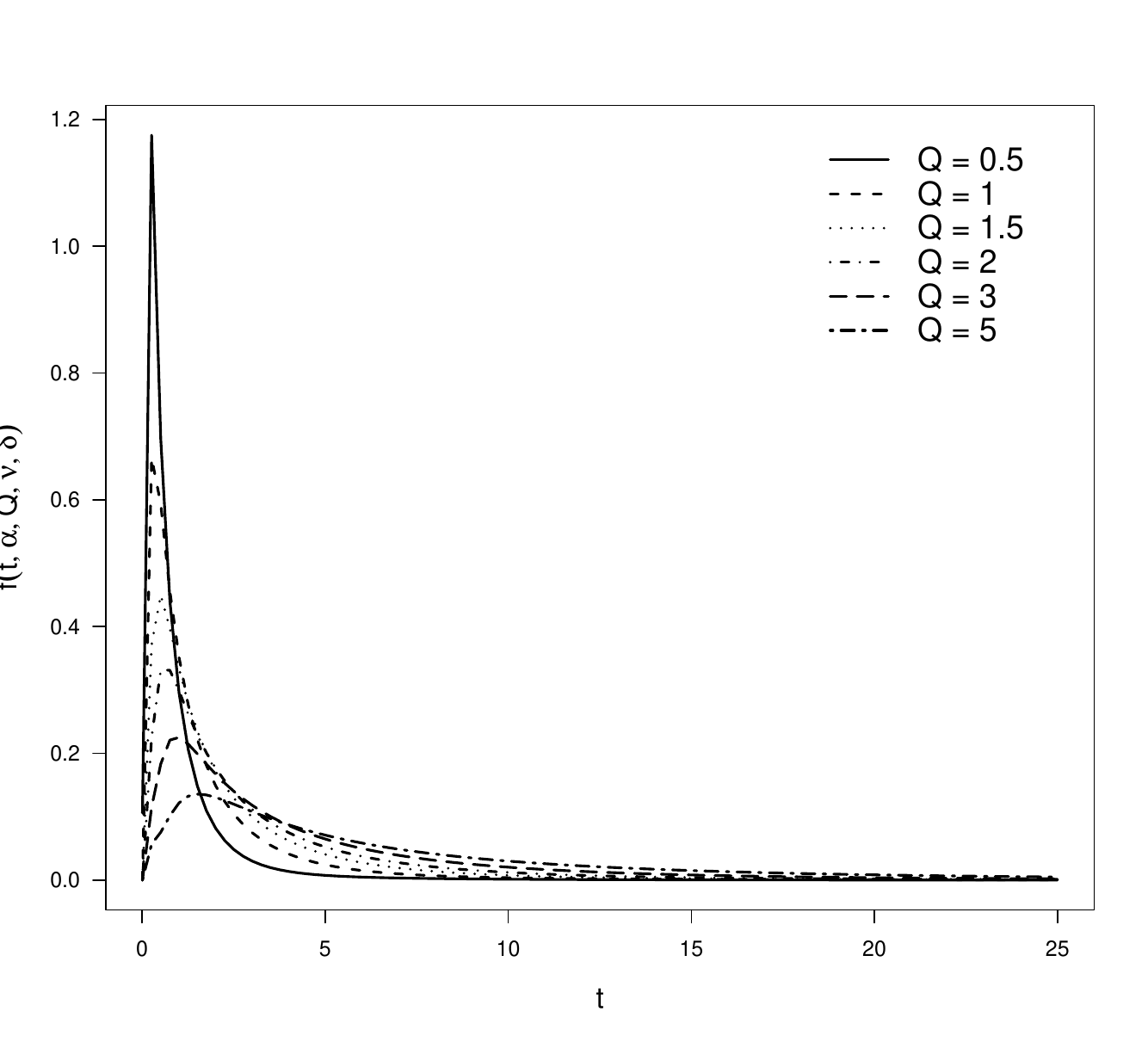}}
	\subfigure[$\text{CN-BS}(\alpha,Q,(0.2,0.2))$, $q=0.75$]{\includegraphics[scale=0.23]{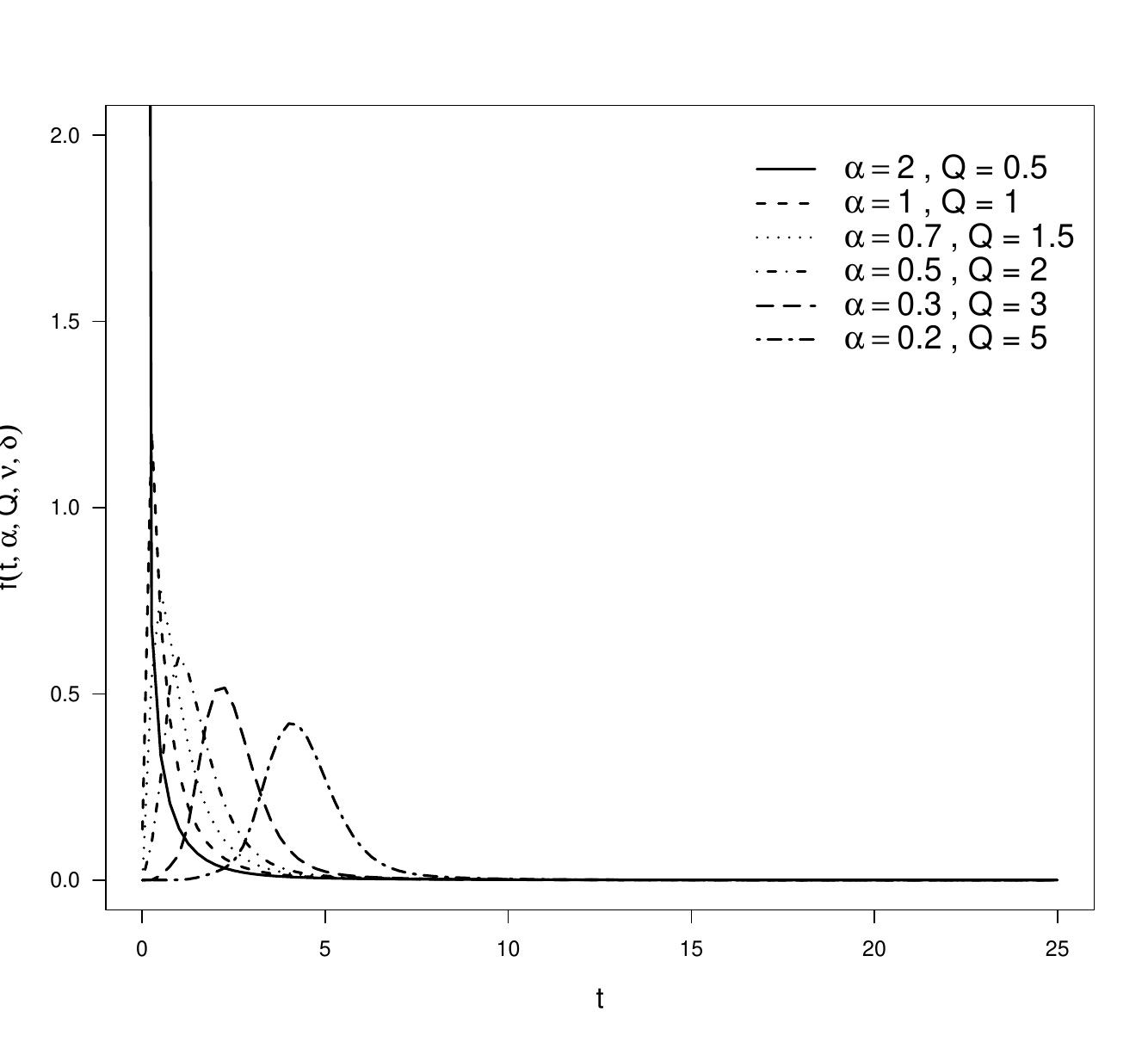}}
	\subfigure[$\text{SL-BS}(\alpha,1,2)$, $q=0.25$]{\includegraphics[scale=0.23]{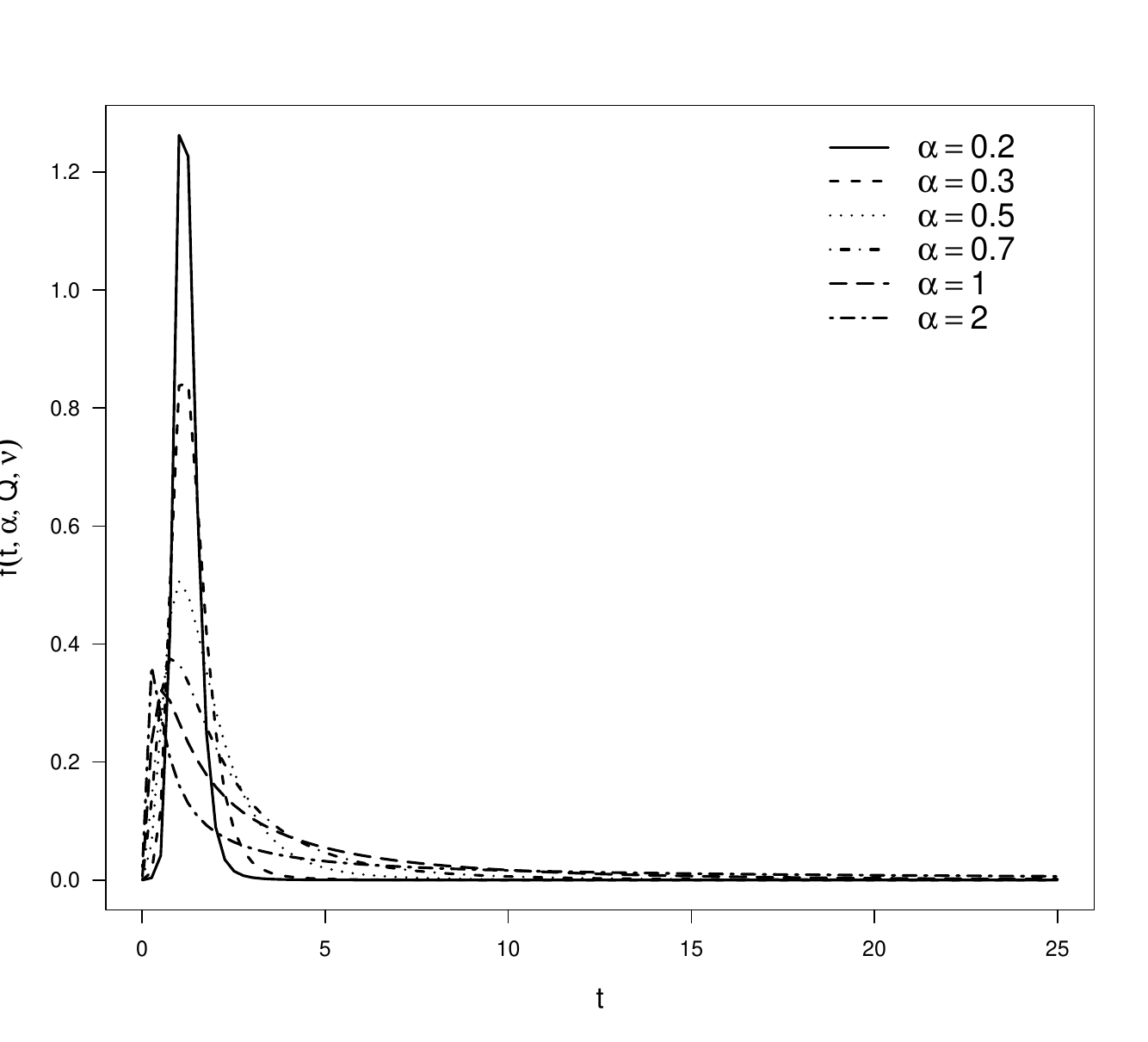}}
	\subfigure[$\text{SL-BS}(1,Q,2)$, $q=0.50$]{\includegraphics[scale=0.23]{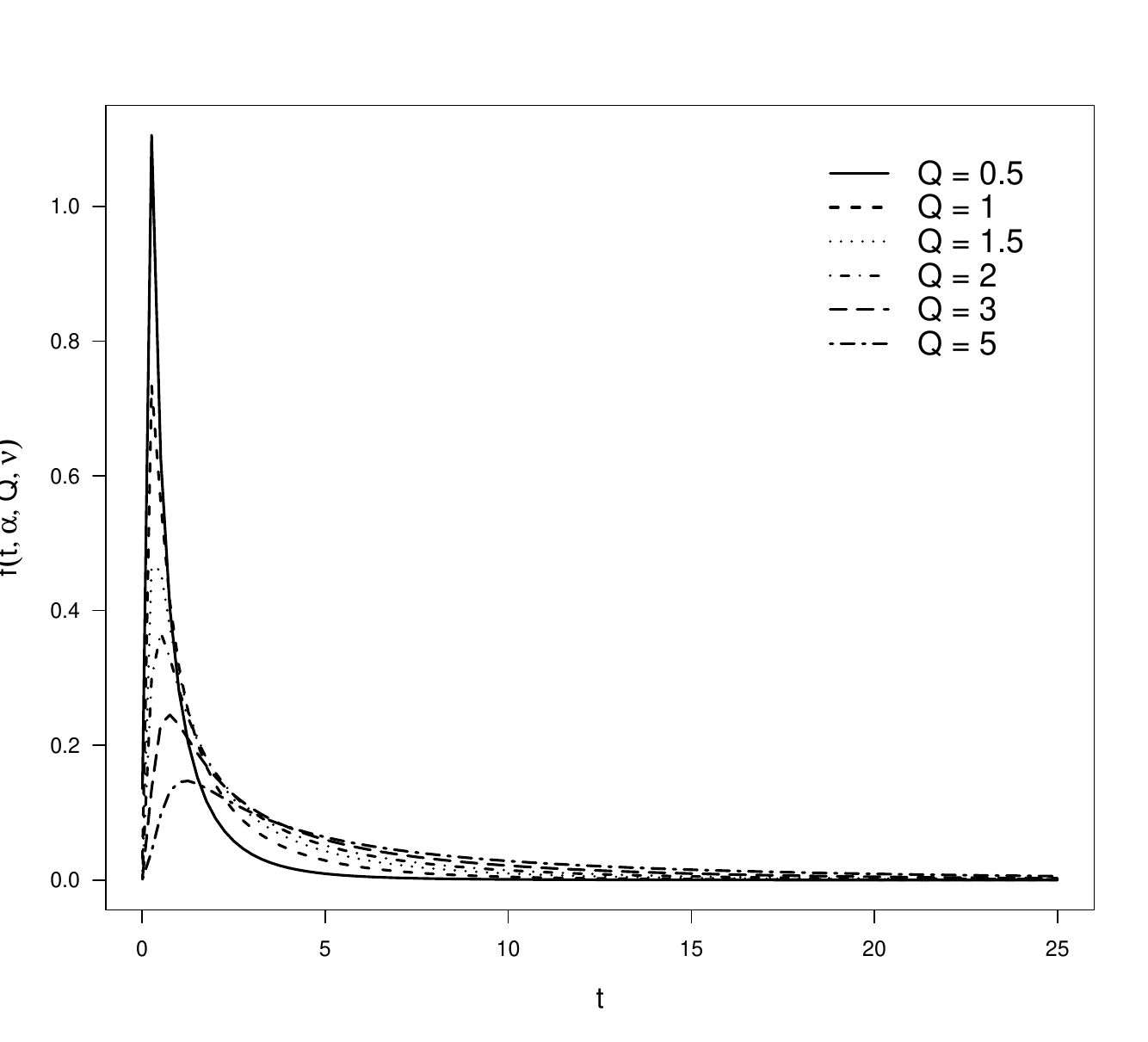}}
	\subfigure[$\text{SL-BS}(\alpha,Q,2)$, $q=0.75$]{\includegraphics[scale=0.23]{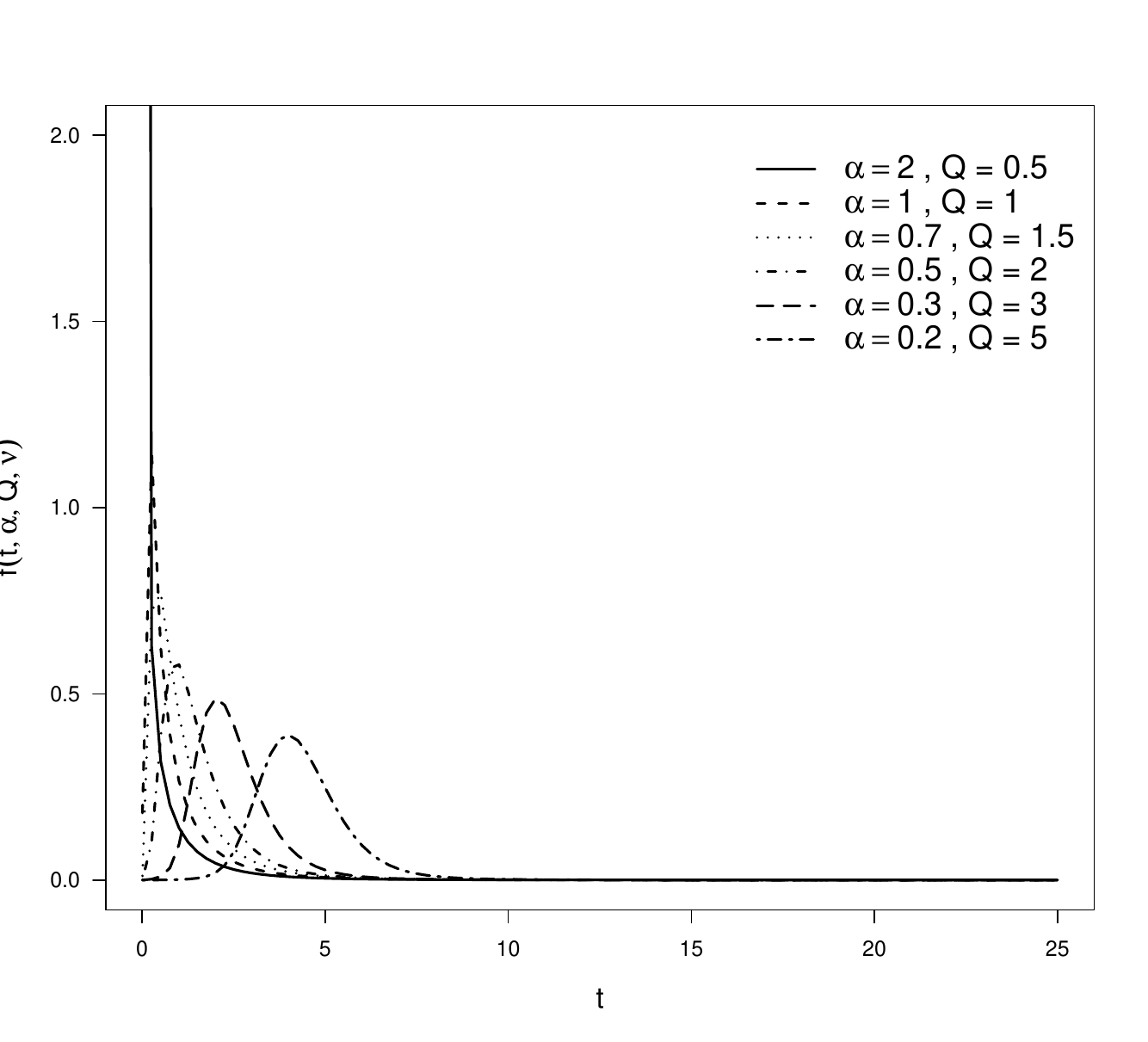}}
	\subfigure[$t_\nu \text{- BS}(\alpha,1,2)$, $q=0.25$]{\includegraphics[scale=0.23]{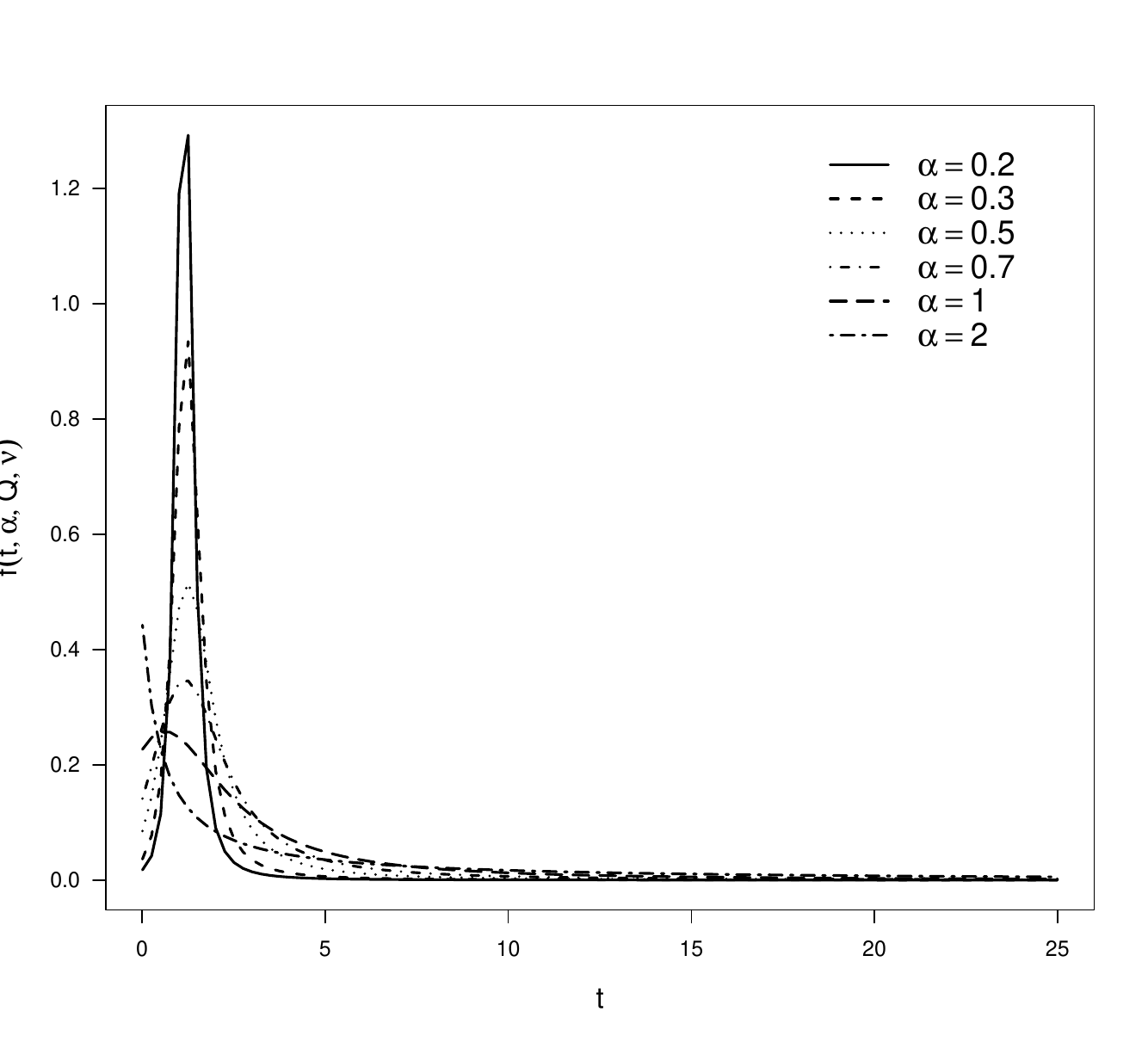}}
	\subfigure[$t_\nu \text{- BS}(1,Q,2)$, $q=0.50$]{\includegraphics[scale=0.23]{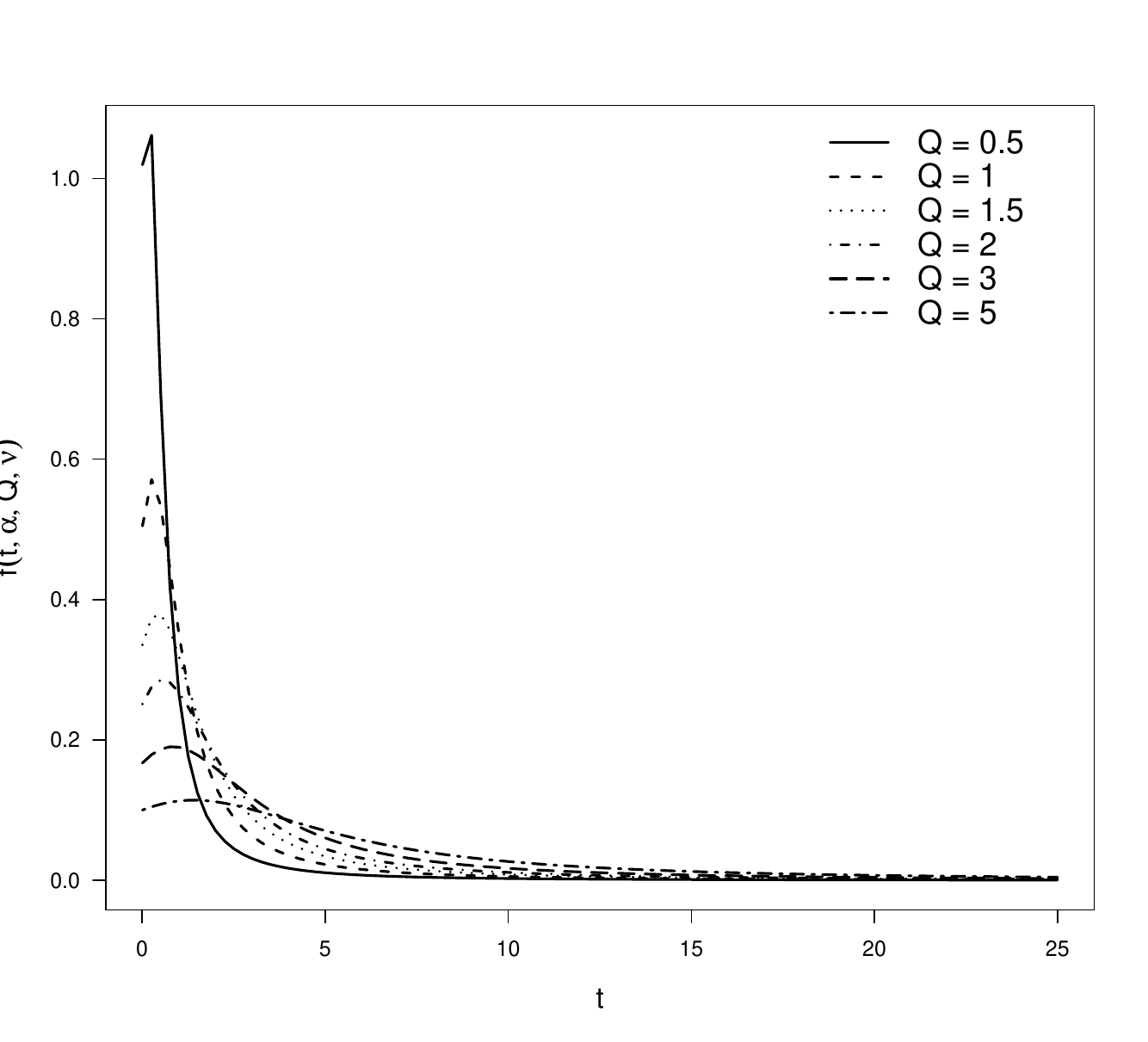}}
	\subfigure[$t_\nu \text{- BS}(\alpha,Q,2)$, $q=0.75$]{\includegraphics[scale=0.23]{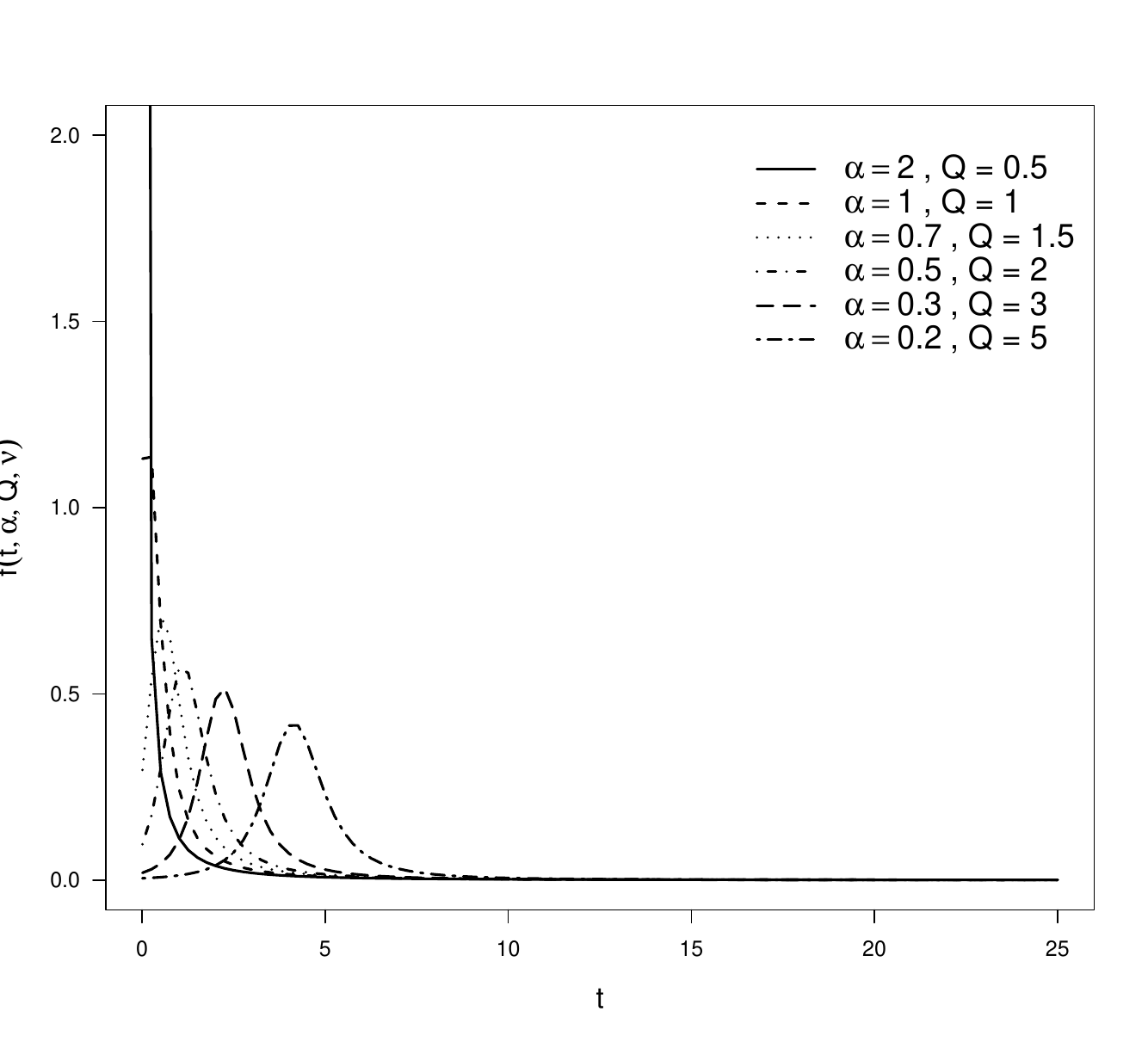}}
	\caption{PDFs for some QSBS family members considering different values of $\alpha$, $Q$ and $q = 0.25, 0.5, 0.75,$ (in the lines) and for the models $t_\nu$-BS, SL-BS and CN-BS (in the columns), with $\bm{\nu} = 2, 2, (0.2,0.2)$, respectively.}
	\label{fig:qsbs}
\end{figure}

Figure \ref{fig:qsbs} displays different shapes of the special cases of the QSBS distribution. We observe that as $q$ increases, the curves become flatter. We also observe that larger the value of $\alpha$, the heavier the tail of the distribution is.

\newpage

\section{QSBS quantile regression model}\label{section:03}

\subsection{Model and EM algorithm}\label{modeloem}

Let ${\bm T }=(T_1,T_2, \ldots,T_n)^{\top}$ be a random sample of size $n$, where $T_i \sim \text{QSBS}(\alpha,Q_i;H)$, for $i = 1,2,\ldots,n$, and let $\bm{t} = (t_1,t_2,\ldots,t_n)^\top$ denote the corresponding realization of $\bm T$. We propose a QSBS quantile regression model with structure for $Q_i$ expressed as
\begin{equation}
h(Q_i) = \eta_i = \bm{x_i^\top \beta}, \; i = 1,2,\ldots,n,
\end{equation}
where $h: \mathbb{R}^+ \rightarrow \mathbb{R}^+$ is a strictly monotonic, positive and at least twice differentiable link function, $Q_i = h^{-1}(\bm{x_i^\top \beta})$, $\bm x_i^\top = (1, x_{i1}, \ldots, x_{ip})$ is the vector of covariates (fixed and known), and $\bm{\beta} = (\beta_0,\beta_1, \ldots,\beta_{p})^\top$ is the corresponding vector of regression coefficients.

 We can implement the EM algorithm \citep{dempster1977} to obtain the ML estimate of $\bm{\theta} = (\alpha,\bm{\beta})^\top$. The parameter $\bm{\nu}$ that indexes the PDF of $h_U(\cdot)$ will be estimated using the profile log-likelihood. Specifically, Let ${\bm t}=(t_{1},\ldots,t_{n})^{\top}$ and ${\bm u}=(u_{1},\ldots,u_{n})^{\top}$ be the observed and missing data, respectively, with ${\bm T }=(T_1,T_2, \ldots,T_n)^{\top}$ and ${\bm U}=(U_1,U_2, \ldots,U_n)^{\top}$ being their corresponding random vectors. Thus, the complete data vector is written as ${\bm y}_{c}=({\bm t}^{\top}, {\bm u}^{\top})^{\top}$. 
{\color{black} From Proposition \ref{prop-TU-conditional},}
\begin{eqnarray}
T_i|U_i = u_i &\overset{\text{ind.}}{\sim}& \text{QBS}(\alpha_u,Q_i), \label{eq:tu}\\
U_i &\overset{\text{ind.}}{\sim}& h_U(u_i), \; i = 1,2,\ldots,n, \label{eq:u}
\end{eqnarray}
where QBS denotes the BS distribution reparameterized by the quantile proposed by \cite{slgs:2020a} and $\alpha_u = \sqrt{g(u)} \alpha$.
The complete data log-likelihood function for the QSBS quantile regression model associated with ${\bm y}_{c}=({\bm t}^{\top}, {\bm u}^{\top})^{\top}$ is given by 
\begin{eqnarray}\label{eq:log:complete:data}
\ell_c(\bm{\theta};\bm{t_c}) &\propto& -n\log(\alpha \gamma_\alpha) - \frac{1}{2} \sum_{i = 1}^{n} \log(Q_i) + \sum_{i = 1}^{n} \log\left( \frac{\gamma_\alpha^2}{2} + \frac{2Q_i}{t_i} \right) \nonumber\\
&& -\frac{2}{(\alpha\gamma_\alpha)^2} \sum_{i = 1}^{n} \frac{1}{g(u_i)} \frac{Q_i}{t_i} \left( \frac{\gamma_\alpha^2 t_i}{4Q_i} - 1 \right)^2.
\end{eqnarray}
By taking the expected value of the complete data log-likelihood function in \eqref{eq:log:complete:data}, conditional on $ \bm U = \bm u$, and letting $\widehat{u}_i = \E[1/g(U_i)|T_i=t_i;\bm{\widehat{\theta}}]$, $i = 1, 2, \ldots, n$, which we obtain from \eqref{eq:meancnbs}, \eqref{eq:meanslbs} and \eqref{eq:meantbs}, we have
\begin{eqnarray} \label{eq:qloglik}
\mathcal{Q}(\bm{\theta};\bm{\widehat{\theta}}) &\propto& -n \log(\alpha\gamma_\alpha) - \frac{1}{2} \sum_{i = 1}^{n} \log(Q_i) + \sum_{i = 1}^{n} \log\left( \gamma_\alpha^2 t_i + 4 Q_i \right) \nonumber\\
&& -\frac{2}{(\alpha\gamma_\alpha)^2} \sum_{i = 1}^{n} \widehat{u}_i \frac{Q_i}{t_i} \left( \frac{\gamma_\alpha^2 t_i}{4Q_i} - 1 \right)^2. 
\end{eqnarray}
Then, the steps to obtain the ML estimates of $\bm{\theta} = (\alpha,\bm{\beta})^\top$ via the EM algorithm are given by 

\paragraph{E-Step.} Compute $\widehat{u}_i$, given $\bm{\theta} = \bm{\widehat{\theta}}$, for $i = 1,2, \ldots, n$.

\paragraph{M-Step.} Update $\bm{\widehat{\theta}}$ by maximizing $\mathcal{Q}(\bm{\theta};\bm{\widehat{\theta}})$ with respect to $\bm \theta$.

The maximization in M-Step can be performed by taking the derivative of Equation \eqref{eq:qloglik} with respect to $\bm{\theta} = (\alpha,\beta_0, \ldots, \beta_{p})$ and then equating it to zero, thereby providing the likelihood equations. The partial derivatives of \eqref{eq:qloglik} with respect to $\bm{\theta}$ are given by
\begin{eqnarray}
\frac{\p \mathcal{Q}(\bm{\theta};\bm{\widehat{\theta}})}{\p \alpha} &=&
-n \left( \frac{\gamma_\alpha + \alpha \gamma'_\alpha}{\alpha \gamma_\alpha} \right) + 2 \gamma_\alpha \gamma'_\alpha \sum_{i=1}^{n} \frac{t_i}{\gamma^2_\alpha t_i + 4Q_i} + \frac{1}{\alpha^3} \sum_{i=1}^{n} \widehat{u}_i \left[ \frac{\gamma^2_\alpha t_i}{4Q_i} + \frac{4Q_i}{\gamma^2_\alpha t_i} - 2  \right] \nonumber \\
&& - \frac{1}{\alpha^2} \sum_{i=1}^{n} \widehat{u}_i \left[ \frac{\gamma_\alpha \gamma'_\alpha t_i}{4Q_i} - \frac{4Q_i \gamma'_\alpha}{\gamma^3_\alpha t_i} \right], \label{eq:s_alpha} \\
\frac{\p \mathcal{Q}(\bm{\theta};\bm{\widehat{\theta}})}{\p \beta_j} &=& \sum_{i=1}^{n} \left\lbrace \left[ -\frac{1}{2Q_i} + \frac{4}{\gamma^2_\alpha t_i + 4Q_i} - \frac{\widehat{u}_i}{\alpha^2} \left( -\frac{\gamma_\alpha^2 t_i}{8Q_i^2} + \frac{2}{\gamma_\alpha^2 t_i} \right)  \right] \left( \frac{\p Q_i}{\p \eta_i} \right) \left( \frac{\p \eta_i}{\p \beta_j} \right) \right\rbrace, \label{eq:s_beta}
\end{eqnarray}
for $j = 0, 1, \ldots, p$, where $\gamma'_\alpha = y_q + \alpha y_q^2/\sqrt{(\alpha y_q)^2 + 4}$ is the first derivative of $ \gamma_\alpha$ with respect to $\alpha$. The system of equations are solved using the BFGS quasi-Newton method; see \cite{mortaza1997,mclachlan2007}. The convergence of the EM algorithm can be established using the following stopping criterion: $|\mathcal{Q}(\bm{\theta}^{(r+1)};\bm{\widehat{\theta}}^ {(r)}) - \mathcal{Q}(\bm{\theta}^{(r)};\bm{\widehat{\theta}}^{(r)})| < \varepsilon$, where $\varepsilon$ is a prespecified tolerance level; see \cite{mclachlan2007}. Starting values are required to initiate the EM procedure, namely, $\widehat{\alpha}^{(0)},\widehat{\beta}_0^{(0)}, \ldots, \widehat{\beta}_{p}^{(0)}$. These can be obtained from the work of \cite{rieck1991}.

A disadvantage of the EM algorithm in relation to Newton-type methods is that we cannot obtain the estimates of the standard errors directly through the Fisher information matrix. There are several approaches proposed to obtain the standard errors of ML estimators; see, for example, \cite{baker1992}, \cite{oakes1999} and \cite{louis1982}, among others.
The approach developed by \cite{louis1982} is based on the missing information principle. In this method, the score function of the incomplete data log-likelihood function is related to the conditional expectation of the complete data log-likelihood function as follows: $S_o(\bm{y};\bm{\theta}) = \E[\p \ell_c(\bm{\theta};\bm{t})/\p\bm{\theta}]$, where $S_o(\bm{y};\bm{\theta}) = \p \ell_o(\bm{\theta};\bm{t})/\p\bm{\theta}$ and $S_c(\bm{y};\bm{\theta}) = \p \ell_c(\bm{\theta};\bm{t})/\p\bm{\theta}$ are the score functions of the incomplete data and complete data, respectively. \cite{meilijson1989} presents a definition of the empirical information matrix as follows:
\begin{equation}
\text{I}_e(\bm{\theta};\bm{t}) = \sum_{i=1}^{n} s(t_i;\bm{\theta}) s^\top(t_i;\bm{\theta}) - \frac{1}{n} S(\bm{t};\bm{\theta}) S^\top(\bm{t};\bm{\theta}), \nonumber
\end{equation}
such that $S(\bm{t};\bm{\theta}) = \sum_{i=1}^{n} s(t_i,\bm{\theta})$ and $s(t_i,\bm{\theta})$ is called the empirical score function of the $i$-th observation, $i = 1,2, \ldots, n$. Replacing $\bm{\theta}$ with its respective ML estimates $\bm{\widehat{\theta}}$, we obtain $S(\bm{t};\bm{\widehat{\theta}} ) = 0$. Then, the empirical information matrix is given by
\begin{equation}
\text{I}_e(\bm{\widehat{\theta}};\bm{t}) = \sum_{i=1}^{n} s(t_i;\bm{\widehat{\theta}})^{(r)} s^\top(t_i;\bm{\widehat{\theta}})^{(r)}, \nonumber
\end{equation}
where $s(t_i;\bm{\widehat{\theta}})^{(r)}$ is the empirical score function obtained in the $r$-th iteration using the EM algorithm, that is,
\begin{equation}
s(t_i;\bm{\widehat{\theta}})^{(r)} = \E[ s(t_i, u_i^{(r)};\bm{\theta}^{(r)})|t_i], \nonumber
\end{equation}
where $u_i^{(r)}$ is the latent variable obtained in the $r$-th iteration with conditional distribution $h_{U|T}(u_i|t_i)$ and the partial derivatives of the complete data log-likelihood function in relation to $\alpha$ and $\bm\beta$ are given, respectively, by \eqref{eq:s_alpha} and \eqref{eq:s_beta}.

By the method of \cite{louis1982}, an approximation to the observed information matrix observed at the $r$-th iteration can be obtained as $\text{I}_e(\bm{\widehat{\theta}};\bm{t})^{(r)} = \sum_{i=1}^{n} s(t_i;\bm{\widehat{\theta}})^{(r)} s^\top(t_i;\bm{\widehat{\theta}})^{(r)}$. Therefore, an approximation of the variance-covariance matrix is obtained by inverting the empirical Fisher information matrix $\text{I}_e(\bm{\widehat{\theta}};\bm{t})^{(r)}$ after convergence.

We estimate the extra parameter $\nu$ (or extra parameter vector $\bm{\nu}$) by using the profile log-likelihood approach. First, we compute the ML estimate of $\bm\theta$ by using the EM algorithm, for each $\nu\in\{a,\ldots,b\}$, where $a$ and $b$ are predefined limits, then the final estimate of $\nu$ is the one that maximizes the log-likelihood function and the associated estimate of $\bm\theta$ is then the final one; see \cite{slnb:21}. The extra parameter  is estimated through the profile log-likelihood approach because it provides robustness to outlying observations under Student-$t$ models. According to \cite{lucas1997}, the robustness only holds if the degree of freedom is fixed rather than directly estimated in the maximization of the log-likelihood function. Moreover, some difficulties in computing the extra parameter can be found in other models.

\subsection{Residual Analysis}

Residuals are important tools to assess goodness-of-fit and departures from the assumptions of the postulated model. In particular, the generalized Cox-Snell (GCS) and randomized quantile (RQ) residuals are widely used for these purposes; see \cite{lee2003}, \cite{dasilva2020} and \cite{ssls:21}. The GCS and RQ residuals are respectively given by 
\begin{equation}
	\widehat{r}_i^{\text{GCS}} = - \log[ 1 - \widehat{\Phi}_{\text{SMN}}(a(t)) ]
	\nonumber
\end{equation}
and
\begin{equation}
	\widehat{r}_i^{\text{RQ}} = \Phi^{-1}[ \Phi_{\text{SMN}}(a(t)) ],
	 \nonumber
\end{equation}
for $i = 1, 2, \ldots, n,$ where $\widehat{\Phi}_{\text{SMN}}(a(t))$ is the CDF of $T \sim \text{QSBS}(\widehat{\alpha}, \bm{\widehat{\beta}}, \bm{\nu})$ fitted to the data, and $\Phi(\cdot)^{-1}$ is the standard normal quantile function. These residuals have asymptotically standard exponential and standard normal distributions when the model is correctly specified. For both residuals, the distributional assumption can be verified using graphical techniques, hypothesis tests and descriptive statistics.

\subsection{Hypothesis testing}

We consider here the Wald, score, likelihood ratio and gradient statistical tests for the QSBS quantile regression model; see \cite{terrell2002} and \cite{ssls:21,slnb:21}. Consider ${\bm \theta}$ to be a $p$-dimensional vector of parameters that index a QSBS quantile regression model. Assume our interest lies in testing the hypothesis ${\cal{H}}_{0}: {\bm \theta}_{1}={\bm \theta}^{(0)}_{1}$
against ${\cal{H}}_{1}: {\bm \theta}_{1}\neq{\bm \theta}^{(0)}_{1}$,
where ${\bm \theta}=({\bm \theta}^{\top}_{1},{\bm \theta}^{\top}_{2})^{\top}$, ${\bm \theta}_{1}$ 
is a $r \times 1$ vector of parameters of interest and ${\bm \theta}_{2}$ is {a} $(p-r) \times 1$ vector of 
nuisance parameters. Then, the Wald (W), score (R), likelihood ratio (LR) and gradient (T) statistics are given, respectively, by
\begin{eqnarray}
	S_W &=& (\bm{\widehat{\theta}} - \widetilde{\bm{\theta}} )^\top \mathcal{J}(\bm{\widehat{\theta}}) (\bm{\widehat{\theta}} - \widetilde{\bm{\theta}}), \nonumber\\	
	S_R &=& S(\widetilde{\bm{\theta}})^\top \mathcal{J}(\widetilde{\bm{\theta}})^{-1} S(\widetilde{\bm{\theta}}), \nonumber\\
	S_{LR} &=& - 2 [\ell(\widetilde{\bm{\theta}}) - \ell(\bm{\widehat{\theta}})], \nonumber\\
	S_T &=& S(\widetilde{\bm{\theta}})^\top (\bm{\widehat{\theta}} - \widetilde{\bm{\theta}}), \nonumber
\end{eqnarray}
where $\ell(\cdot)$ is the log-likelihood function, $S(\cdot)$ is the score function, $\mathcal{J}(\cdot)$ is the Fisher information matrix, and $\widehat{\bm \theta}=(\widehat{\bm \theta}^{\top}_{1},\widehat{\bm \theta}^{\top}_{2})^{\top}$ and
$\widetilde{\bm \theta}=({\bm \theta}^{(0)\top}_{1},\widetilde{\bm \theta}^{\top}_{2})^{\top}$
are the unrestricted and restricted ML estimators of ${\bm \theta}$, respectively. In particular, we use the EM algorithm to obtain the ML estimates. The log-likelihood function in the $S_{LR}$ statistic is replaced by the expected value of the complete data log-likelihood function. Moreover, for the $S_W$ , $S_R$ and $S_T$ statistics, the score function and Fisher information matrix are approximated using their respective empirical versions presented in Subsection \ref{modeloem}. In regular cases, we have, under ${\cal{H}}_{0}$ and as $n\rightarrow\infty$, the Wald, score, likelihood ratio and gradient statistical tests converging in distribution to $\chi^{2}_{r}$. We then reject ${\cal{H}}_{0}$ at nominal 
level $\alpha$ if the test statistic is larger than $\chi^{2}_{1-\alpha,r}$, the upper $\alpha$ quantile of the $\chi^{2}_{r}$ distribution.

\section{Monte Carlo simulation}\label{section:04}

In this section, the results from two Monte Carlo simulation studies are presented. In the first study, we evaluate the performances of the EM algorithm for ML estimation and residuals. In the second study, we evaluate
the performances of the aforementioned statistical tests. The simulations were performed using the \texttt{R} software; see \cite{rteam:20}. The simulation scenario considers sample size $n = \{50, 100, 200\}$ and quantiles $q =\{0.25, 0.5, 0.75\}$, with a logarithmic link function for $Q_i$, and 5,000 Monte Carlo replications.

\subsection{ML estimation and model selection}
In this study, the data generating model has two covariates and is given by
\begin{equation}\label{model:simu}
	\log(Q_i) = \beta_0 + \beta_1 x_{1i} + \beta_2 x_{2i}, \, i = 1,\ldots, n,
\end{equation}
where $x_{1}$ and $x_{2}$ are covariates obtained from a uniform distribution in the interval (0,1), $\beta_0 = 2.5$, $\beta_1 = 3$ and $\beta_2 = 0.9$. Moreover, the simulation scenario considers shape parameters $\alpha = 0.2, 0.5, 1.0$ and extra parameters $\bm{\nu} = (0.1,0.3)$ (CN-BS), $\nu = 4$ (SL-BS)  and $\nu = 11$ ($t_\nu$-BS). The response observations, $t_1,\ldots,t_n$, are then generated by using \eqref{eq:qsbs} and \eqref{model:simu}.

We evaluate the performance of the EM algorithm for obtaining the ML estimates using the bias, mean square error (MSE) and coverage probability (CP) of 95\% asymptotic confidence interval. The Monte Carlo estimates of these quantities are given, respectively, by
\begin{eqnarray*}
	\text{Bias}(\widehat{\varphi}) &=& \frac{1}{M} \sum_{r = 1}^{M} (\widehat{\varphi}^{(r)} - \varphi), \, \\
	\text{MSE}(\widehat{\varphi}) &=& \frac{1}{M} \sum_{r = 1}^{M} (\widehat{\varphi}^{(r)} - \varphi)^2, \, \\
		\text{CP}(\widehat{\varphi}) &=& \frac{1}{M} \sum_{r = 1}^{M} I(\varphi \in [\widehat{\varphi}_L^{(r)}, \widehat{\varphi}_U^{(r)}]), \nonumber
\end{eqnarray*}
where $M$ is the number of Monte Carlo replications, $\varphi$ is the true parameter, $\widehat{\varphi}^{(r)}$ is the $r$-th ML estimate of $\varphi$, and $I(\cdot)$ is an indicator function of $\varphi$ belonging to the $r$-th asymptotic interval $[\widehat{\varphi}_L^{(r)}, \widehat{\varphi} _U^{(r)}]$ with $\widehat{\varphi}_L^{(r)}$ and $\widehat{\varphi}_U^{(r)}$ denoting the lower and upper bounds, respectively.

In addition to analyzing the performance of the EM algorithm, this Monte Carlo simulation study also investigates the empirical distribution of the GCS and RQ residuals, which are assessed by the empirical mean (MN), median (MD), standard deviation (SD), coefficient of skewness (CS) and coefficient of (excess) kurtosis (CK).

{Table \ref{tab:sim-slbs} reports the empirical bias, MSE and CP for the quantile regression models based on the SL-BS distributions (due to space limitations we do not present the CN-BS and $t_\nu$-BS results). A look at the results in Table \ref{tab:sim-slbs} allows us to conclude that as the sample size increases the bias and MSE both decreases, as one would expect. Moreover, these quantities increase as $\alpha$ increases. Finally, the CP approaches the 95\% nominal level as the sample size increases.}

Tables \ref{tab:gcs-res}-\ref{tab:rq-res} present the empirical MN, MD, SD, CS and CK. Note that these values are expected to be 1, 0.69, 1, 2 and 6, respectively, for the GCS residuals, and  0, 0, 1, 0 and 0, respectively, for the RQ residuals.
From Tables \ref{tab:gcs-res}-\ref{tab:rq-res}, note that for the CN-BS and $t_\nu$-BS cases, as the sample size increases the values of the empirical MN, MD, SD, CS and CK approaches these values of the reference EXP(1) and N(0,1) distributions. Therefore, the GCS and RQ residuals conform well with the reference distributions for the CN-BS and $t_\nu$-BS cases. Nevertheless, in the results of the SL-BS model the considered residuals do not conform well with the reference distributions.

\begin{table}[H]
	\centering
	\caption{Empirical bias, MSE and CP from simulated data for the indicated ML estimates of the SL-BS quantile regression model ($\nu =4$).}
	\adjustbox{max height=\dimexpr\textheight-3.5cm\relax,
		max width=\textwidth}{
		\begin{tabular}{lccccccccccccc}
			\toprule
			& & & \multicolumn{3}{c}{$n = 50$} & & \multicolumn{3}{c}{$n = 100$} & & \multicolumn{3}{c}{$n = 200$} \\
			\cline{4-6} \cline{8-10} \cline{12-14}
			$\alpha$ & $q$ & & Bias & MSE & CP & & Bias & MSE & CP & & Bias & MSE & CP \\
			\hline
			\multirow{13}{*}{0.2} & \multirow{4}{*}{0.25} & $\widehat{\alpha}$ & -0.0063 & 0.0005 & 0.9536 && -0.0035 & 0.0002 & 0.9502 && -0.0016 & 0.0001 & 0.9460 \\
			& & $\widehat{\beta}_0$ & 0.0035 & 0.0067 & 0.9258 &&  0.0041 & 0.0051 & 0.9338 && 0.0007 & 0.0018 & 0.9452 \\
			& & $\widehat{\beta}_1$ & 0.0010 & 0.0105 & 0.9478 && -0.0014 & 0.0067 & 0.9472 && -0.0003 & 0.0033 & 0.9448 \\
			& & $\widehat{\beta}_2$ & 0.0008 & 0.0131 & 0.9516 && -0.0015 & 0.0061 & 0.9462 && 0.0008 & 0.0031 & 0.9504 \\
			\cline{2-14}
			& \multirow{4}{*}{0.5} & $\widehat{\alpha}$ & -0.0072 & 0.0005 & 0.9542 && -0.0033 & 0.0002 & 0.9498 && -0.0019 & 0.0001 & 0.9530 \\
			& & $\widehat{\beta}_0$ & -0.0013 & 0.0067 & 0.9198 && -0.0006 & 0.0037 & 0.9382 && -0.0003 & 0.0023 & 0.9422 \\
			& & $\widehat{\beta}_1$ & 0.0035 & 0.0118 & 0.9456 && 0.0016 & 0.0067 & 0.9500 && -0.0001 & 0.0037 & 0.9506 \\
			& & $\widehat{\beta}_2$ & -0.0001 & 0.0117 & 0.9468 && 0.0000 & 0.0068 & 0.9516 && 0.0008 & 0.0033 & 0.9472 \\
			\cline{2-14}
			& \multirow{4}{*}{0.75} & $\widehat{\alpha}$ & -0.0071 & 0.0005 & 0.9470 && -0.0036 & 0.0002 & 0.9522 && -0.0017 & 0.0001 & 0.9540 \\
			& & $\widehat{\beta}_0$ & -0.0041 & 0.0107 & 0.9212 && -0.0038 & 0.0033 & 0.9316 && -0.0013 & 0.0018 & 0.9418 \\
			& & $\widehat{\beta}_1$ & -0.0017 & 0.0176 & 0.9496 && 0.0003 & 0.0064 & 0.9470 && -0.0002 & 0.0031 & 0.9522 \\
			& & $\widehat{\beta}_2$ & -0.0020 & 0.0140 & 0.9498 && 0.0005 & 0.0067 & 0.9486 && -0.0006 & 0.0031 & 0.9554 \\
			\hline
			\multirow{13}{*}{0.5} & \multirow{4}{*}{0.25} & $\widehat{\alpha}$ & -0.0180 & 0.0031 & 0.9486 && -0.0096 & 0.0015 & 0.9514 && -0.0044 & 0.0007 & 0.9504 \\
			& & $\widehat{\beta}_0$ & 0.0158 & 0.0421 & 0.9244 && 0.0074 & 0.0228 & 0.9346 && 0.0039 & 0.0108 & 0.9392 \\
			& & $\widehat{\beta}_1$ & -0.0068 & 0.0647 & 0.9404 && 0.0012 & 0.0327 & 0.9464 && -0.0007 & 0.0178 & 0.9506 \\
			& & $\widehat{\beta}_2$ & 0.0034 & 0.0734 & 0.9568 && 0.0001 & 0.0382 & 0.9494 && 0.0012 & 0.0193 & 0.9484 \\
			\cline{2-14}
			& \multirow{4}{*}{0.5} & $\widehat{\alpha}$ & -0.0183 & 0.0031 & 0.9538 && -0.0083 & 0.0015 & 0.9508 && -0.0043 & 0.0007 & 0.9552 \\
			& & $\widehat{\beta}_0$ & -0.0013 & 0.0459 & 0.9170 &&  0.0004 & 0.0282 & 0.9356 && 0.0004 & 0.0119 & 0.9354 \\
			& & $\widehat{\beta}_1$ & -0.0006 & 0.0813 & 0.9506 && -0.0009 & 0.0383 & 0.9530 && -0.0004 & 0.0208 & 0.9520 \\
			& & $\widehat{\beta}_2$ & -0.0015 & 0.0649 & 0.9492  && 0.0002 & 0.0427 & 0.9548 && -0.0001 & 0.0180 & 0.9446 \\ 
			\cline{2-14}
			& \multirow{4}{*}{0.75} & $\widehat{\alpha}$ & -0.0182 & 0.0030 & 0.9460 && -0.0087 & 0.0014 & 0.9508 && -0.0049 & 0.0007 & 0.9516 \\
			& & $\widehat{\beta}_0$ & -0.0122 & 0.0671 & 0.9264 && -0.0075 & 0.0264 & 0.9378 && -0.0048 & 0.0108 & 0.9432 \\
			& & $\widehat{\beta}_1$ & 0.0009 & 0.0893 & 0.9442 && 0.0006 & 0.0456 & 0.9514 && -0.0015 & 0.0164 & 0.9522 \\
			& & $\widehat{\beta}_2$ & -0.0047 & 0.0767 & 0.9504 && 0.0030 & 0.0353 & 0.9466 && 0.0029 & 0.0190 & 0.9484 \\
			\hline
			\multirow{13}{*}{1} & \multirow{4}{*}{0.25} & $\widehat{\alpha}$ & -0.0414 & 0.0127 & 0.9616 && -0.0206 & 0.0061 & 0.9580 && -0.0103 & 0.0029 & 0.9530 \\
			& & $\widehat{\beta}_0$ & 0.0278 & 0.1608 & 0.9196 && 0.0120 & 0.0731 & 0.9284 && 0.0081 & 0.0361 & 0.9442 \\
			& & $\widehat{\beta}_1$ & 0.0031 & 0.2456 & 0.9554 && 0.0035 & 0.1498 & 0.9566 && -0.0028 & 0.0620 & 0.9524 \\
			& & $\widehat{\beta}_2$ & 0.0041 & 0.2423 & 0.9606 && 0.0039 & 0.1439 & 0.9580 && 0.0017 & 0.0641 & 0.9536 \\
			\cline{2-14}
			& \multirow{4}{*}{0.5} & $\widehat{\alpha}$ & -0.0372 & 0.0129 & 0.9680 && -0.0177 & 0.0057 & 0.9580 && -0.0086 & 0.0029 & 0.9536 \\
			& & $\widehat{\beta}_0$ & -0.0057 & 0.1361 & 0.9200 && -0.0002 & 0.0736 & 0.9388 && 0.0025 & 0.0385 & 0.9394 \\
			& & $\widehat{\beta}_1$ & -0.0088 & 0.2754 & 0.9582 && -0.0013 & 0.1190 & 0.9532 && -0.0064 & 0.0595 & 0.9524 \\
			& & $\widehat{\beta}_2$ & 0.0192 & 0.2526 & 0.9618 && 0.0037 & 0.1142 & 0.9554 && 0.0013 & 0.0621 & 0.9528 \\
			\cline{2-14}
			& \multirow{4}{*}{0.75} & $\widehat{\alpha}$ & -0.0398 & 0.0130 & 0.9634 && -0.0170 & 0.0059 & 0.9544 && -0.0091 & 0.0029 & 0.9528 \\
			& & $\widehat{\beta}_0$ & -0.0289 & 0.1368 & 0.9158 && -0.0088 & 0.0708 & 0.9372 && -0.0096 & 0.0337 & 0.9446 \\
			& & $\widehat{\beta}_1$ & -0.0062 & 0.2156 & 0.9482 && -0.0076 & 0.1237 & 0.9532 && -0.0001 & 0.0592 & 0.9492 \\
			& & $\widehat{\beta}_2$ & 0.0062 & 0.2486 & 0.9608 && -0.0019 & 0.1315 & 0.9524 && 0.0053 & 0.0575 & 0.9542 \\
			\bottomrule
		\end{tabular}
	}
\label{tab:sim-slbs}
\end{table}

\begin{table}[H]
	\centering
	\caption{Summary statistics for the GCS residuals.}
	\adjustbox{max height=\dimexpr\textheight-3.5cm\relax,
		max width=\textwidth}{
		\begin{tabular}{lccccccccccccc}
			\toprule
			& & & \multicolumn{3}{c}{$n = 50$} & & \multicolumn{3}{c}{$n = 100$} & & \multicolumn{3}{c}{$n = 200$} \\
			\cline{4-6} \cline{8-10} \cline{12-14}
			$\alpha$ & $q$ & & CN-BS & SL-BS & $t_\nu$-BS & & CN-BS & SL-BS & $t_\nu$-BS & & CN-BS & SL-BS & $t_\nu$-BS \\
			\hline
			\multirow{15}{*}{0.2} & \multirow{5}{*}{0.25} & MN & 1.0007 & 0.2249 & 1.0002 && 1.0003 & 3.4373 & 1.0000 && 1.0002 & 2.7172 & 1.0001 \\
			& & MD & 0.6980 & 0.0000 & 0.6964 && 0.6956 & 0.7111 & 0.6952 && 0.6941 & 0.2370 & 0.6940 \\
			& & SD & 1.0008 & 0.7546 & 1.0007 && 0.9994 & 4.4814 & 0.9993 && 1.0000 & 3.8889 & 0.9998 \\
			& & CS & 1.6378 & 4.0803 & 1.6428 && 1.7848 & 1.0108 & 1.7883 && 1.8767 & 1.2936 & 1.8774 \\
			& & CK & 3.1251 & 17.7753 & 3.1549 && 4.1353 & -0.4425 & 4.1577 && 4.8594 & 0.4671 & 4.8577 \\
			\cline{2-14}
			& \multirow{5}{*}{0.5} & MN & 1.0004 & 6.9641 & 0.9999 && 1.0003 & 1.1698 & 1.0000 && 1.0002 & 2.8319 & 1.0001 \\
			& & MD & 0.6974 & 4.5335 & 0.6968 && 0.6956 & 0.0004 & 0.6952 && 0.6941 & 0.5567 & 0.6940 \\
			& & SD & 0.9997 & 6.6111 & 0.9995 && 0.9994 & 2.5135 & 0.9993 && 1.0000 & 3.8723 & 0.9999 \\
			& & CS & 1.6322 & 0.5019 & 1.6373 && 1.7848 & 2.3886 & 1.7883 && 1.8767 & 1.2494 & 1.8774 \\
			& & CK & 3.0959 & -1.2344 & 3.1217 && 4.1354 & 4.9818 & 4.1578 && 4.8594 & 0.3308 & 4.8576 \\
			\cline{2-14}
			& \multirow{5}{*}{0.75} & MN & 1.0004 & 3.4180 & 0.9999 && 1.0003 & 5.4234 & 1.0000 && 1.0002 & 13.7272 & 1.0001 \\
			& & MD & 0.6974 & 1.1057 & 0.6968 && 0.6956 & 3.1935 & 0.6952 && 0.6941 & 14.9827 & 0.6940 \\
			& & SD & 0.9997 & 4.3039 & 0.9995 && 0.9994 & 5.6753 & 0.9993 && 1.0000 & 6.0437 & 0.9999 \\
			& & CS & 1.6322 & 0.9856 & 1.6373 && 1.7849 & 0.5817 & 1.7883 && 1.8768 & -0.5336 & 1.8774 \\
			& & CK & 3.0961 & -0.4622 & 3.1218 && 4.1356 & -1.1559 & 4.1579 && 4.8596 & -0.6473 & 4.8578 \\
			\hline
			\multirow{15}{*}{0.5} & \multirow{5}{*}{0.25} & MN & 1.0002 & 6.2318 & 0.9996 && 1.0001 & 5.0268 & 0.9998 && 1.0001 & 1.0149 & 1.0000 \\
			& & MD & 0.6976 & 5.4538 & 0.6970 && 0.6957 & 3.8996 & 0.6953 && 0.6941 & 0.2530 & 0.6939 \\
			& & SD & 0.9978 & 4.8210 & 0.9976 && 0.9984 & 4.3076 & 0.9983 && 0.9995 & 1.6239 & 0.9993 \\
			& & CS & 1.6229 & 0.4127 & 1.6271 && 1.7793 & 0.6738 & 1.7821 && 1.8734 & 2.3450 & 1.8738 \\
			& & CK & 3.0526 & -1.0874 & 3.0718 && 4.1048 & -0.6312 & 4.1224 && 4.8365 & 6.2482 & 4.8344 \\
			\cline{2-14}
			& \multirow{5}{*}{0.5} & MN & 1.0001 & 3.4666 & 0.9996 && 1.0001 & 3.4666 & 0.9998 && 1.0001 & 3.2460 & 1.0000 \\
			& & MD & 0.6976 & 2.4009 & 0.6970 && 0.6957 & 2.4009 & 0.6953 && 0.6941 & 2.0450 & 0.6939 \\
			& & SD & 0.9978 & 3.3556 & 0.9976 && 0.9984 & 3.3556 & 0.9983 && 0.9995 & 3.3414 & 0.9993 \\
			& & CS & 1.6229 & 0.9513 & 1.6271 && 1.7793 & 0.9513 & 1.7821 && 1.8734 & 1.1899 & 1.8738 \\
			& & CK & 3.0526 & 0.0021 & 3.0717 && 4.1049 & 0.0021 & 4.1224 && 4.8366 & 0.6885 & 4.8344 \\
			\cline{2-14}
			& \multirow{5}{*}{0.75} & MN & 1.0001 & 1.3331 & 0.9996 && 1.0001 & 0.7800 & 0.9998 && 1.0001 & 2.4466 & 1.0000 \\
			& & MD & 0.6976 & 0.4540 & 0.6970 && 0.6957 & 0.2002 & 0.6953 && 0.6941 & 1.0708 & 0.6939 \\
			& & SD & 0.9978 & 1.9673 & 0.9976 && 0.9984 & 1.3305 & 0.9983 && 0.9995 & 3.0264 & 0.9993 \\
			& & CS & 1.6229 & 1.9690 & 1.6271 && 1.7794 & 2.6832 & 1.7822 && 1.8734 & 1.4214 & 1.8738 \\
			& & CK & 3.0527 & 3.8322 & 3.0718 && 4.1049 & 8.5732 & 4.1224 && 4.8366 & 1.2562 & 4.8345 \\
			\hline
			\multirow{15}{*}{1} & \multirow{5}{*}{0.25} & MN & 0.9993 & 1.7884 & 0.9989 && 0.9997 & 0.8007 & 0.9994 && 0.9997 & 2.6806 & 0.9997 \\
			& & MD & 0.6986 & 1.0537 & 0.6979 && 0.6961 & 0.3817 & 0.6957 && 0.6942 & 1.7059 & 0.6941 \\
			& & SD & 0.9921 & 2.0649 & 0.9918 && 0.9956 & 1.1435 & 0.9953 && 0.9978 & 2.7691 & 0.9976 \\
			& & CS & 1.5961 & 1.7839 & 1.5976 && 1.7637 & 2.6123 &  1.7642 && 1.8639 & 1.6326 & 1.8633 \\
			& & CK & 2.9246 & 3.3013 & 2.9240 && 4.0165 & 8.8216 & 4.0183 && 4.7733 & 2.5941 & 4.7663 \\
			\cline{2-14}
			& \multirow{5}{*}{0.5} & MN & 0.9993 & 1.3375 & 0.9989 && 0.9997 & 1.4969 & 0.9994 && 0.9997 & 0.6229 & 0.9997 \\
			& & MD & 0.6986 & 0.7459 & 0.6979 && 0.6961 & 0.8467 & 0.6957 && 0.6942 & 0.2700 & 0.6941 \\
			& & SD & 0.9921 & 1.6592 & 0.9918 && 0.9956 & 1.8269 & 0.9953 && 0.9978 & 0.9235 & 0.9976 \\
			& & CS & 1.5961 & 1.9805 & 1.5976 && 1.7637 & 2.0812 & 1.7642 && 1.8639 & 2.7777 & 1.8633 \\
			& & CK & 2.9246 & 4.4033 & 2.9239 && 4.0165 & 5.0290 & 4.0183 && 4.7733 & 10.5620 & 4.7663 \\
			\cline{2-14}
			& \multirow{5}{*}{0.75} & MN & 0.9993 & 1.9122 & 0.9989 && 0.9997 & 3.4851 & 0.9994 && 0.9997 & 0.9887 & 0.9997 \\
			& & MD & 0.6986 & 1.0797 & 0.6979 && 0.6961 & 2.3979 & 0.6957 && 0.6942 & 0.4844 & 0.6941 \\
			& & SD & 0.9921 & 2.2636 & 0.9918 && 0.9956 & 3.2461 & 0.9953 && 0.9978 & 1.3787 & 0.9976 \\
			& & CS & 1.5961 & 1.7622 & 1.5976 && 1.7637 & 1.2772 & 1.7642 && 1.8639 & 2.6184 & 1.8633 \\
			& & CK & 2.9247 & 3.0441 & 2.9240 && 4.0165 & 1.1415 & 4.0184 && 4.7733 & 8.8776 & 4.7663 \\
			\bottomrule
		\end{tabular}
	}
\label{tab:gcs-res}
\end{table}

\begin{table}[H]
	\centering
	\caption{Summary statistics for the RQ residuals.}
	\adjustbox{max height=\dimexpr\textheight-3.5cm\relax,
		max width=\textwidth}{
		\begin{tabular}{lccccccccccccc}
			\toprule
			& & & \multicolumn{3}{c}{$n = 50$} & & \multicolumn{3}{c}{$n = 100$} & & \multicolumn{3}{c}{$n = 200$} \\
			\cline{4-6} \cline{8-10} \cline{12-14}
			$\alpha$ & $q$ & & CN-BS & SL-BS & $t_\nu$-BS & & CN-BS & SL-BS & $t_\nu$-BS & & CN-BS & SL-BS & $t_\nu$-BS \\
			\hline
			\multirow{15}{*}{0.2} & \multirow{5}{*}{0.25} & MN & -0.0007 & 3.6621 & -0.0003 && -0.0004 & -0.0222 & -0.0001 && -0.0002 & 0.5812 & -0.0002 \\
			& & MD & -0.0018 & 4.4962 & -0.0002 && -0.0010 & 0.0091 & -0.0007 && -0.0002 & 0.8356 & -0.0001 \\
			& & SD & 1.0093 & 2.1911 & 1.0093 && 1.0045 & 3.1032 & 1.0044 && 1.0022 & 3.1333 & 1.0021 \\
			& & CS & -0.0054 & -0.9419 & -0.0057 && -0.0023 &  -0.0018 & -0.0028 && -0.0035 & -0.1484 & -0.0033 \\
			& & CK & -0.2145 & -0.1194 & -0.2044 && -0.1149 & -1.4337 & -0.1129 && -0.0611 & -1.4539 & -0.0606 \\
			\cline{2-14}
			& \multirow{5}{*}{0.5} & MN & -0.0006 & -2.1173 & -0.0001 && -0.0004 & 2.2921 & -0.0001 && -0.0002 & 0.2326 & -0.0002 \\
			& & MD & -0.0012 & -2.2758 & -0.0008 && -0.0010 & 3.4442 & -0.0007 && -0.0002 & 0.2018 & -0.0001 \\
			& & SD & 1.0092 & 2.6909 & 1.0092 && 1.0045 & 2.8829 & 1.0044 && 1.0022 & 2.9603 & 1.0021 \\
			& & CS & -0.0027 & 0.3635 & -0.0030 && -0.0023 &  -0.7669 & -0.0028 && -0.0035 & -0.0082 & -0.0033 \\
			& & CK & -0.2150 & -1.0596 & -0.2074 && -0.1149 & -0.7860 & -0.1129 && -0.0611 & -1.3435 & -0.0606 \\
			\cline{2-14}
			& \multirow{5}{*}{0.75} & MN & -0.0006 & -0.2725 & -0.0001 && -0.0004 & -1.1287 & -0.0001 && -0.0002 & -4.5000 & -0.0002 \\
			& & MD & -0.0012 & -0.4018 & -0.0008 && -0.0010 & -1.7211 & -0.0007 && -0.0002 & -4.9832 & -0.0001 \\
			& & SD & 1.0092 & 2.8834 & 1.0092 && 1.0045 & 3.1633 & 1.0044 && 1.0022 & 1.5547 & 1.0021 \\
			& & CS & -0.0027 & 0.0591 & -0.0030 && -0.0023 & 0.3594 & -0.0029 && -0.0035 & 1.2891 & -0.0033 \\
			& & CK & -0.2149 & -1.3973 & -0.2074 && -0.1149 & -1.2754 & -0.1129 && -0.0611 & 1.2414 & -0.0606 \\
			\hline
			\multirow{15}{*}{0.5} & \multirow{5}{*}{0.25} & MN & -0.0006 & -2.3923 & -0.0001 && -0.0004 & -1.9300 & -0.0001 && -0.0001 & 0.7844 & -0.0001 \\
			& & MD & -0.0012 & -2.5996 & -0.0007 && -0.0010 & -2.0355 & -0.0006 && -0.0001 & 0.7682 & 0.0000 \\
			& & SD & 1.0088 & 1.8164 & 1.0088 && 1.0042 & 1.8304 & 1.0042 && 1.0020 & 1.8929 & 1.0020 \\
			& & CS & -0.0021 & 0.3022 & -0.0023 && -0.0023 & 0.2409 & -0.0027 && -0.0035 & -0.0576 & -0.0033 \\
			& & CK & -0.2258 & -1.0039 & -0.2194 && -0.1212 & -0.8613 & -0.1196 && -0.0649 & -0.7816 & -0.0645 \\
			\cline{2-14}
			& \multirow{5}{*}{0.5} & MN & -0.0006 & -0.2474 & -0.0001 && -0.0004 & -1.2189 & -0.0001 && -0.0001 & -1.1277 & -0.0000 \\
			& & MD & -0.0012 & -0.3091 & -0.0007 && -0.0010 & -1.3267 & -0.0006 && -0.0001 & -1.1230 & 0.0000 \\
			& & SD & 1.0088 & 1.9014 & 1.0088 && 1.0042 & 1.8326 & 1.0042 && 1.0020 & 1.7712 & 1.0020 \\
			& & CS & -0.0021 & 0.1187 & -0.0023 && -0.0023 &  0.2699 & -0.0027 && -0.0035 & 0.0708 & -0.0033 \\
			& & CK & -0.2258 & -0.7672 & -0.2194 && -0.1212 & -0.6692 & -0.1196 && -0.0649 & -0.6857 & -0.0645 \\
			\cline{2-14}
			& \multirow{5}{*}{0.75} & MN & -0.0006 & 0.3435 & -0.0001 && -0.0004 & 0.9719 & -0.0001 && -0.0001 & -0.4675 & -0.0000 \\
			& & MD & -0.0012 & 0.3910 & -0.0007 && -0.0010 & 0.9259 & -0.0006 && -0.0001 & -0.3988 & 0.0000 \\
			& & SD & 1.0088 & 1.8131 & 1.0088 && 1.0042 & 1.7567 & 1.0042 && 1.0020 & 1.9726 & 1.0020 \\
			& & CS & -0.0021 & -0.1176 & -0.0023 && -0.0023 & -0.0706 & -0.0027 && -0.0035 & -0.0207 & -0.0033 \\
			& & CK & -0.2258 & -0.7264 & -0.2194 && -0.1212 & -0.6657 & -0.1196 && -0.0649 & -0.8315 & -0.0645 \\
			\hline
			\multirow{15}{*}{1} & \multirow{5}{*}{0.25} & MN & -0.0007 & -0.4018 & -0.0002 && -0.0004 & 0.5441 & -0.0000 && 0.0000 & -1.0096 & 0.0001 \\
			& & MD & -0.0012 & -0.3610 & -0.0008 && -0.0009 & 0.4899 & -0.0007 && 0.0001 & -0.9013 & 0.0002 \\
			& & SD & 1.0071 & 1.4377 & 1.0074 && 1.0034 & 1.3864 & 1.0035 && 1.0016 & 1.4481 & 1.0016 \\
			& & CS & -0.0004 & -0.0746 & -0.0004 && -0.0020 &  0.1054 & -0.0022 && -0.0034 & -0.2055 & -0.0032 \\
			& & CK & -0.2586 & -0.2157 & -0.2562 && -0.1399 & -0.0067 & -0.1402 && -0.0758 & -0.2410 & -0.0763 \\
			\cline{2-14}
			& \multirow{5}{*}{0.5} & MN & -0.0007 & -0.0396 & -0.0002 && -0.0004 & -0.1923 & -0.0000 && 0.0000 & 0.8391 & 0.0001 \\
			& & MD & -0.0012 & -0.0447 & -0.0008 && -0.0009 & -0.1726 & -0.0007 && 0.0001 & 0.7316 & 0.0002 \\
			& & SD & 1.0071 & 1.4219 & 1.0074 && 1.0034 & 1.4142 & 1.0035 && 1.0016 & 1.4253 & 1.0016 \\
			& & CS & -0.0004 & 0.0318 & -0.0004 && -0.0020 &  -0.0342 & -0.0022 && -0.0034 & 0.2282 & -0.0032 \\
			& & CK & -0.2587 & -0.0933 & -0.2562 && -0.1399 & -0.0014 & -0.1402 && -0.0758 & -0.1461 & -0.0763 \\
			\cline{2-14}
			& \multirow{5}{*}{0.75} & MN & -0.0007 & -0.4664 & -0.0002 && -0.0004 & -1.4258 & -0.0000 && 0.0000 & 0.3369 & 0.0001 \\
			& & MD & -0.0012 & -0.4011 & -0.0008 && -0.0009 & -1.3223 & -0.0007 && 0.0001 & 0.2991 & 0.0002 \\
			& & SD & 1.0071 & 1.4966 & 1.0074 && 1.0034 & 1.4880 & 1.0035 && 1.0016 & 1.4284 & 1.0016 \\
			& & CS & -0.0005 & -0.1262 & -0.0004 && -0.0020 &  -0.1390 & -0.0022 && -0.0034 & 0.0640 & -0.0032 \\
			& & CK & -0.2586 & -0.2693 & -0.2562 && -0.1399 & -0.5011 & -0.1402 && -0.0758 & 0.0339 & -0.0763 \\
			\bottomrule
\end{tabular}
}
\label{tab:rq-res}
\end{table}

\subsection{Hypothesis tests}

We now present Monte Carlo simulation studies to evaluate the performances of the Wald, score, likelihood ratio and gradient tests.  We considered two measures: null rejection rate (size) and non-null rejection rate (power). The data generating model has three covariates and is given by
\begin{equation}
	\log(Q_i) = \beta_0 + \beta_1 x_{1i} + \beta_2 x_{2i} + \beta_3 x_{3i}, \; i = 1,2, \ldots, n, \nonumber
\end{equation}
where the values of the coefficients $\beta_j$, $j = 0,1, \ldots, 3$, not fixed in $H_0$ are all equal to 1. The covariate values were obtained as U(0,1) random draws. The interest lies in testing $H_0: \beta_{3 - \kappa + 1} = \cdots = \beta_3 = 0$, with $\kappa = 1, 3$ against $H_0: \beta_{3 - \kappa + 1} = \cdots = \beta_3 \neq 0$, with $\kappa = 1, 3$. The tests' nominal levels used are $\alpha = 0.01, 0.05, 0.1$.

Tables \ref{tab:null-cnbs}-\ref{tab:null-tbs} present the simulation results of null rejection rates for $H_0: \beta_3 = 0$ and in the CN-BS, SL-BS and $ t_\nu$-BS quantile regression models, respectively. From these tables, we observe that as the sample size increases the test sizes for all considered test statistics tends to approach their respective nominal values. In particular, the Wald and gradient statistics present null rejection rates closer to the nominal levels even for small sample sizes. Finally, in general, the tests' null rejection rates do not seem to be affected by the shape parameter $\alpha$. Tables \ref{tab:null-3-cnbs}-\ref{tab:null-3-tbs} (Appendix~\ref{app_simu_hyp}) present the simulation results of null rejection rates for $H_0: \beta_1 = \beta_2 = \beta_3 = 0$ in the CN-BS, SL-BS and $ t_\nu$-BS quantile regression models, respectively. Analogously to the case where $\kappa = 1$,  the tests' null rejection rates does not change according to the value of $\alpha$ and the Wald and gradient statistics showed greater control over the error type I when compared to the other test statistics that we analyzed. On the other hand, the score statistic is the one with the highest null rejection rates among all considered test statistics.

\begin{table}[H]
	\centering
	\caption{Null rejection rates for $H_0: \beta_3 = 0$ in the CN-BS quantile regression model ($\nu = 0.1$, $\delta = 0.3$).}
	\adjustbox{max height=\dimexpr\textheight-3.5cm\relax,
		max width=\textwidth}{
		\begin{tabular}{lccccccccccccc}
			\toprule
			& & & \multicolumn{3}{c}{$n = 50$} & & \multicolumn{3}{c}{$n = 100$} & & \multicolumn{3}{c}{$n = 200$} \\
			\cline{4-6} \cline{8-10} \cline{12-14}
			$\alpha$ & $q$ & & 1\% & 5\% & 10\% & & 1\% & 5\% & 10\% & & 1\% & 5\% & 10\% \\
			\hline
			\multirow{13}{*}{0.2} & \multirow{4}{*}{0.25} & $S_W$ & 0.0126 & 0.0516 & 0.1026 && 0.015 & 0.0566 & 0.1024 && 0.0078 & 0.0488 & 0.104 \\
			& & $S_{LR}$ & 0.0214 & 0.0768 & 0.1324 && 0.0196 & 0.0728 & 0.126 && 0.012 & 0.0668 & 0.1262 \\
			& & $S_R$ & 0.0386 & 0.108 & 0.1764 && 0.0204 & 0.0782 & 0.1356 && 0.0142 & 0.063 & 0.1232 \\
			& & $S_T$ & 0.012 & 0.0618 & 0.1156 && 0.0114 & 0.0502 & 0.1038 && 0.01 & 0.0534 & 0.1026 \\
			\cline{2-14}
			& \multirow{4}{*}{0.5} & $S_W$ & 0.0126 & 0.0514 & 0.1026 && 0.015 & 0.0566 & 0.1022 && 0.0078 & 0.0488 & 0.1038 \\
			& & $S_{LR}$ & 0.0186 & 0.0786 & 0.1386 && 0.0196 & 0.073 & 0.126 && 0.012 & 0.0668 & 0.1262 \\
			& & $S_R$ & 0.0258 & 0.0894 & 0.1486 && 0.0198 & 0.074 & 0.1354 && 0.0116 & 0.06 & 0.1102 \\
			& & $S_T$ & 0.0086 & 0.0486 & 0.1012 && 0.0082 & 0.0504 & 0.102 && 0.0126 & 0.0544 & 0.104 \\
			\cline{2-14}
			& \multirow{4}{*}{0.75} & $S_W$ & 0.0126 & 0.0514 & 0.1024 && 0.0148 & 0.0566 & 0.102 && 0.0078 & 0.049 & 0.1042 \\
			& & $S_{LR}$ & 0.0186 & 0.0786 & 0.1386 && 0.0196 & 0.0728 & 0.126 && 0.012 & 0.0668 & 0.1262 \\
			& & $S_R$ & 0.038 & 0.1068 & 0.1824 && 0.0198 & 0.0758 & 0.137 && 0.0164 & 0.0736 & 0.132 \\
			& & $S_T$ & 0.0124 & 0.0544 & 0.1112 && 0.0088 & 0.0508 & 0.1012 && 0.0108 & 0.0506 & 0.1038 \\
			\hline
			\multirow{13}{*}{0.5} & \multirow{4}{*}{0.25} & $S_W$ & 0.013 & 0.0482 & 0.0964 && 0.0142 & 0.0554 & 0.1024 && 0.0074 & 0.0488 & 0.1014 \\
			& & $S_{LR}$ & 0.0178 & 0.078 & 0.1374 && 0.02 & 0.0716 & 0.1252 && 0.0114 & 0.066 & 0.1236 \\
			& & $S_R$ & 0.0378 & 0.1104 & 0.1834 && 0.0236 & 0.076 & 0.142 && 0.012 & 0.0602 & 0.1112 \\
			& & $S_T$ & 0.0128 & 0.0592 & 0.111 && 0.0108 & 0.048 & 0.1018 && 0.0086 & 0.0526 & 0.094 \\
			\cline{2-14}
			& \multirow{4}{*}{0.5} & $S_W$ & 0.013 & 0.0482 & 0.0964 && 0.0142 & 0.0554 & 0.1024 && 0.0074 & 0.0488 & 0.1012 \\
			& & $S_{LR}$ & 0.0178 & 0.078 & 0.1374 && 0.02 & 0.0716 & 0.1252 && 0.0114 & 0.066 & 0.1236 \\
			& & $S_R$ & 0.0216 & 0.0834 & 0.1486 && 0.0202 & 0.0704 & 0.1236 && 0.0122 & 0.056 & 0.1056 \\
			& & $S_T$ & 0.0062 & 0.0468 & 0.0982 && 0.0122 & 0.0518 & 0.1076 && 0.0092 & 0.0524 & 0.1044 \\
			\cline{2-14}
			& \multirow{4}{*}{0.75} & $S_W$ & 0.013 & 0.0482 & 0.0964 && 0.0142 & 0.0556 & 0.1024 && 0.0074 & 0.0488 & 0.1012 \\
			& & $S_{LR}$ & 0.0178 & 0.078 & 0.1374 && 0.02 & 0.0716 & 0.1252 && 0.0114 & 0.066 & 0.1236 \\
			& & $S_R$ & 0.0442 & 0.1186 & 0.1932 && 0.02 & 0.0716 & 0.1342 && 0.015 & 0.0668 & 0.118 \\
			& & $S_T$ & 0.0144 & 0.0604 & 0.1122 && 0.0108 & 0.0556 & 0.1102 && 0.0134 & 0.0588 & 0.1158 \\
			\hline
			\multirow{13}{*}{1} & \multirow{4}{*}{0.25} & $S_W$ & 0.0108 & 0.0418 & 0.083 && 0.0122 & 0.05 & 0.0932 && 0.0062 & 0.046 & 0.098 \\
			& & $S_{LR}$ & 0.018 & 0.0728 & 0.1322 && 0.0176 & 0.0632 & 0.116 && 0.011 & 0.0634 & 0.1168 \\
			& & $S_R$ & 0.0424 & 0.1252 & 0.189 && 0.0256 & 0.0834 & 0.1422 && 0.0158 & 0.0624 & 0.1162 \\
			& & $S_T$ & 0.0092 & 0.0544 & 0.1114 && 0.0084 & 0.0556 & 0.1104 && 0.0108 & 0.0494 & 0.1066 \\
			\cline{2-14}
			& \multirow{4}{*}{0.5} & $S_W$ & 0.0108 & 0.0418 & 0.083 && 0.0122 & 0.05 & 0.0932 && 0.0062 & 0.046 & 0.098 \\
			& & $S_{LR}$ & 0.018 & 0.0728 & 0.1322 && 0.0178 & 0.0632 & 0.116 && 0.011 & 0.0634 & 0.1168 \\
			& & $S_R$ & 0.0238 & 0.0898 & 0.1594 && 0.0182 & 0.0724 & 0.1286 && 0.0136 & 0.0582 & 0.1096 \\
			& & $S_T$ & 0.0086 & 0.0484 & 0.1018 && 0.007 & 0.0444 & 0.0972 && 0.0096 & 0.0458 & 0.0928 \\
			\cline{2-14}
			& \multirow{4}{*}{0.75} & $S_W$ & 0.0108 & 0.0418 & 0.083 && 0.0122 & 0.05 & 0.0932 && 0.0062 & 0.046 & 0.098 \\
			& & $S_{LR}$ & 0.018 & 0.0728 & 0.1322 && 0.0176 & 0.0632 & 0.116 && 0.011 & 0.0634 & 0.1168 \\
			& & $S_R$ & 0.0384 & 0.1166 & 0.1834 && 0.021 & 0.0774 & 0.1342 && 0.012 & 0.06 & 0.1154 \\
			& & $S_T$ & 0.013 & 0.0538 & 0.107 && 0.0112 & 0.0516 & 0.1052 && 0.0106 & 0.052 & 0.1066 \\
			\bottomrule
		\end{tabular}
	}
\label{tab:null-cnbs}
\end{table}

\begin{table}[H]
	\centering
	\caption{Null rejection rates for  $H_0: \beta_3 = 0$ in the SL-BS quantile regression model ($\nu = 4$).}
	\adjustbox{max height=\dimexpr\textheight-3.5cm\relax,
		max width=\textwidth}{
		\begin{tabular}{lccccccccccccc}
			\toprule
			& & & \multicolumn{3}{c}{$n = 50$} & & \multicolumn{3}{c}{$n = 100$} & & \multicolumn{3}{c}{$n = 200$} \\
			\cline{4-6} \cline{8-10} \cline{12-14}
			$\alpha$ & $q$ & & 1\% & 5\% & 10\% & & 1\% & 5\% & 10\% & & 1\% & 5\% & 10\% \\
			\hline
			\multirow{13}{*}{0.2} & \multirow{4}{*}{0.25} & $S_W$ & 0.0116 & 0.0478 & 0.0898 && 0.0112 & 0.0504 & 0.0996 && 0.0104 & 0.049 & 0.098 \\
			& & $S_{LR}$ & 0.0146 & 0.064 & 0.122 && 0.0114 & 0.056 & 0.11 && 0.012 & 0.0534 & 0.1118 \\
			& & $S_R$ & 0.0366 & 0.1116 & 0.1876 && 0.0234 & 0.0774 & 0.1334 && 0.0126 & 0.065 & 0.1178 \\
			& & $S_T$ & 0.0078 & 0.0522 & 0.1034 && 0.0122 & 0.0554 & 0.106 && 0.0098 & 0.06 & 0.113 \\
			\cline{2-14}
			& \multirow{4}{*}{0.5} & $S_W$ & 0.0106 & 0.049 & 0.0908 && 0.0112 & 0.0502 & 0.0996 && 0.0104 & 0.049 & 0.098 \\
			& & $S_{LR}$ & 0.017 & 0.0634 & 0.117 && 0.0124 & 0.0608 & 0.1112 && 0.0108 & 0.0484 & 0.1006 \\
			& & $S_R$ & 0.0284 & 0.0926 & 0.1524 && 0.016 & 0.0672 & 0.1192 && 0.0106 & 0.051 & 0.1022 \\
			& & $S_T$ & 0.0088 & 0.0452 & 0.0998 && 0.0102 & 0.05 & 0.1016 && 0.0102 & 0.0516 & 0.0966 \\
			\cline{2-14}
			& \multirow{4}{*}{0.75} & $S_W$ & 0.0106 & 0.049 & 0.0908 && 0.0112 & 0.0502 & 0.0996 && 0.0104 & 0.049 & 0.0978 \\
			& & $S_{LR}$ & 0.0148 & 0.0672 & 0.1272 && 0.0116 & 0.0602 & 0.1114 && 0.0132 & 0.0534 & 0.1004 \\
			& & $S_R$ & 0.031 & 0.098 & 0.1684 && 0.022 & 0.0766 & 0.1306 && 0.0148 & 0.0604 & 0.122 \\
			& & $S_T$ & 0.0082 & 0.0506 & 0.108 && 0.0086 & 0.0482 & 0.1014 && 0.0114 & 0.052 & 0.0996 \\
			\hline
			\multirow{13}{*}{0.5} & \multirow{4}{*}{0.25} & $S_W$ & 0.01 & 0.0452 & 0.0852 && 0.0106 & 0.0486 & 0.098 && 0.0102 & 0.0482 & 0.097 \\
			& & $S_{LR}$ & 0.0158 & 0.0668 & 0.1232 && 0.0122 & 0.0566 & 0.1082 && 0.01 & 0.053 & 0.1064 \\
			& & $S_R$ & 0.0386 & 0.112 & 0.1766 && 0.0288 & 0.0892 & 0.15 && 0.0152 & 0.0608 & 0.1144 \\
			& & $S_T$ & 0.0078 & 0.0486 & 0.0984 && 0.0092 & 0.0514 & 0.099 && 0.0086 & 0.047 & 0.1004 \\
			\cline{2-14}
			& \multirow{4}{*}{0.5} & $S_W$ & 0.01 & 0.0452 & 0.0852 && 0.0106 & 0.0486 & 0.098 && 0.012 & 0.0498 & 0.0918 \\
			& & $S_{LR}$ & 0.0122 & 0.057 & 0.1136 && 0.0136 & 0.058 & 0.1162 && 0.0098 & 0.0502 & 0.0998 \\
			& & $S_R$ & 0.0258 & 0.0884 & 0.1528 && 0.0168 & 0.0674 & 0.125 && 0.013 & 0.063 & 0.111 \\
			& & $S_T$ & 0.008 & 0.0466 & 0.103 && 0.0076 & 0.0492 & 0.1016 && 0.009 & 0.0532 & 0.1022 \\
			\cline{2-14}
			& \multirow{4}{*}{0.75} & $S_W$ & 0.01 & 0.0452 & 0.0852 && 0.0106 & 0.0486 & 0.098 && 0.008 & 0.0446 & 0.0926 \\
			& & $S_{LR}$ & 0.0142 & 0.059 & 0.1112 && 0.0126 & 0.0644 & 0.1264 && 0.0108 & 0.0562 & 0.1096 \\
			& & $S_R$ & 0.0348 & 0.1076 & 0.182 && 0.0248 & 0.0836 & 0.145 && 0.0162 & 0.0606 & 0.1132 \\
			& & $S_T$ & 0.0086 & 0.0522 & 0.108 && 0.0104 & 0.0492 & 0.1022 && 0.0102 & 0.045 & 0.0938 \\
			\hline
			\multirow{13}{*}{1} & \multirow{4}{*}{0.25} & $S_W$ & 0.0072 & 0.031 & 0.0696 && 0.0088 & 0.0422 & 0.0852 && 0.0076 & 0.0434 & 0.0926 \\
			& & $S_{LR}$ & 0.014 & 0.0624 & 0.1206 && 0.0108 & 0.0546 & 0.1038 && 0.0088 & 0.0486 & 0.095 \\
			& & $S_R$ & 0.0402 & 0.1184 & 0.1794 && 0.0196 & 0.073 & 0.1378 && 0.0178 & 0.0616 & 0.1146 \\
			& & $S_T$ & 0.0084 & 0.051 & 0.1066 && 0.0124 & 0.0606 & 0.1114 && 0.0094 & 0.0528 & 0.1042 \\
			\cline{2-14}
			& \multirow{4}{*}{0.5} & $S_W$ & 0.0072 & 0.031 & 0.0696 && 0.0088 & 0.042 & 0.0852 && 0.0076 & 0.0434 & 0.0926 \\
			& & $S_{LR}$ & 0.0108 & 0.053 & 0.1098 && 0.0144 & 0.0574 & 0.1122 && 0.011 & 0.0442 & 0.0944 \\
			& & $S_R$ & 0.0246 & 0.0874 & 0.15 && 0.0186 & 0.0684 & 0.1252 && 0.013 & 0.0642 & 0.1162 \\
			& & $S_T$ & 0.008 & 0.0474 & 0.0986 && 0.0076 & 0.0476 & 0.0998 && 0.008 & 0.047 & 0.0994 \\
			\cline{2-14}
			& \multirow{4}{*}{0.75} & $S_W$ & 0.0072 & 0.031 & 0.0696 && 0.0088 & 0.0422 & 0.0852 && 0.0076 & 0.0434 & 0.0926 \\
			& & $S_{LR}$ & 0.0124 & 0.0568 & 0.1102 && 0.013 & 0.055 & 0.105 && 0.01 & 0.0454 & 0.0894 \\
			& & $S_R$ & 0.0344 & 0.1048 & 0.1744 && 0.023 & 0.0816 & 0.1424 && 0.0144 & 0.0596 & 0.1218 \\
			& & $S_T$ & 0.0104 & 0.0638 & 0.113 && 0.011 & 0.0524 & 0.1036 && 0.013 & 0.052 & 0.1066 \\
			\bottomrule
		\end{tabular}
	}
	\label{tab:null-slbs}
\end{table}

\begin{table}[H]
	\centering
	\caption{Null rejection rates for $H_0: \beta_3 = 0$ in the $t_\nu$-BS quantile regression model ($\nu = 11$).}
	\adjustbox{max height=\dimexpr\textheight-3.5cm\relax,
		max width=\textwidth}{
		\begin{tabular}{lccccccccccccc}
			\toprule
			& & & \multicolumn{3}{c}{$n = 50$} & & \multicolumn{3}{c}{$n = 100$} & & \multicolumn{3}{c}{$n = 200$} \\
			\cline{4-6} \cline{8-10} \cline{12-14}
			$\alpha$ & $q$ & & 1\% & 5\% & 10\% & & 1\% & 5\% & 10\% & & 1\% & 5\% & 10\% \\
			\hline
			\multirow{13}{*}{0.2} & \multirow{4}{*}{0.25} & $S_W$ & 0.0126 & 0.0534 & 0.1036 && 0.0156 & 0.0546 & 0.1062 && 0.007 & 0.0502 & 0.1052 \\
			& & $S_{LR}$ & 0.0162 & 0.0696 & 0.1296 && 0.0182 & 0.0664 & 0.123 && 0.0092 & 0.0598 & 0.1164 \\
			& & $S_R$ & 0.0332 & 0.0966 & 0.1634 && 0.0204 & 0.0786 & 0.1428 && 0.014 & 0.0666 & 0.1222 \\
			& & $S_T$ & 0.0096 & 0.0534 & 0.11 && 0.0086 & 0.0544 & 0.1032 && 0.0104 & 0.049 & 0.1054 \\
			\cline{2-14}
			& \multirow{4}{*}{0.5} & $S_W$ & 0.0126 & 0.0536 & 0.1038 && 0.0156 & 0.0546 & 0.1062 && 0.007 & 0.0502 & 0.1052 \\
			& & $S_{LR}$ & 0.0162 & 0.0696 & 0.1296 && 0.0182 & 0.0662 & 0.123 && 0.0092 & 0.0598 & 0.1162 \\
			& & $S_R$ & 0.021 & 0.0812 & 0.1446 && 0.015 & 0.0646 & 0.1208 && 0.0126 & 0.0568 & 0.107 \\
			& & $S_T$ & 0.0086 & 0.048 & 0.101 && 0.0116 & 0.0484 & 0.1036 && 0.0122 & 0.0494 & 0.1054 \\
			\cline{2-14}
			& \multirow{4}{*}{0.75} & $S_W$ & 0.0126 & 0.0536 & 0.1038 && 0.0156 & 0.0546 & 0.1064 && 0.007 & 0.0502 & 0.1052 \\
			& & $S_{LR}$ & 0.0162 & 0.0694 & 0.1296 && 0.0184 & 0.0662 & 0.1232 && 0.0092 & 0.0598 & 0.116 \\
			& & $S_R$ & 0.0424 & 0.1118 & 0.1776 && 0.0202 & 0.0742 & 0.1394 && 0.0184 & 0.0676 & 0.1274 \\
			& & $S_T$ & 0.0106 & 0.0548 & 0.1082 && 0.0102 & 0.0536 & 0.1028 && 0.0098 & 0.052 & 0.1052 \\
			\hline
			\multirow{13}{*}{0.5} & \multirow{4}{*}{0.25} & $S_W$ & 0.0124 & 0.0514 & 0.098 && 0.0138 & 0.056 & 0.1024 && 0.0066 & 0.0482 & 0.1046 \\
			& & $S_{LR}$ & 0.016 & 0.0666 & 0.1274 && 0.0178 & 0.063 & 0.1216 && 0.0088 & 0.0576 & 0.1126 \\
			& & $S_R$ & 0.032 & 0.1006 & 0.1736 && 0.0212 & 0.0748 & 0.14 && 0.0146 & 0.0578 & 0.1162 \\
			& & $S_T$ & 0.012 & 0.0608 & 0.1174 && 0.0112 & 0.0562 & 0.1036 && 0.0104 & 0.0528 & 0.0974 \\
			\cline{2-14}
			& \multirow{4}{*}{0.5} & $S_W$ & 0.0124 & 0.0514 & 0.098 && 0.0138 & 0.056 & 0.1024 && 0.0066 & 0.0482 & 0.1046 \\
			& & $S_{LR}$ & 0.016 & 0.0666 & 0.1274 && 0.0178 & 0.063 & 0.1216 && 0.0088 & 0.0576 & 0.1126 \\
			& & $S_R$ & 0.0218 & 0.084 & 0.1466 && 0.0144 & 0.063 & 0.1178 && 0.01 & 0.0576 & 0.1072 \\
			& & $S_T$ & 0.0092 & 0.053 & 0.1026 && 0.0114 & 0.0504 & 0.1034 && 0.0098 & 0.051 & 0.098 \\
			\cline{2-14}
			& \multirow{4}{*}{0.75} & $S_W$ & 0.0124 & 0.0514 & 0.098 && 0.0138 & 0.056 & 0.1024 && 0.0066 & 0.0482 & 0.1046 \\
			& & $S_{LR}$ & 0.016 & 0.0666 & 0.1274 && 0.0178 & 0.063 & 0.1216 && 0.0088 & 0.0576 & 0.1126 \\
			& & $S_R$ & 0.0326 & 0.103 & 0.1688 && 0.018 & 0.069 & 0.1326 && 0.0126 & 0.0648 & 0.1234 \\
			& & $S_T$ & 0.0106 & 0.0576 & 0.1136 && 0.0146 & 0.0586 & 0.105 && 0.0104 & 0.0528 & 0.1044 \\
			\hline
			\multirow{13}{*}{1} & \multirow{4}{*}{0.25} & $S_W$ & 0.0104 & 0.0426 & 0.0846 && 0.012 & 0.051 & 0.0952 && 0.0064 & 0.0438 & 0.0992 \\ 
			& & $S_{LR}$ & 0.0136 & 0.063 & 0.123 && 0.0154 & 0.0584 & 0.1144 && 0.0078 & 0.0508 & 0.1072 \\
			& & $S_R$ & 0.0338 & 0.1046 & 0.1722 && 0.0196 & 0.0744 & 0.1364 && 0.0124 & 0.0592 & 0.1208 \\
			& & $S_T$ & 0.0112 & 0.0586 & 0.1092 && 0.012 & 0.0526 & 0.1014 && 0.0112 & 0.0548 & 0.1054 \\
			\cline{2-14}
			& \multirow{4}{*}{0.5} & $S_W$ & 0.0104 & 0.0426 & 0.0846 && 0.012 & 0.051 & 0.0952 && 0.0064 & 0.0438 & 0.0992 \\
			& & $S_{LR}$ & 0.0136 & 0.063 & 0.123 && 0.0154 & 0.0584 & 0.1144 && 0.0078 & 0.0508 & 0.1072 \\
			& & $S_R$ & 0.0246 & 0.0888 & 0.1424 && 0.0188 & 0.0654 & 0.1188 && 0.0122 & 0.0566 & 0.1104 \\
			& & $S_T$ & 0.007 & 0.0498 & 0.1016 && 0.0094 & 0.0492 & 0.1016 && 0.0108 & 0.0502 & 0.1028 \\
			\cline{2-14}
			& \multirow{4}{*}{0.75} & $S_W$ & 0.0104 & 0.0426 & 0.0846 && 0.012 & 0.051 & 0.0952 && 0.0064 & 0.0438 & 0.0992 \\
			& & $S_{LR}$ & 0.0136 & 0.063 & 0.123 && 0.0154 & 0.0584 & 0.1144 && 0.0078 & 0.0508 & 0.1072 \\
			& & $S_R$ & 0.036 & 0.1148 & 0.1822 && 0.025 & 0.0818 & 0.1356 && 0.014 & 0.0632 & 0.1212 \\
			& & $S_T$ & 0.0124 & 0.0582 & 0.1082 && 0.012 & 0.0568 & 0.1076 && 0.0096 & 0.047 & 0.1002 \\
			\bottomrule
		\end{tabular}
	}
	\label{tab:null-tbs}
\end{table}

\begin{figure}[H]
	\centering
	\subfigure[CN-BS, $\kappa = 1$]{\includegraphics[scale=0.35]{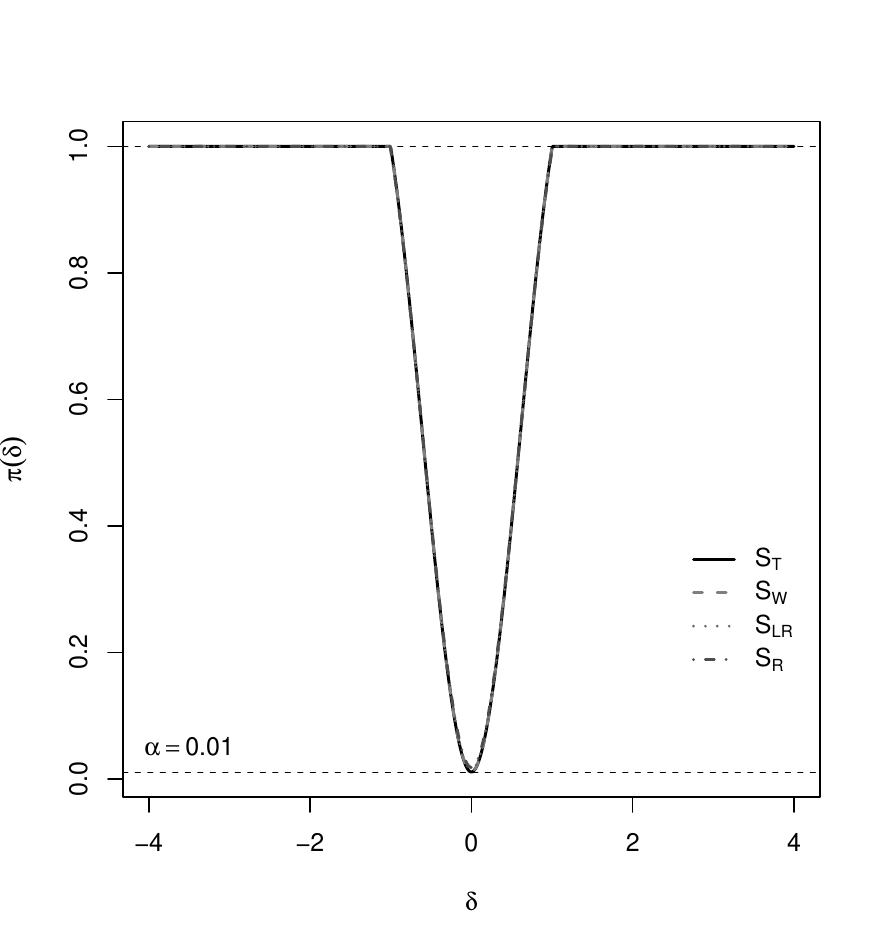}}
	\subfigure[SL-BS, $\kappa = 1$]{\includegraphics[scale=0.35]{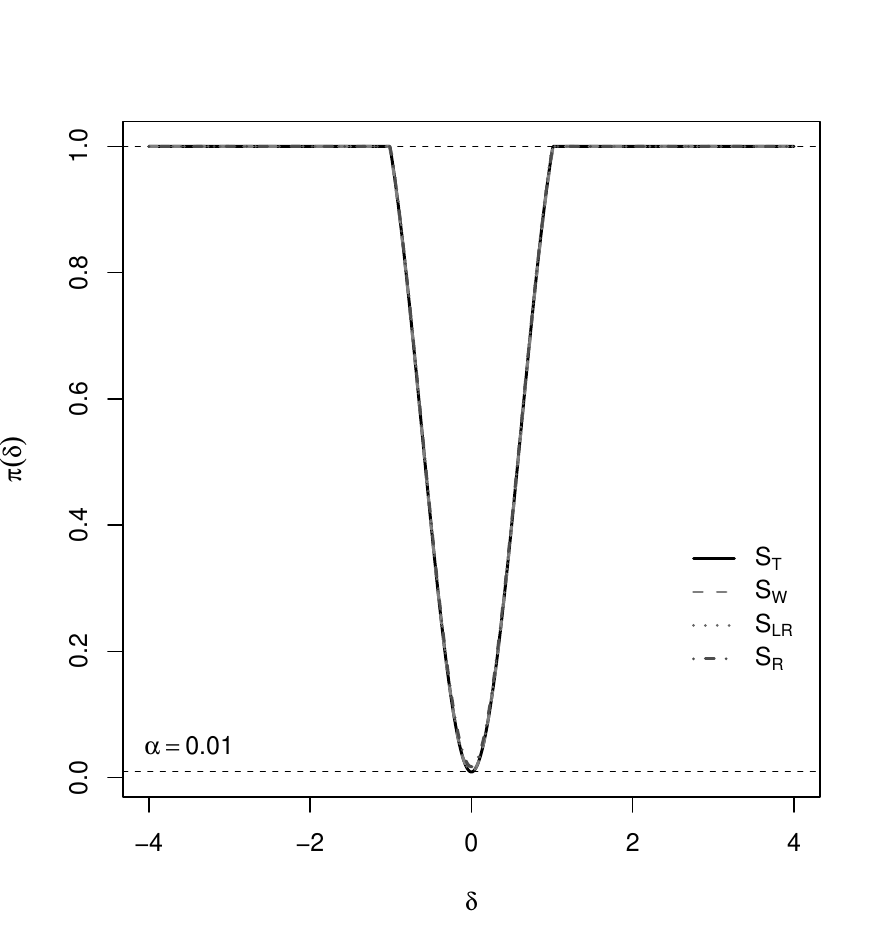}}
	\subfigure[$t_\nu$-BS, $\kappa = 1$]{\includegraphics[scale=0.35]{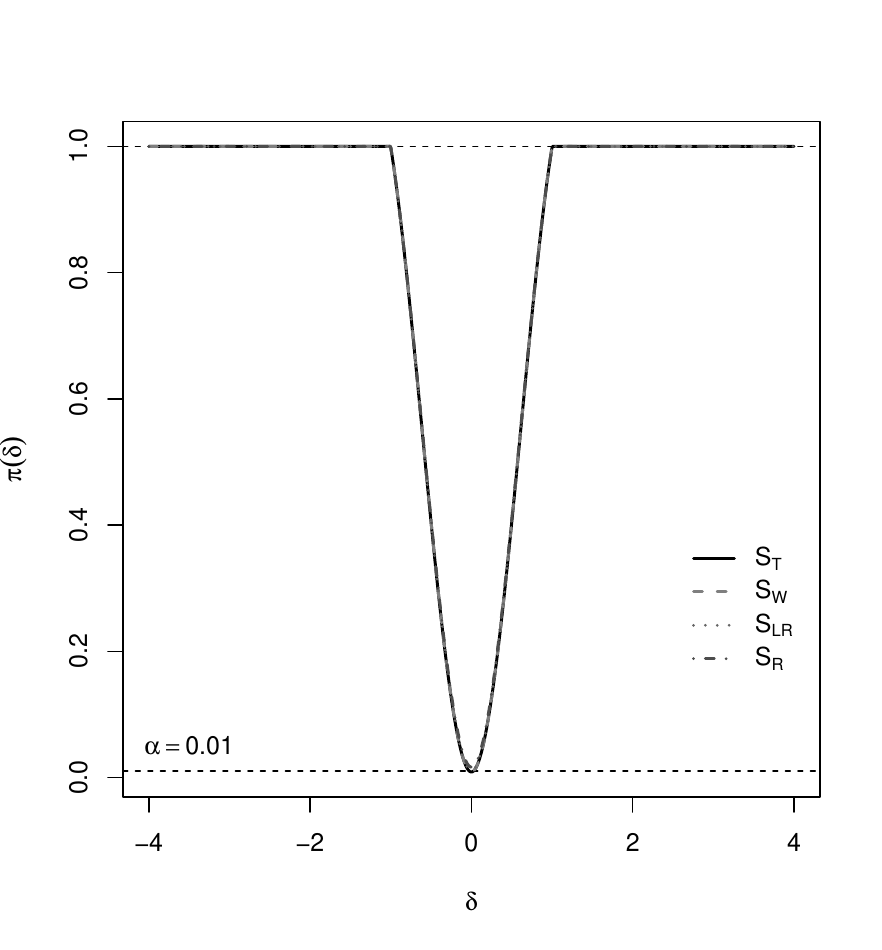}}
	\subfigure[CN-BS, $\kappa = 3$]{\includegraphics[scale=0.35]{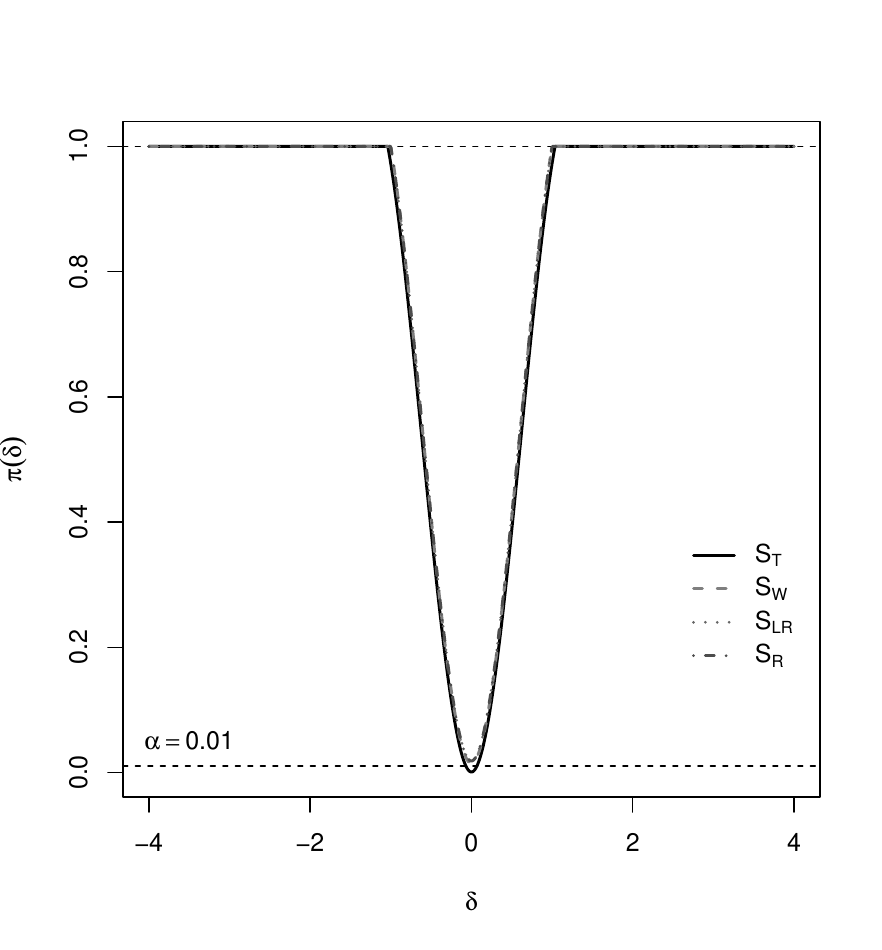}}
	\subfigure[SL-BS, $\kappa = 3$]{\includegraphics[scale=0.35]{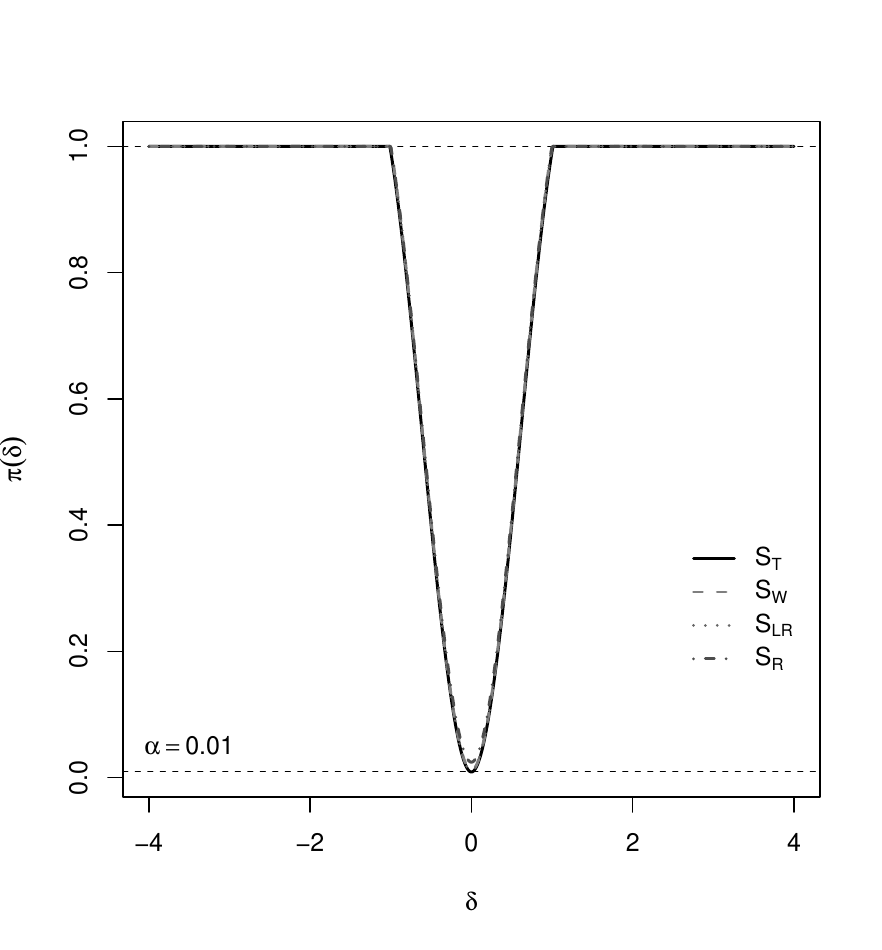}}
	\subfigure[$t_\nu$-BS, $\kappa = 3$]{\includegraphics[scale=0.35]{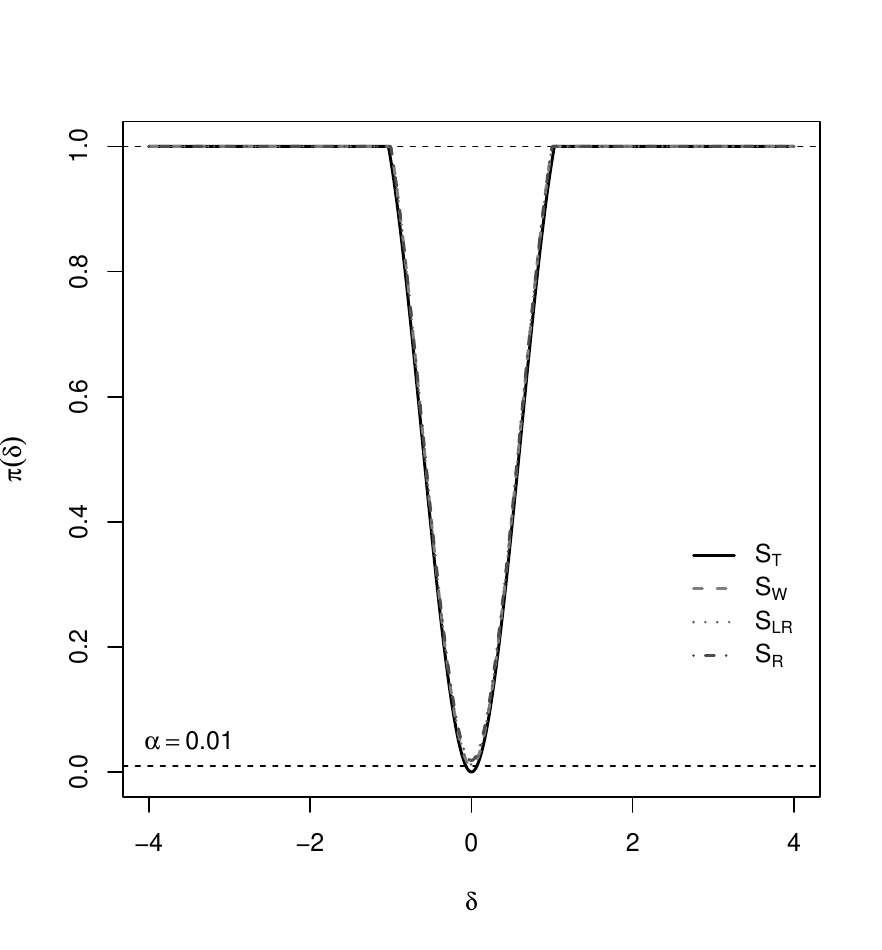}}
	\caption{Power curves of the $S_W, S_{LR}, S_R$ and $S_T$ tests for the CN-BS, SL-BS and $t_\nu$-BS quantile regression models (nominal level = 1\%, $n = 100$, $q = 0.5$).}
	\label{fig:power_stats}
\end{figure}

Figure \ref{fig:power_stats} displays the test power curves for the CN-BS, SL-BS and $t_\nu$-BS models considering the hypotheses $H_0: \beta_{3 - \kappa + 1} = \ldots = \beta_3 = 0$ against $H_1: \beta_{3 - \kappa + 1} = \ldots = \beta_3 = \delta$ with $|\delta| = 0, 1, 2, 3, 4$ and $\kappa = 1, 3$, where $\pi(\delta)$ denotes the power function. The significance level considered was 1\% with $n = 100$ and $q = 0.5$ (other settings present similar results). From this figure, we observe that there are no changes in the power performances of the tests, as well as for any of the CN-BS, SL-BS and $t_\nu$-BS models, so that we have $\pi (\delta) = 1$ for $|\delta| \geq 1$.

\section{Application to personal accident insurance data}\label{section:05}

QSBS quantile regression models are now used to analyze a data set related to the personal injury insurance claims that correspond to amounts of paid money by an insurance policy in Australian dollars (response variable,  \texttt{amount}). The data is collected from 767 individuals and the claims occurred between January 1998 and January 1999; see \cite{dejong2008}. The covariates considered in the study are: \texttt{optime}, denoting the operating time in percentage; \texttt{legrep}, with (1) or without (0) legal representation; and \texttt{month}, denoting the month of accident occurrence. Due to lack of correlation between the covariate \texttt{month} and the response variable \texttt{amount}, we removed this covariate from our analysis.

\begin{table}[H]
	\centering
	\caption{Summary statistics for the personal accident insurance data.}
	\begin{tabular}{cccccccccc}
		\toprule
		MN  &MD &  SD & CV & CS & CK & Range & Min & Max & $n$ \\
		\hline
		7820.59& 6000   & 8339.26 & 106.63\% & 5.09 & 47.77 & 116556.7 & 30 & 116586.7 & 767 \\
		\bottomrule
	\end{tabular}
\label{tab:desc}
\end{table}

\begin{figure}[H]
	\centering
	\subfigure[Density estimate]{\includegraphics[scale = 0.35]{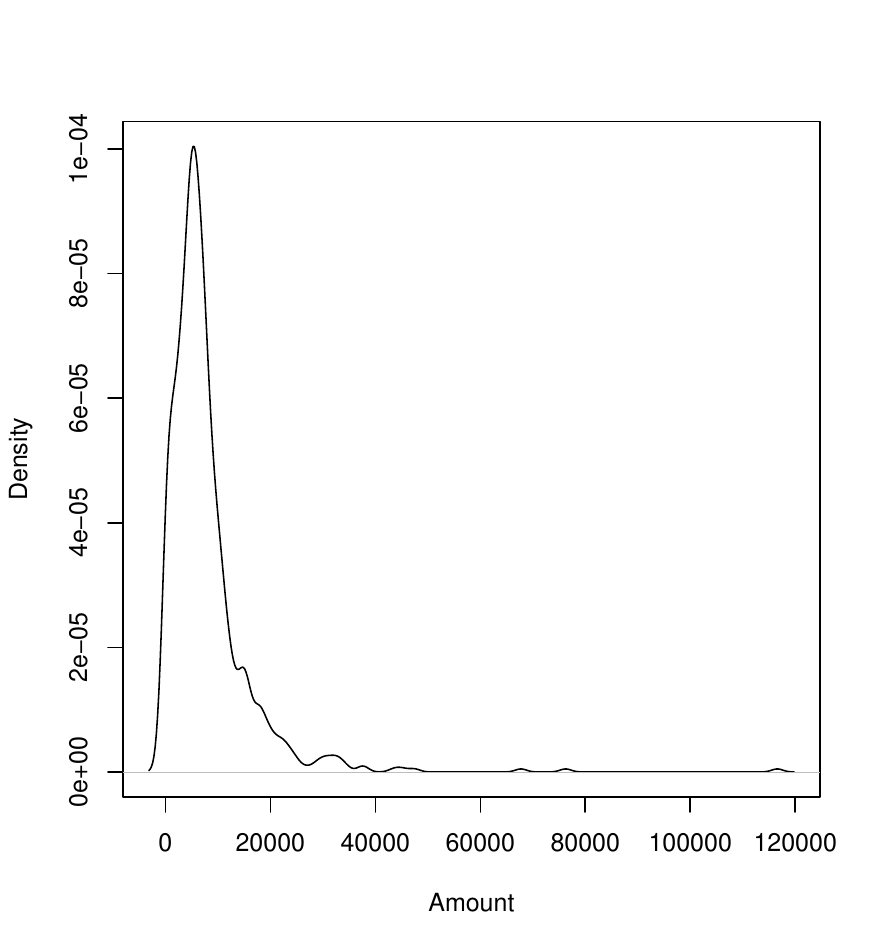}}
	\subfigure[Usual boxplot]{\includegraphics[scale = 0.35]{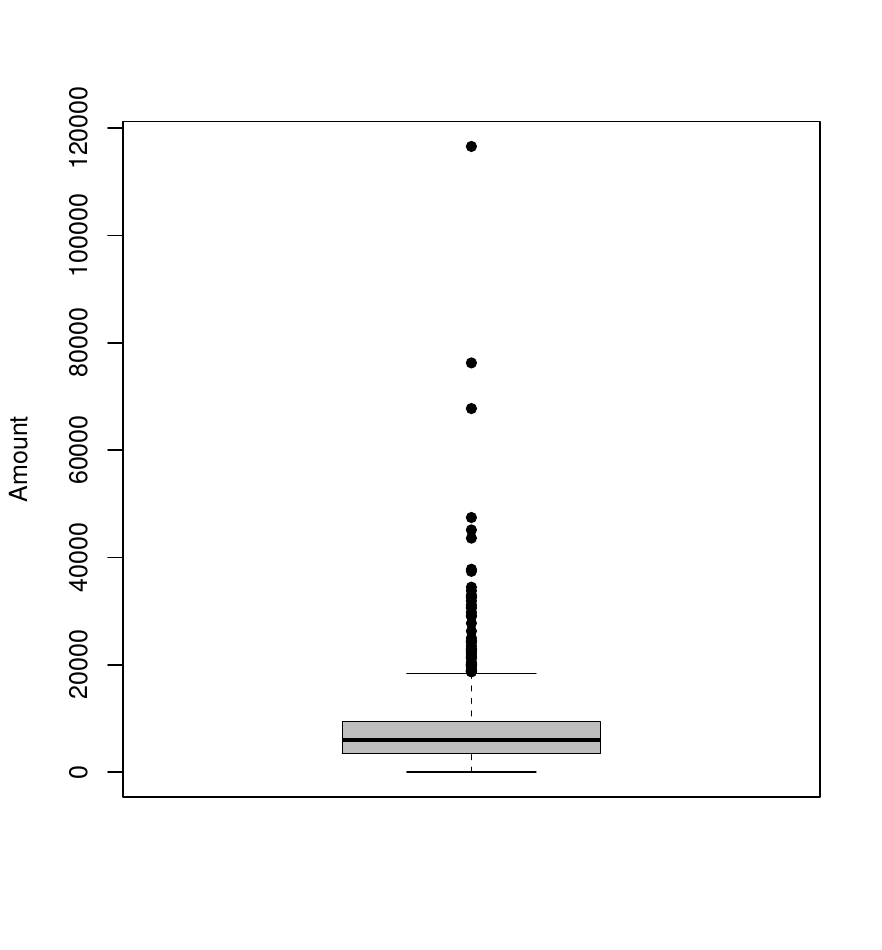}}
	\subfigure[Adjusted boxplot]{\includegraphics[scale = 0.35]{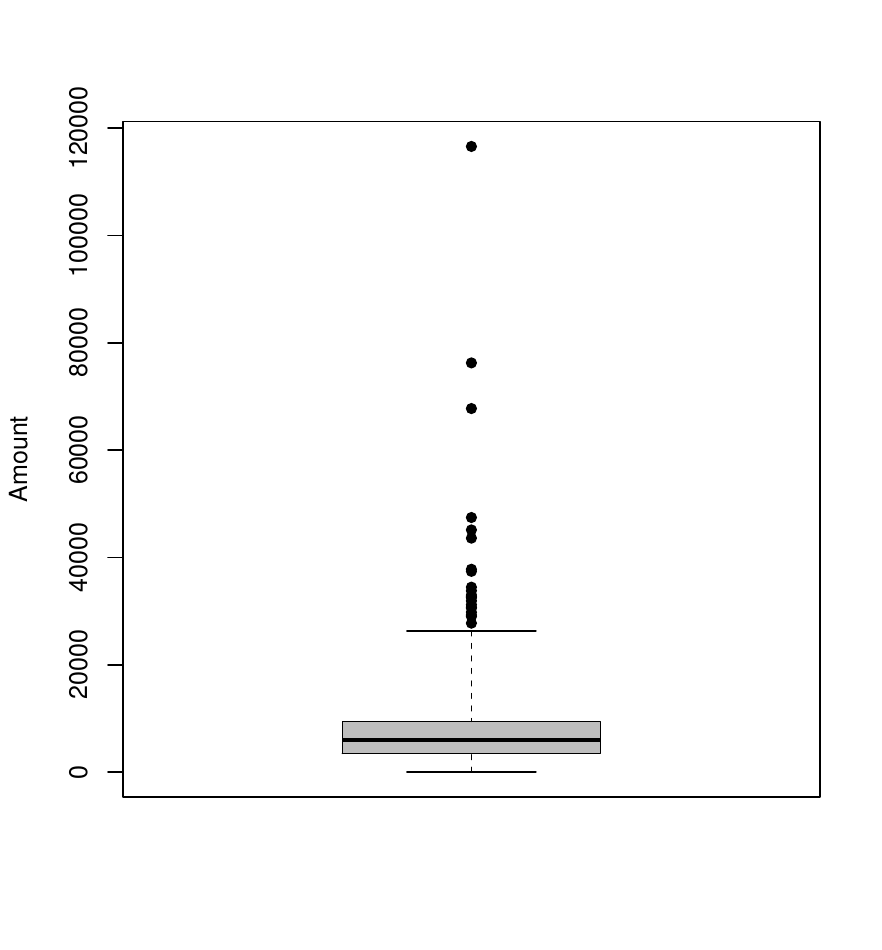}}
\caption{Density estimate and boxplots for the personal accident insurance data.}
\label{fig:amount}
\end{figure}

Table \ref{tab:desc} reports the descriptive statistics of the observed amounts of paid money that includes MN, MD, SD, coefficient of variation (CV), CS, CK (excess) and range.  From this table, we observe skewed and high kurtosis features in the data. Figure \ref{fig:amount}(a) confirms the skewness observed in Table \ref{tab:desc}. Figure \ref{fig:amount}(b) displays the usual and adjusted boxplots \citep{sauloetal:19}, where the latter is useful when the data is distributed as skewedd. Note that the adjusted boxplot indicates that some potential outliers identified by the usual boxplot are not outliers. 

We then analyze the personal accident insurance data using the QSBS quantile regression model, expressed as
\begin{equation}
	\log(Q_i) = \beta_0 + \beta_1 \texttt{optime} + \beta_2 \texttt{legrep}, \; i = 1,2, \ldots, 767.
\end{equation}

\begin{figure}[H]
	\centering
	\subfigure[$\widehat{\beta}_0$]{\includegraphics[scale = 0.35]{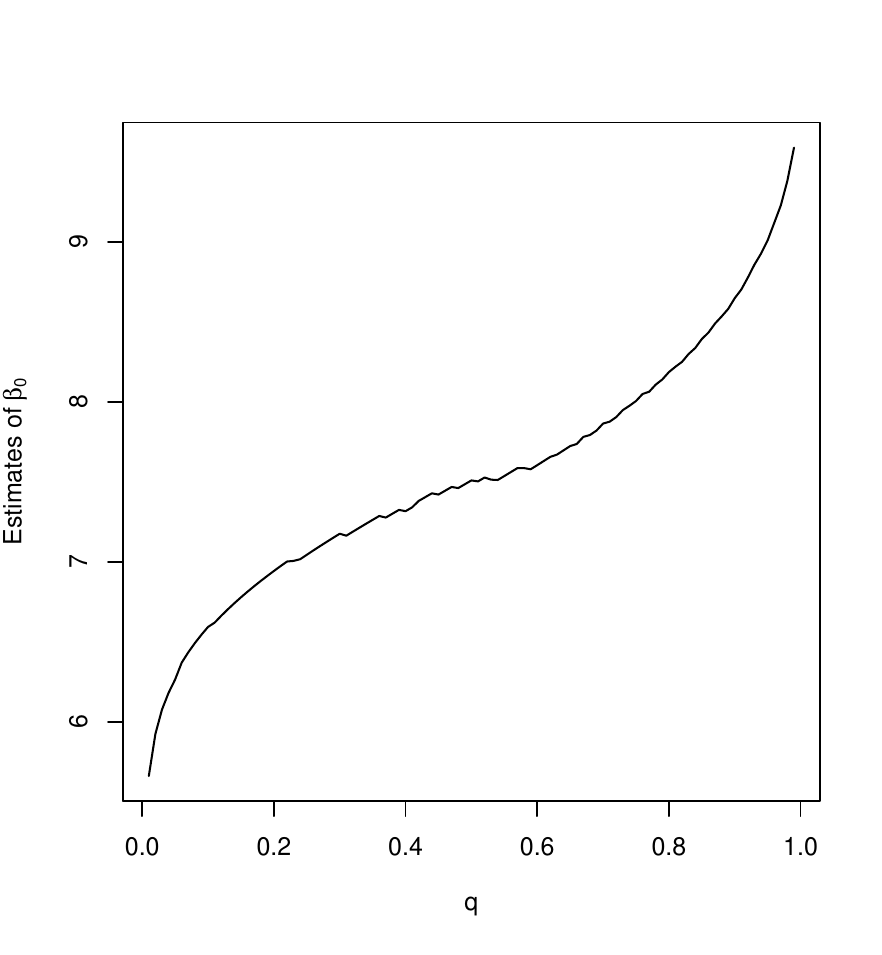}}
	\subfigure[$\widehat{\beta}_1$]{\includegraphics[scale = 0.35]{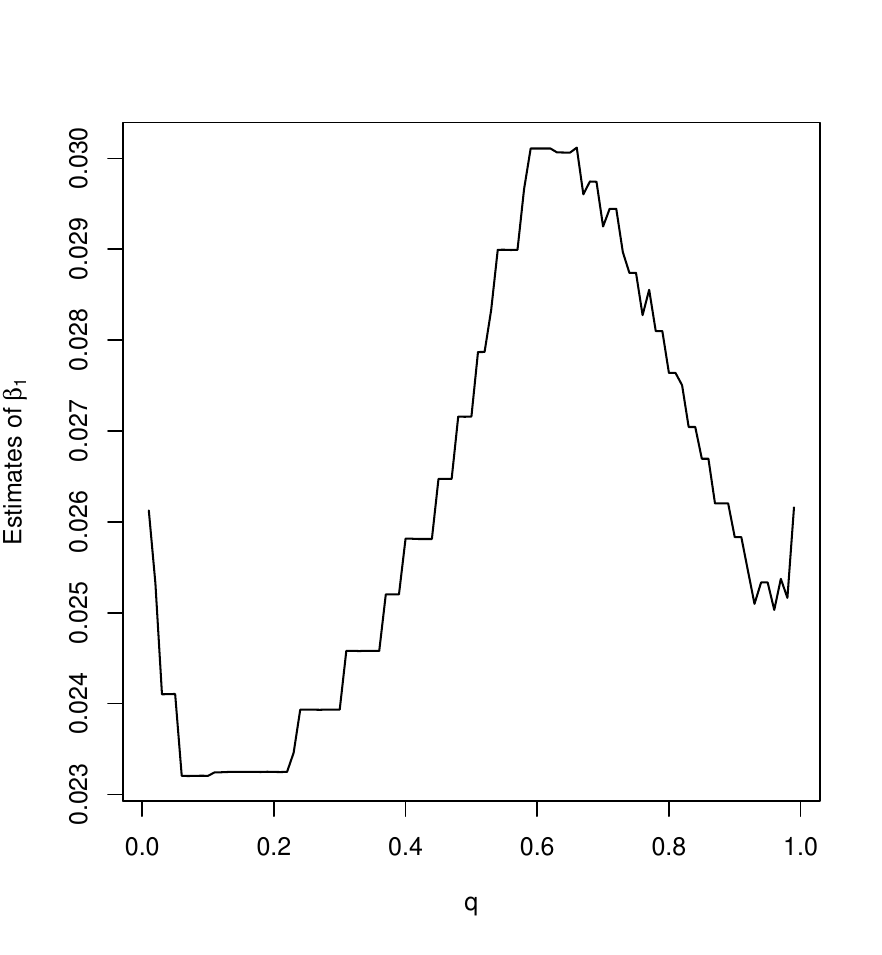}}
	\subfigure[$\widehat{\beta}_2$]{\includegraphics[scale = 0.35]{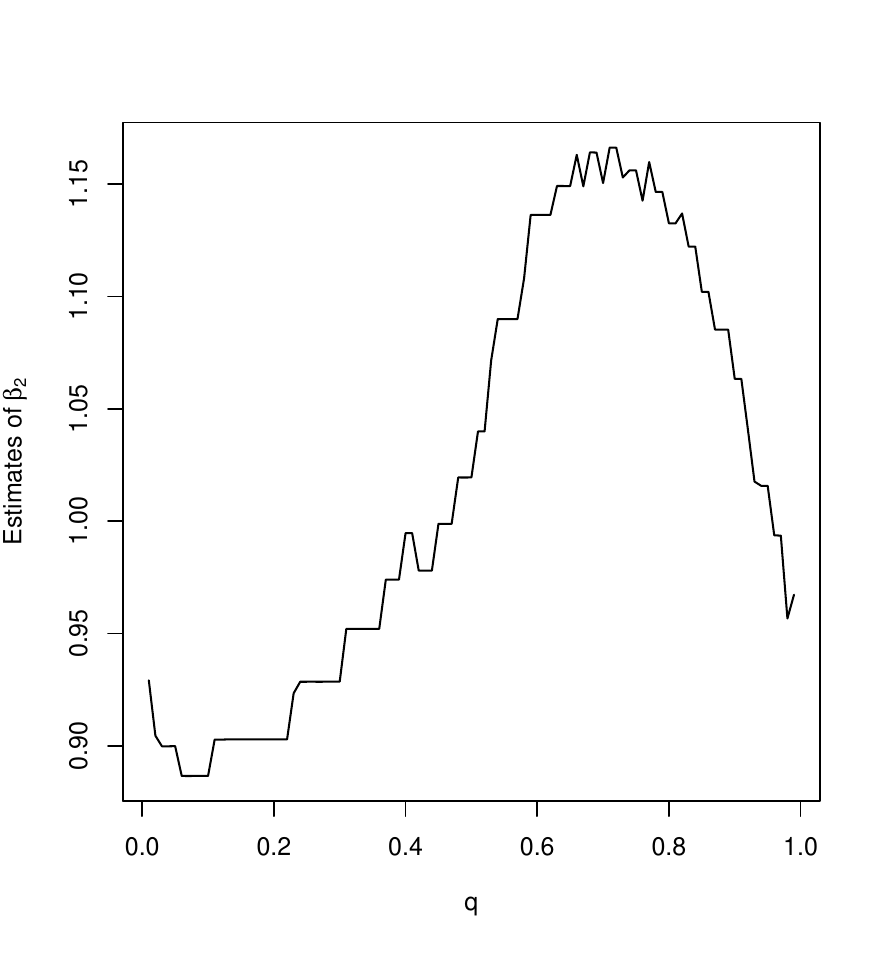}}
	\subfigure[$\widehat{\alpha}$]{\includegraphics[scale = 0.35]{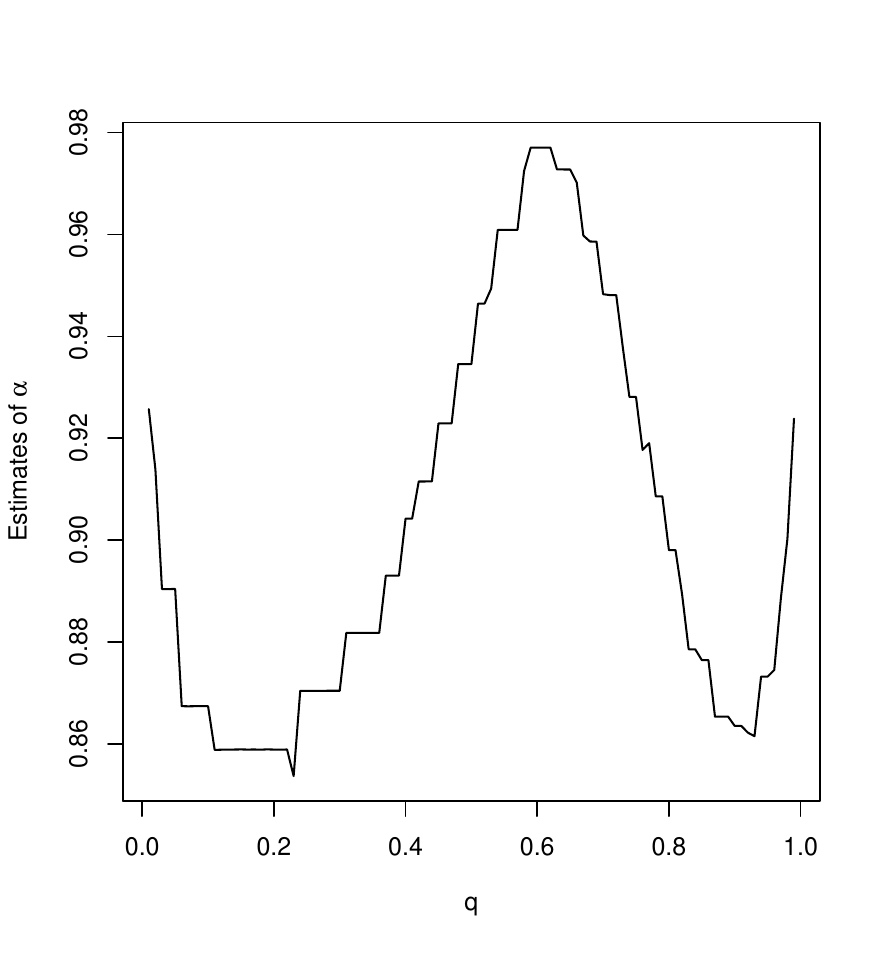}}
	\subfigure[$\widehat{\nu}$]{\includegraphics[scale = 0.35]{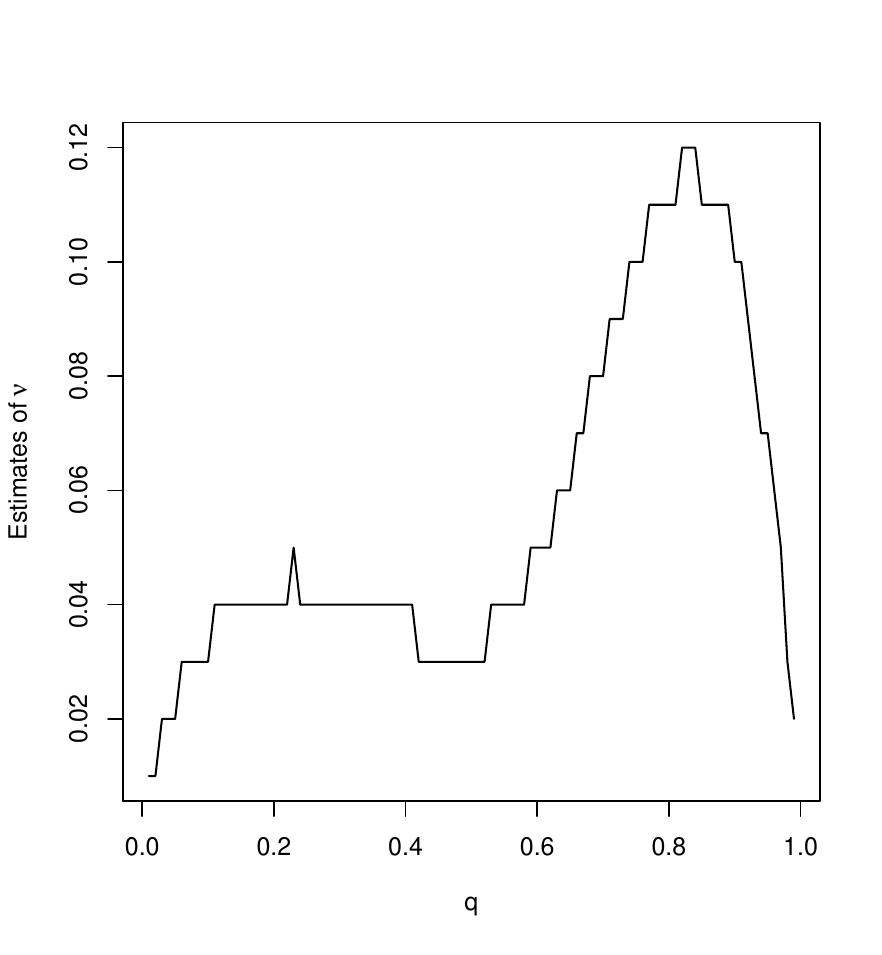}}
	\subfigure[$\widehat{\delta}$]{\includegraphics[scale = 0.35]{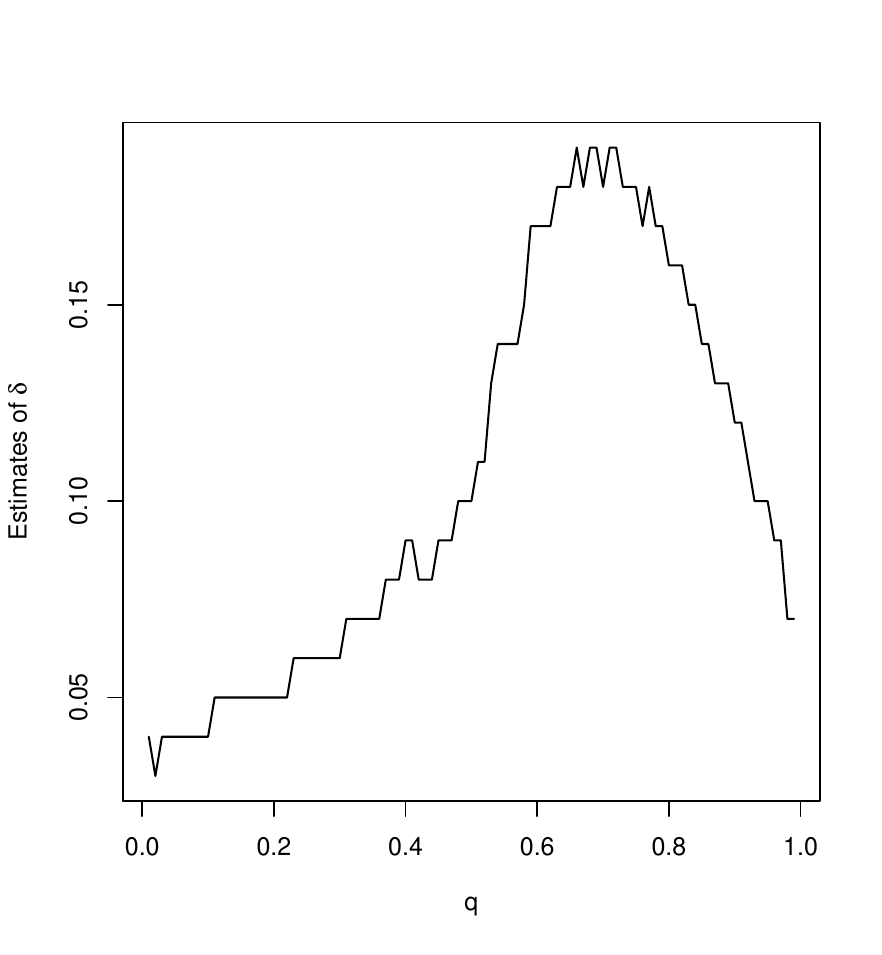}}
	\caption{Estimated parameters in the CN-BS quantile regression models across $q$ for the personal accident insurance data.}
	\label{fig:estimates}
\end{figure}

Table \ref{tab:criterions} reports the the averages of the AIC, BIC, AICc and HQIC values based on $q\in\{0.01,0.02,\ldots,0.99\}$ for the CN-BS, SL-BS and $t_\nu$-BS quantile regression models. The results indicate that the lowest values of AIC, BIC, AICc and HQIC are those based on the CN-BS model. Figure \ref{fig:estimates} displays the estimates of the CN-BS model parameters across $q\in\{0.01,0.02,\ldots,0.99\}$. From this figure, we observe that the estimate of $\beta_0$ tends to increase with an increase in $q$. Moreover, the estimates of ${\beta}_1$, ${\beta}_2$, $\alpha$ and $\nu$ present a cyclic behavior. These results show the relevance of considering a quantile approach rather than traditional mean/median approaches.

\begin{table}[H]
	\centering
	\caption{Averages of the AIC, BIC, AICc and HQIC values based on $q\in\{0.01,0.02,\ldots,0.99\}$ for different models for the personal accident insurance data.}
	\begin{tabular}{cccccc}
		\toprule
		Model & AIC & BIC & AICc & HQIC  \\
		\hline
		CN-BS & 8204.634 & 8223.204 & 8204.686 & 8211.781  \\
		SL-BS & 8220.061 & 8238.631 & 8220.114 & 8227.209  \\
		$t_\nu$-BS & 8213.111 & 8231.681 & 8213.163 & 8220.259  \\
		\bottomrule
	\end{tabular}
\label{tab:criterions}
\end{table}

Table \ref{tab:estimates} presents the ML estimates computed by the EM algorithm and SEs of the CN-BS quantile regression model parameters considering the quantiles $q = 0.025, 0.25, 0.5, 0.75, 0.975$. For the quantiles analyzed, we can observe that the impact of operational time and legal representation is greater for the quantile $q = 0.75$, the same is true with the estimates of ${\alpha}$ and ${\beta}_0$.

\begin{table}[H]
	\centering
	\caption{ML estimates (with SE in parentheses) for the CN-BS quantile regression model for the personal accident insurance data.}
	\begin{tabular}{ccccccc}
		\toprule
		\multirow{2}{*}{$q$} & \multicolumn{6}{c}{Estimate (SE)} \\
		\cline{2-7}
		& $\widehat{\beta}_0$ & $\widehat{\beta}_1$ & $\widehat{\beta}_2$ & $\widehat{\alpha}$ & $\widehat{\nu}$ & $\widehat{\delta}$ \\
		\hline
		0.025 & 5.9965(0.0541) & 0.0253(0.0033) & 0.9045(0.0630) & 0.9138(0.0205) & 0.01 & 0.03 \\
		0.25 & 7.0447(0.0543) & 0.0239(0.0033) & 0.9286(0.0638) & 0.8704(0.0217) & 0.04 & 0.06 \\
		0.5 & 7.5092(0.0543) & 0.0272(0.0033) & 1.0195(0.0671) & 0.9356(0.0220) & 0.03 & 0.10 \\
		0.75 & 8.0053(0.0614) & 0.0287(0.0030) & 1.1561(0.0694) & 0.9281(0.0217) & 0.10 & 0.18 \\
		0.975 & 9.3056(0.0698) & 0.0254(0.0033) & 0.9936(0.0658) & 0.8885(0.0220) & 0.05 & 0.09 \\
		\bottomrule
	\end{tabular}
	\label{tab:estimates}
\end{table}

Table \ref{tab:tests} presents the results for the CN-BS quantile regression model to test the null hypotheses $H_0: \beta_1 = 0$ and $H_0:\beta_2 = 0$ using the $S_W$, $S_{LR}$, $S_R$ and $S_T$ test statistics. At a 5\% significance level, we reject the null hypothesis in all cases tested in Table \ref{tab:tests}, indicating that coefficients associated with the operating time and legal representation are significant. In addition, from Table \ref{tab:tests}, we observe that the values of $S_W$, $S_{LR}$, $S_R$ and $S_T$ statistics increase as $q$ increases.

\begin{table}[H]
	\centering
	\caption{Observed values of the indicated test statistics and their $p$-values (in parentheses) for the CN-BS quantile regression model for the personal accident insurance data.}
	\begin{tabular}{cccccc}
		\toprule
		\multirow{2}{*}{$q$} & & \multicolumn{4}{c}{Test statistics} \\
		\cline{3-6}
		& Hypothesis & $S_W$ & $S_{LR}$ & $S_R$ & $S_T$ \\
		\hline
		\multirow{2}{*}{0.25} & $H_0: \beta_1 = 0$ & 53.471($<0.0001$) & 61.971($<0.0001$) & 149.8($<0.0001$) & 144.23($<0.0001$) \\
		& $H_0: \beta_2 = 0$ & 212.12($<0.0001$) & 13.438(0.0002) & 36.392($<0.0001$) & 40.174($<0.0001$) \\
		\hline
		\multirow{2}{*}{0.5} & $H_0: \beta_1 = 0$ & 68.507($<0.0001$) & 159.35($<0.0001$) & 202.04($<0.0001$) & 165.38($<0.0001$) \\
		& $H_0: \beta_2 = 0$ & 230.97($<0.0001$) & 17.197($<0.0001$) & 42.467($<0.0001$) & 47.923($<0.0001$) \\
		\hline
		\multirow{2}{*}{0.75} & $H_0: \beta_1 = 0$ & 90.682($<0.0001$) & 231.28($<0.0001$) & 262.37($<0.0001$) & 190.9($<0.0001$) \\
		& $H_0: \beta_2 = 0$ & 275.56($<0.0001$) & 36.009($<0.0001$) & 40.564($<0.0001$) & 51.648($<0.0001$) \\
		\bottomrule
	\end{tabular}
	\label{tab:tests}
\end{table}

\begin{figure}[H]
	\centering
	\subfigure[\texttt{legrep = Yes}]{\includegraphics[scale = 0.35]{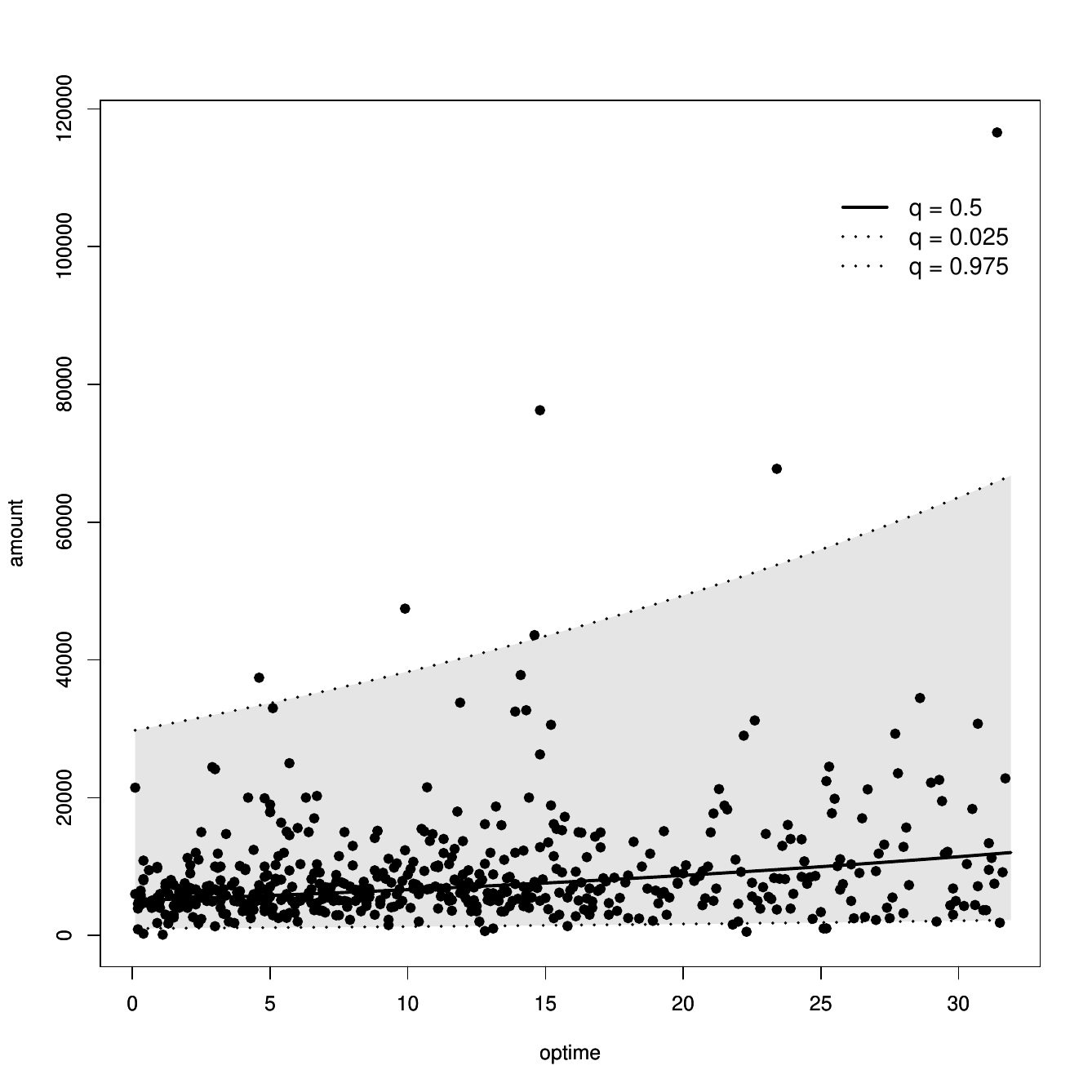}}
	\subfigure[\texttt{legrep = No}]{\includegraphics[scale = 0.35]{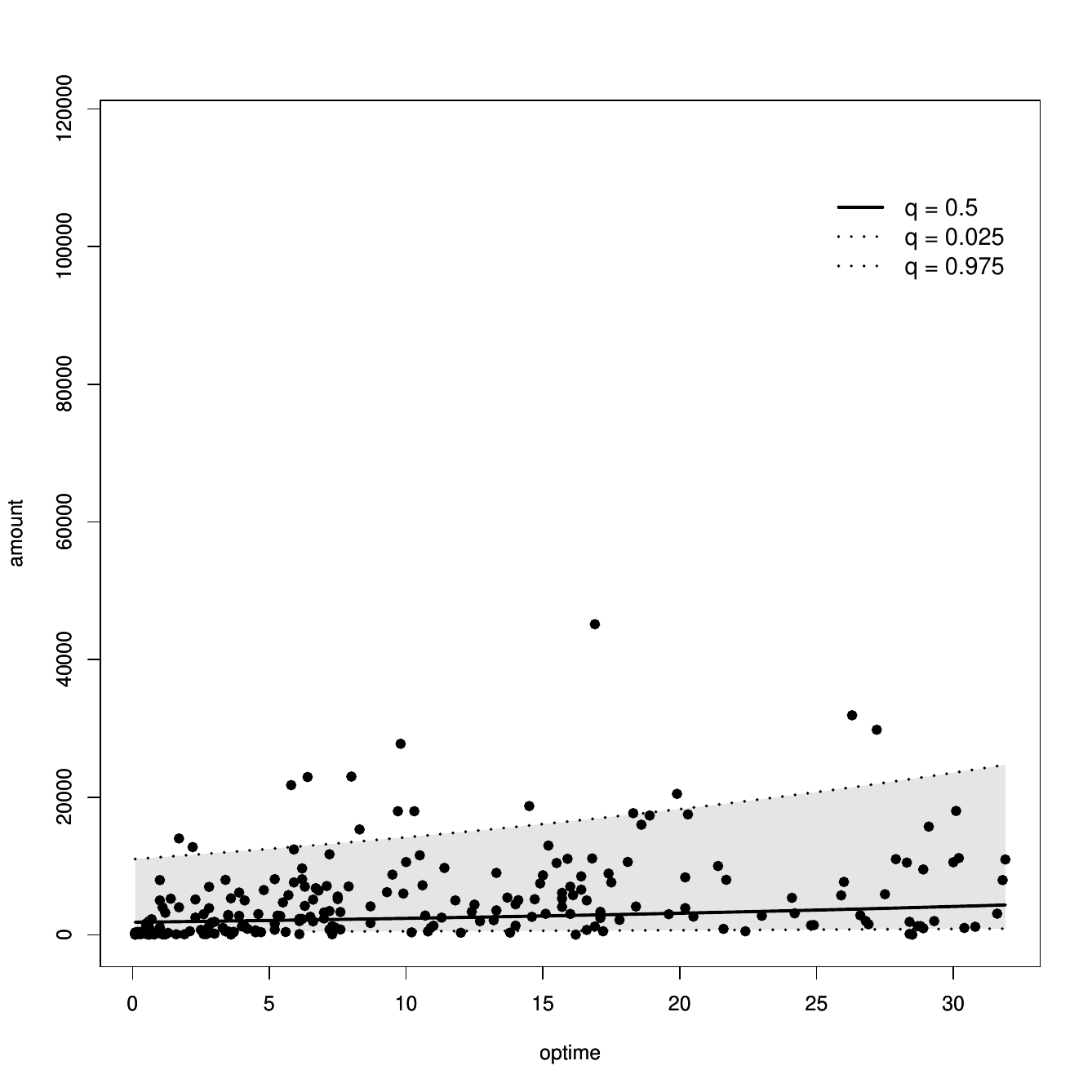}}
	\caption{95\% confidence bands in the CN-BS quantile regression model for the personal accident insurance data.}
	\label{fig:cband}
\end{figure}

Figure~\ref{fig:cband} presents the scatterplot of the amount against the optime according to the group with legal representation and without legal representation with 95\% confidence bands in the CN-BS quantile regression model for the personal accident insurance data. We use ML estimates based on the quantiles 2.5\% and 97.5\% to build these confidence bands. We can see that the upper bound of the confidence bands increases exponentially.

\begin{figure}[H]
	\centering
	\subfigure[$q = 0.25$]{\includegraphics[scale = 0.35]{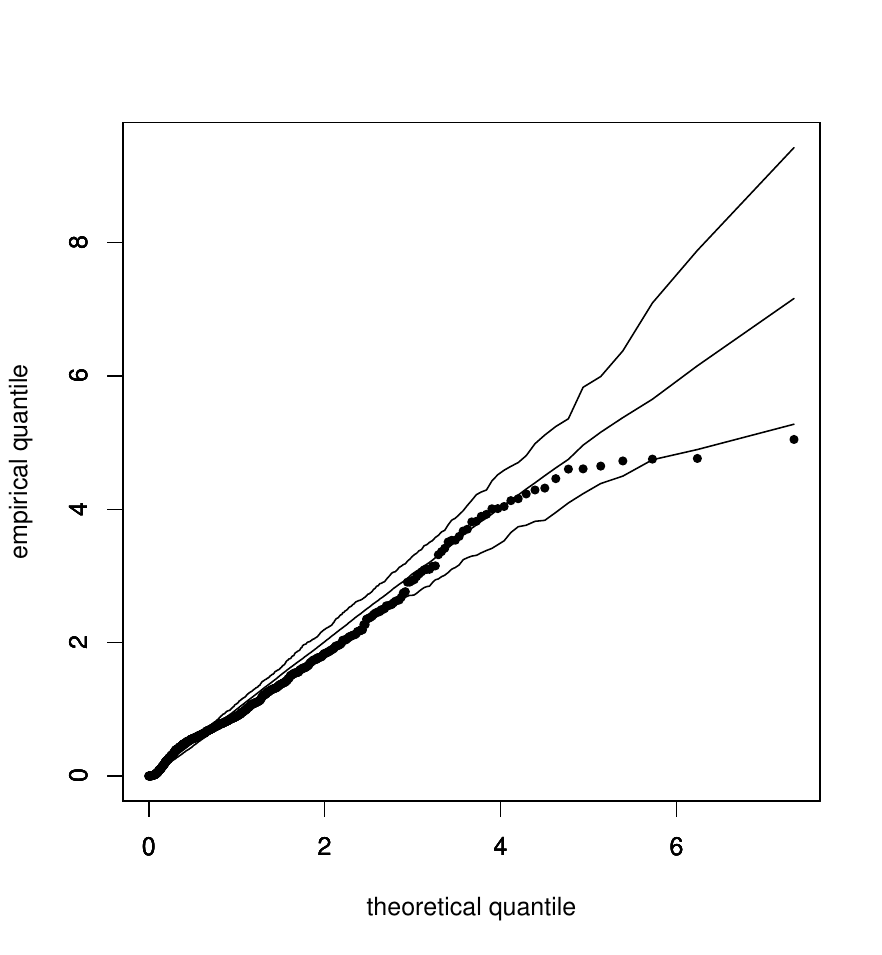}}
	\subfigure[$q = 0.5$]{\includegraphics[scale = 0.35]{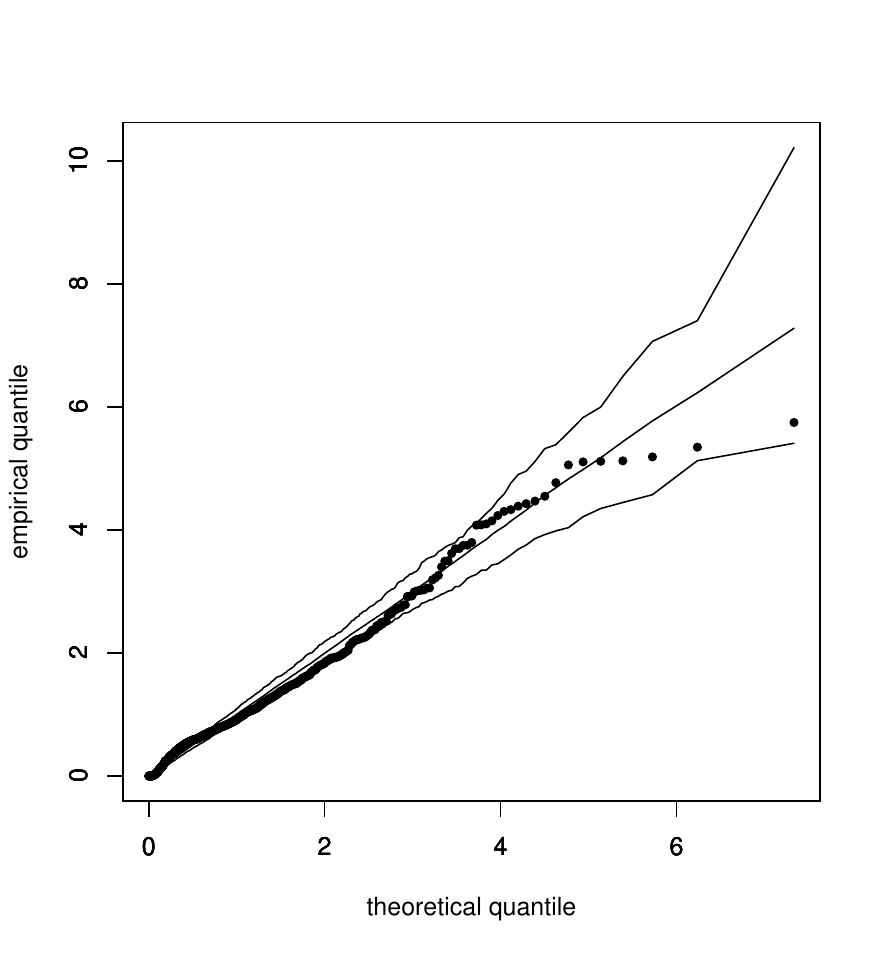}}
	\subfigure[$q = 0.75$]{\includegraphics[scale = 0.35]{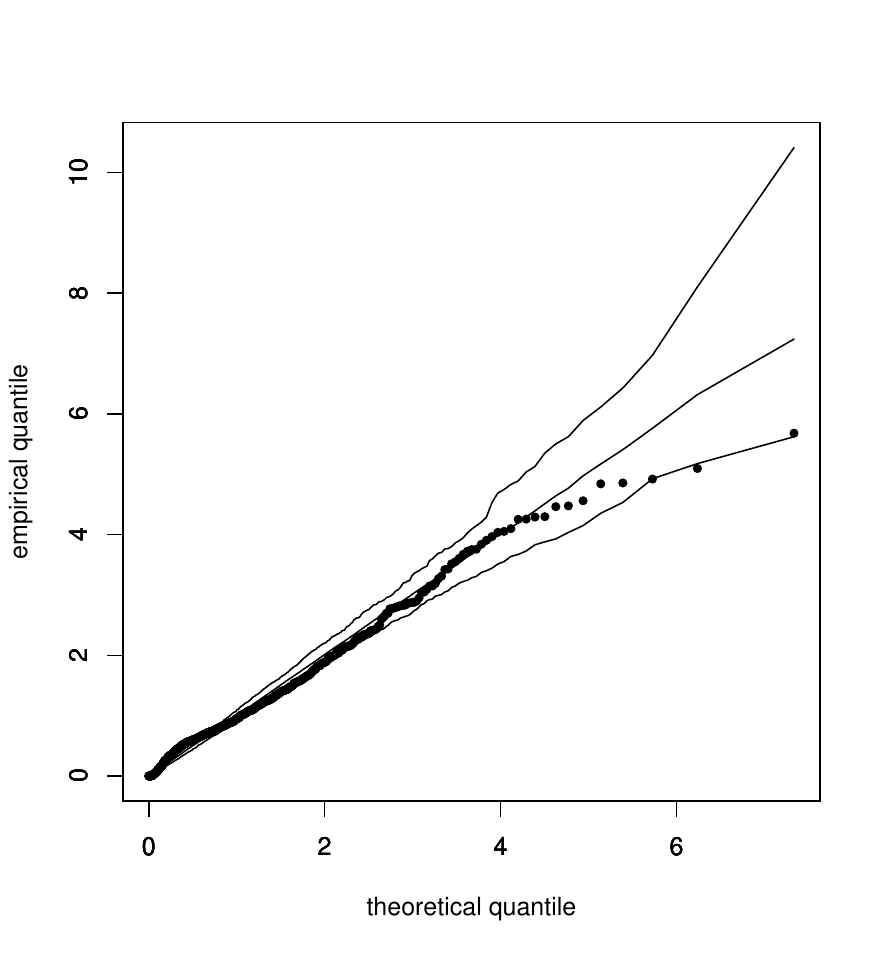}}
	\subfigure[$q = 0.25$]{\includegraphics[scale = 0.35]{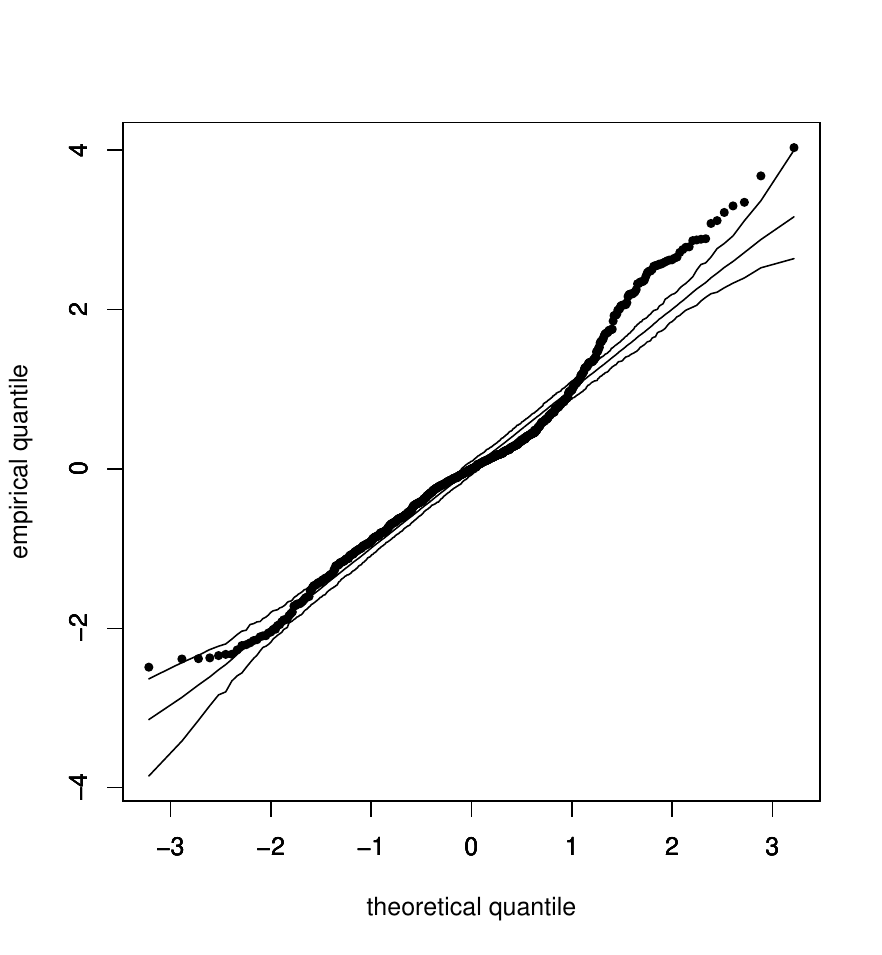}}
	\subfigure[$q = 0.5$]{\includegraphics[scale = 0.35]{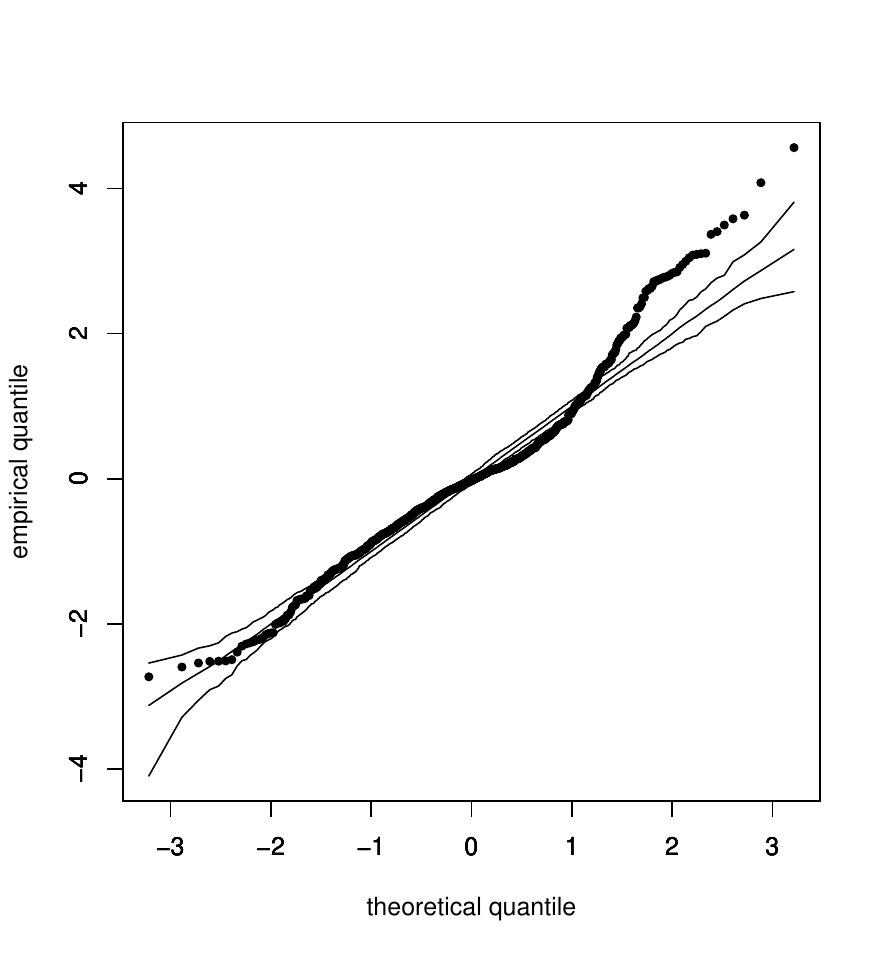}}
	\subfigure[$q = 0.75$]{\includegraphics[scale = 0.35]{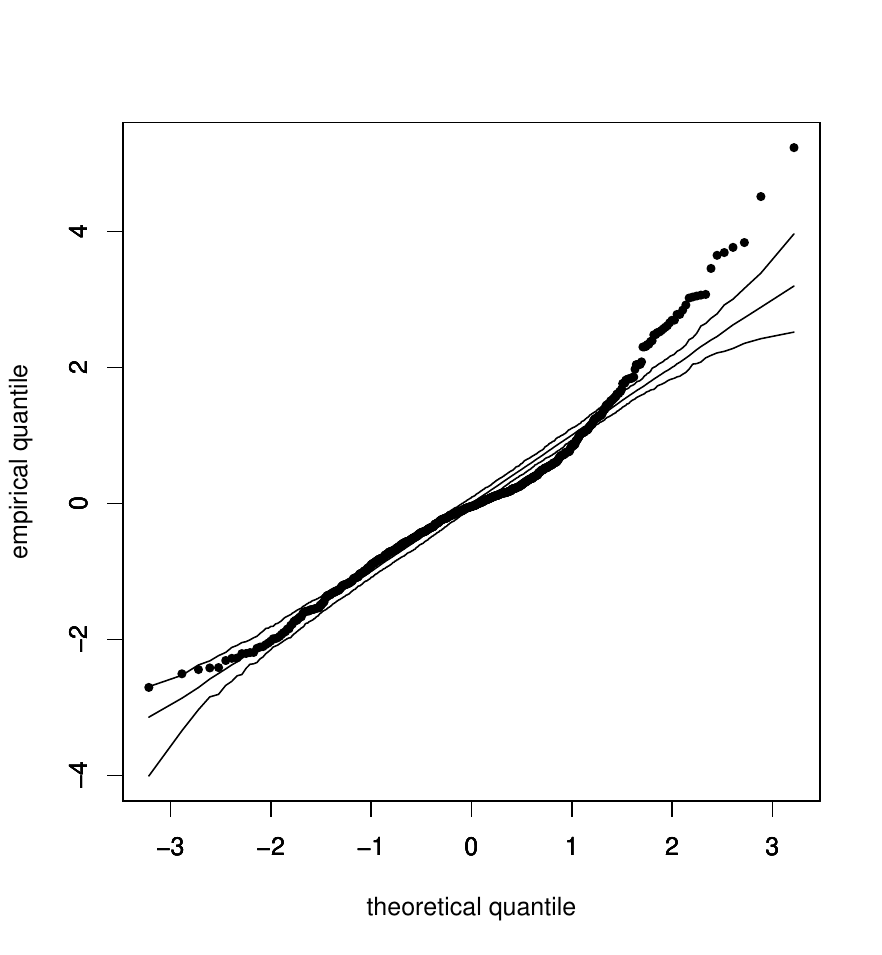}}
	\caption{QQ plots and their envelopes for the GCS ((a), (b) and (c)) and RQ ((d), (e) and (f)) residuals in the CN-BS quantile regression model for the personal accident insurance data.}
	\label{fig:residuals}
\end{figure}

Figure~\ref{fig:residuals} shows the QQ plots with simulated envelope of the GCS and RQ residuals for the CN-BS quantile regression model considered in Table \ref{tab:estimates}. We can see clearly that the CN-BS model provides a good fit based on the GCS residual. On the other hand, the plots for the RQ residuals present some more extreme points outside the region delimited by the confidence bands. 

\section{Concluding remarks}\label{section:06}

In this paper, we have proposed a class of quantile regression models based on the reparameterized scale-mixture Birnbaum-Saunders distributions, which have the distribution quantile as one of their parameters. The relevance of this work lies in the proposal of a new quantile regression model that presents a new alternative capable of providing a very flexible fit for strictly positive and asymmetric data. The results obtained in the Monte Carlo simulation studies showed that (a) the maximum likelihood estimates obtained by the EM algorithm, in terms of bias, MSE and CP, perform well; (b) the GCS and RQ residuals conform well with their respective reference distributions with the exception of the SL-BS model; and (c) the Wald, likelihood ratio, score and gradient tests have similar performances. However, in terms of power, the Wald and gradient tests have null rejection rates closer to their nominal levels. Thus, the Wald and gradient tests are preferable to the other tests analyzed. We have applied the proposed models to a real data set related to personal accident insurance. The application results showed the flexibility of the proposed model by providing a richer characterization of the effects of the covariates on the dependent variable. As part of future research, it will be of interest to study influence diagnostic tools as well as multivariate versions. Furthermore, Bartlett and Bartlett-type corrections can be used to attenuate the size distortion of hypothesis tests. Work on these problems is currently in progress and we hope to report these findings in future papers.

\paragraph{Acknowledgements }
Helton Saulo and Roberto Vila gratefully acknowledges financial support from CNPq, CAPES, and FAP-DF, Brazil.

\paragraph{Disclosure statement}
There are no conflicts of interest to disclose.


\appendix

\section{Simulation results: Hypothesis tests}\label{app_simu_hyp}

\begin{table}[H]
	\centering
	\caption{Null rejection rates for $H_0: \beta_1 = \beta_2 = \beta_3 = 0$ in the CN-BS quantile regression model ($\nu = 0.1$, $\delta = 0.3$).}
	\adjustbox{max height=\dimexpr\textheight-3.5cm\relax,
		max width=\textwidth}{
		\begin{tabular}{lccccccccccccc}
			\toprule
			& & & \multicolumn{3}{c}{$n = 50$} & & \multicolumn{3}{c}{$n = 100$} & & \multicolumn{3}{c}{$n = 200$} \\
			\cline{4-6} \cline{8-10} \cline{12-14}
			$\alpha$ & $q$ & & 1\% & 5\% & 10\% & & 1\% & 5\% & 10\% & & 1\% & 5\% & 10\% \\
			\hline
			\multirow{13}{*}{0.2} & \multirow{4}{*}{0.25} & $S_W$ & 0.0156 & 0.057 & 0.1082 && 0.0124 & 0.054 & 0.1046 && 0.0108 & 0.0512 & 0.1028 \\
			& & $S_{LR}$ & 0.021 & 0.0792 & 0.1402 && 0.0202 & 0.0778 & 0.1344 && 0.0188 & 0.0742 & 0.1362 \\
			& & $S_R$ & 0.0388 & 0.1204 & 0.1972 && 0.0232 & 0.0836 & 0.1488 && 0.015 & 0.0662 & 0.1214 \\
			& & $S_T$ & 0.0066 & 0.0476 & 0.1024 && 0.0102 & 0.0472 & 0.0978 && 0.009 & 0.0468 & 0.1012 \\
			\cline{2-14}
			& \multirow{4}{*}{0.5} & $S_W$ & 0.019 & 0.0608 & 0.1078 && 0.0158 & 0.0558 & 0.1034 && 0.011 & 0.0478 & 0.0956 \\
			& & $S_{LR}$ & 0.02 & 0.085 & 0.1402 && 0.0146 & 0.0726 & 0.129 && 0.0142 & 0.0666 & 0.122 \\
			& & $S_R$ & 0.0294 & 0.1032 & 0.1704 && 0.0188 & 0.0706 & 0.1266 && 0.014 & 0.0646 & 0.1284 \\
			& & $S_T$ & 0.0066 & 0.0446 & 0.0944 && 0.0094 & 0.0458 & 0.0958 && 0.0084 & 0.046 & 0.1 \\
			\cline{2-14}
			& \multirow{4}{*}{0.75} & $S_W$ & 0.0184 & 0.0614 & 0.1066 && 0.0128 & 0.0536 & 0.1046 && 0.0142 & 0.0504 & 0.0998 \\
			& & $S_{LR}$ & 0.0212 & 0.0764 & 0.1332 && 0.0166 & 0.07 & 0.132 && 0.0172 & 0.0728 & 0.139 \\
			& & $S_R$ & 0.0496 & 0.1384 & 0.2166 && 0.0284 & 0.1004 & 0.1636 && 0.019 & 0.061 & 0.1178 \\
			& & $S_T$ & 0.0074 & 0.0436 & 0.1002 && 0.0106 & 0.045 & 0.101 && 0.0092 & 0.047 & 0.1008 \\
			\hline
			\multirow{13}{*}{0.5} & \multirow{4}{*}{0.25} & $S_W$ & 0.017 & 0.0544 & 0.099 && 0.0124 & 0.051 & 0.0994 && 0.0094 & 0.0492 & 0.1022 \\
			& & $S_{LR}$ & 0.0192 & 0.0798 & 0.141 && 0.0166 & 0.0662 & 0.1226 && 0.0142 & 0.071 & 0.123 \\
			& & $S_R$ & 0.0664 & 0.1646 & 0.245 && 0.0228 & 0.0814 & 0.143 && 0.0134 & 0.0596 & 0.116 \\
			& & $S_T$ & 0.0074 & 0.0464 & 0.098 && 0.0102 & 0.0474 & 0.0978 && 0.0086 & 0.047 & 0.1002 \\
			\cline{2-14}
			& \multirow{4}{*}{0.5} & $S_W$ & 0.0128 & 0.0568 & 0.1044 && 0.0114 & 0.0536 & 0.1014 && 0.0104 & 0.0514 & 0.1002 \\
			& & $S_{LR}$ & 0.025 & 0.0832 & 0.1364 && 0.0166 & 0.0656 & 0.1252 && 0.0144 & 0.0638 & 0.1182 \\
			& & $S_R$ & 0.0266 & 0.0994 & 0.1678 && 0.0176 & 0.0668 & 0.1266 && 0.0138 & 0.0638 & 0.1202 \\
			& & $S_T$ & 0.0068 & 0.0446 & 0.0944 && 0.0094 & 0.0458 & 0.0956 && 0.0084 & 0.0458 & 0.0996 \\
			\cline{2-14}
			& \multirow{4}{*}{0.75} & $S_W$ & 0.0154 & 0.0636 & 0.1052 && 0.0136 & 0.0532 & 0.1004 && 0.0118 & 0.053 & 0.0996 \\
			& & $S_{LR}$ & 0.0166 & 0.0746 & 0.1338 && 0.0154 & 0.0782 & 0.135 && 0.0166 & 0.0728 & 0.13 \\
			& & $S_R$ & 0.0406 & 0.1244 & 0.1922 && 0.0258 & 0.0834 & 0.141 && 0.013 & 0.0594 & 0.1104 \\
			& & $S_T$ & 0.0076 & 0.043 & 0.1008 && 0.0108 & 0.0444 & 0.0992 && 0.0088 & 0.0468 & 0.1012 \\
			\hline
			\multirow{13}{*}{1} & \multirow{4}{*}{0.25} & $S_W$ & 0.0128 & 0.0452 & 0.0866 && 0.0108 & 0.046 & 0.0862 && 0.0084 & 0.0448 & 0.0954 \\
			& & $S_{LR}$ & 0.017 & 0.071 & 0.1252 && 0.015 & 0.0642 & 0.116 && 0.0146 & 0.0632 & 0.1136 \\
			& & $S_R$ & 0.07 & 0.1586 & 0.241 && 0.0224 & 0.0798 & 0.1482 && 0.0142 & 0.0668 & 0.1256 \\
			& & $S_T$ & 0.0074 & 0.046 & 0.0988 && 0.0104 & 0.0472 & 0.098 && 0.0088 & 0.0462 & 0.1016 \\
			\cline{2-14}
			& \multirow{4}{*}{0.5} & $S_W$ & 0.0122 & 0.0478 & 0.0908 && 0.013 & 0.047 & 0.0938 && 0.0104 & 0.0488 & 0.0944 \\
			& & $S_{LR}$ & 0.0206 & 0.0772 & 0.1318 && 0.0138 & 0.0622 & 0.122 && 0.0144 & 0.0578 & 0.1072 \\
			& & $S_R$ & 0.0292 & 0.1054 & 0.174 && 0.0186 & 0.0718 & 0.1276 && 0.014 & 0.0644 & 0.1208 \\
			& & $S_T$ & 0.0068 & 0.0444 & 0.0948 && 0.0094 & 0.0462 & 0.0962 && 0.0084 & 0.0454 & 0.1004 \\
			\cline{2-14}
			& \multirow{4}{*}{0.75} & $S_W$ & 0.0194 & 0.0534 & 0.0922 && 0.0106 & 0.0486 & 0.0888 && 0.0092 & 0.0514 & 0.101 \\
			& & $S_{LR}$ & 0.0208 & 0.0726 & 0.1276 && 0.0138 & 0.0622 & 0.115 && 0.0158 & 0.0642 & 0.113 \\
			& & $S_R$ & 0.0454 & 0.1348 & 0.207 && 0.0208 & 0.0866 & 0.1506 && 0.0126 & 0.0592 & 0.1128 \\
			& & $S_T$ & 0.0078 & 0.043 & 0.1012 && 0.011 & 0.0462 & 0.1022 && 0.0086 & 0.0464 & 0.1014 \\
			\bottomrule
		\end{tabular}
	}
	\label{tab:null-3-cnbs}
\end{table}

\begin{table}[H]
	\centering
	\caption{Null rejection rates for $H_0: \beta_1 = \beta_2 = \beta_3 = 0$ in the SL-BS quantile regression model ($\nu = 4$).}
	\adjustbox{max height=\dimexpr\textheight-3.5cm\relax,
		max width=\textwidth}{
		\begin{tabular}{lccccccccccccc}
			\toprule
			& & & \multicolumn{3}{c}{$n = 50$} & & \multicolumn{3}{c}{$n = 100$} & & \multicolumn{3}{c}{$n = 200$} \\
			\cline{4-6} \cline{8-10} \cline{12-14}
			$\alpha$ & $q$ & & 1\% & 5\% & 10\% & & 1\% & 5\% & 10\% & & 1\% & 5\% & 10\% \\
			\hline
			\multirow{13}{*}{0.2} & \multirow{4}{*}{0.25} & $S_W$ & 0.0146 & 0.0584 & 0.1 && 0.0118 & 0.0506 & 0.0974 && 0.0096  & 0.0468 & 0.0956 \\
			& & $S_{LR}$ & 0.016 & 0.0658 & 0.125 && 0.013 & 0.0616 & 0.117 && 0.0126 & 0.0542 & 0.1038 \\
			& & $S_R$ & 0.0446 & 0.1258 & 0.2072 && 0.0222 & 0.0892 & 0.152 && 0.017 & 0.0702 & 0.1286 \\
			& & $S_T$ & 0.0072 & 0.0436 & 0.0916 && 0.0078 & 0.0436 & 0.096 && 0.009 & 0.0524 & 0.101 \\
			\cline{2-14}
			& \multirow{4}{*}{0.5} & $S_W$ & 0.0148 & 0.0496 & 0.0902 && 0.0118 & 0.0506 & 0.0974 && 0.0096 & 0.0468 & 0.0954 \\
			& & $S_{LR}$ & 0.0154 & 0.0666 & 0.1284 && 0.0164 & 0.0632 & 0.119 && 0.0118 & 0.0548 & 0.1092 \\
			& & $S_R$ & 0.0664 & 0.1626 & 0.2434 && 0.0188 & 0.0736 & 0.1406 && 0.0148 & 0.0664 & 0.1254 \\
			& & $S_T$ & 0.0066 & 0.0426 & 0.0914 && 0.0082 & 0.0494 & 0.1004 && 0.0108 & 0.045 & 0.0948 \\
			\cline{2-14}
			& \multirow{4}{*}{0.75} & $S_W$ & 0.015 & 0.0496 & 0.0902 && 0.0118 & 0.0508 & 0.0974 && 0.0096 & 0.0468 & 0.0952 \\
			& & $S_{LR}$ & 0.015 & 0.0676 & 0.1236 && 0.0118 & 0.0574 & 0.1106 && 0.0104 & 0.0572 & 0.1054 \\
			& & $S_R$ & 0.0422 & 0.124 & 0.1978 && 0.026 & 0.0882 & 0.1554 && 0.0176 & 0.064 & 0.1146 \\
			& & $S_T$ & 0.007 & 0.0474 & 0.1028 && 0.0062 & 0.0458 & 0.0926 && 0.0086 & 0.0482 & 0.096 \\
			\hline
			\multirow{13}{*}{0.5} & \multirow{4}{*}{0.25} & $S_W$ & 0.0136 & 0.0446 & 0.0826 && 0.0114 & 0.047 & 0.0916 && 0.0096 & 0.0454 & 0.0904 \\
			& & $S_{LR}$ & 0.016 & 0.0708 & 0.129 && 0.0152 & 0.0544 & 0.1066 && 0.0102 & 0.05 & 0.1042 \\
			& & $S_R$ & 0.0418 & 0.1262 & 0.2064 && 0.027 & 0.089 & 0.1538 && 0.0158 & 0.0684 & 0.1182 \\
			& & $S_T$ & 0.0058 & 0.0424 & 0.095 && 0.0078 & 0.0488 & 0.0974 && 0.0094 & 0.0486 & 0.1 \\
			\cline{2-14}
			& \multirow{4}{*}{0.5} & $S_W$ & 0.0136 & 0.0446 & 0.0826 && 0.0114 & 0.047 & 0.0916 && 0.0096 & 0.0454 & 0.0902 \\
			& & $S_{LR}$ & 0.0144 & 0.0666 & 0.121 && 0.0134 & 0.0584 & 0.109 && 0.013 & 0.0566 & 0.1048 \\
			& & $S_R$ & 0.0436 & 0.1268 & 0.2104 && 0.0332 & 0.1036 & 0.169 && 0.013 & 0.0602 & 0.1174 \\
			& & $S_T$ & 0.0042 & 0.0438 & 0.0994 && 0.0104 & 0.0456 & 0.0928 && 0.0112 & 0.0476 & 0.1008 \\
			\cline{2-14}
			& \multirow{4}{*}{0.75} & $S_W$ & 0.0136 & 0.0446 & 0.0824 && 0.0114 & 0.047 & 0.0916 && 0.0096 & 0.0454 & 0.0902 \\
			& & $S_{LR}$ & 0.0156 & 0.0608 & 0.1194 && 0.0124 & 0.0574 & 0.1072 && 0.012 & 0.0566 & 0.1064 \\
			& & $S_R$ & 0.0366 & 0.1098 & 0.1852 && 0.026 & 0.0896 & 0.156 && 0.0178 & 0.0694 & 0.1288 \\
			& & $S_T$ & 0.0062 & 0.046 & 0.0984 && 0.007 & 0.0408 & 0.0944 && 0.0094 & 0.0562 & 0.103 \\
			\hline
			\multirow{13}{*}{1} & \multirow{4}{*}{0.25} & $S_W$ & 0.0088 & 0.031 & 0.0608 && 0.0102 & 0.0366 & 0.0738 && 0.0088 & 0.0394 & 0.0834 \\
			& & $S_{LR}$ & 0.013 & 0.0618 & 0.113 && 0.009 & 0.0504 & 0.0992 && 0.01 & 0.045 & 0.0946 \\
			& & $S_R$ & 0.045 & 0.1288 & 0.21 && 0.027 & 0.0942 & 0.1568 && 0.0122 & 0.0638 & 0.1268 \\
			& & $S_T$ & 0.008 & 0.0478 & 0.0928 && 0.008 & 0.0508 & 0.1056 && 0.0112 & 0.0486 & 0.0966 \\
			\cline{2-14}
			& \multirow{4}{*}{0.5} & $S_W$ & 0.0088 & 0.0308 & 0.0608 && 0.0102 & 0.0366 & 0.0738 && 0.0088 & 0.0394 & 0.0834 \\
			& & $S_{LR}$ & 0.013 & 0.0578 & 0.112 && 0.0114 & 0.05 & 0.0992 && 0.0084 & 0.0428 & 0.0934 \\
			& & $S_R$ & 0.0338 & 0.108 & 0.183 && 0.0206 & 0.0762 & 0.1396 && 0.0114 & 0.0602 & 0.1142 \\
			& & $S_T$ & 0.0064 & 0.046 & 0.0974 && 0.0076 & 0.047 & 0.0974 && 0.0088 & 0.0502 & 0.104 \\
			\cline{2-14}
			& \multirow{4}{*}{0.75} & $S_W$ & 0.0088 & 0.0308 & 0.0608 && 0.0102 & 0.0366 & 0.0738 && 0.0088 & 0.0394 & 0.0834 \\
			& & $S_{LR}$ & 0.0144 & 0.0594 & 0.1148 && 0.0084 & 0.0468 & 0.0894 && 0.0094 & 0.0528 & 0.0978 \\
			& & $S_R$ & 0.0586 & 0.1494 & 0.231 && 0.0216 & 0.0826 & 0.1496 && 0.0128 & 0.0644 & 0.1218 \\
			& & $S_T$ & 0.0054 & 0.0436 & 0.0916 && 0.0088 & 0.0468 & 0.0934 && 0.0086 & 0.0498 & 0.0994 \\
			\bottomrule
		\end{tabular}
	}
	\label{tab:null-3-slbs}
\end{table}

\begin{table}[H]
	\centering
	\caption{Null rejection rates for $H_0: \beta_1 = \beta_2 = \beta_3 = 0$ in the $t_\nu$-BS quantile regression model ($\nu = 11$).}
	\adjustbox{max height=\dimexpr\textheight-3.5cm\relax,
		max width=\textwidth}{
		\begin{tabular}{lccccccccccccc}
			\toprule
			& & & \multicolumn{3}{c}{$n = 50$} & & \multicolumn{3}{c}{$n = 100$} & & \multicolumn{3}{c}{$n = 200$} \\
			\cline{4-6} \cline{8-10} \cline{12-14}
			$\alpha$ & $q$ & & 1\% & 5\% & 10\% & & 1\% & 5\% & 10\% & & 1\% & 5\% & 10\% \\
			\hline
			\multirow{13}{*}{0.2} & \multirow{4}{*}{0.25} & $S_W$ & 0.016 & 0.063 & 0.1128 && 0.0112 & 0.0522 & 0.0982 && 0.0104 & 0.0494 & 0.0984 \\
			& & $S_{LR}$ & 0.0178 & 0.0712 & 0.1304 && 0.0156 & 0.0668 & 0.1214 && 0.0138 & 0.0604 & 0.1164 \\
			& & $S_R$ & 0.0452 & 0.1308 & 0.2092 && 0.0154 & 0.0704 & 0.129 && 0.015 & 0.0692 & 0.1254 \\
			& & $S_T$ & 0.0086 & 0.0498 & 0.1016 && 0.0068 & 0.043 & 0.0918 && 0.0082 & 0.0478 & 0.0962 \\
			\cline{2-14}
			& \multirow{4}{*}{0.5} & $S_W$ & 0.0156 & 0.0662 & 0.1232 && 0.016 & 0.0566 & 0.1074 && 0.012 & 0.0538 & 0.106 \\
			& & $S_{LR}$ & 0.0192 & 0.079 & 0.1418 && 0.0146 & 0.066 & 0.124 && 0.013 & 0.058 & 0.1102 \\
			& & $S_R$ & 0.0422 & 0.1188 & 0.1908 && 0.015 & 0.0662 & 0.123 && 0.0122 & 0.0568 & 0.1126 \\
			& & $S_T$ & 0.0058 & 0.0394 & 0.0886 && 0.0094 & 0.0454 & 0.0934 && 0.0096 & 0.0506 & 0.102 \\
			\cline{2-14}
			& \multirow{4}{*}{0.75} & $S_W$ & 0.0214 & 0.0674 & 0.1206 && 0.0138 & 0.0572 & 0.1084 && 0.0116 & 0.058 & 0.1032 \\
			& & $S_{LR}$ & 0.018 & 0.0676 & 0.1364 && 0.0146 & 0.0594 & 0.1156 && 0.0138 & 0.063 & 0.1176 \\
			& & $S_R$ & 0.0604 & 0.1486 & 0.2198 && 0.0168 & 0.0668 & 0.127 && 0.0156 & 0.0618 & 0.1204 \\
			& & $S_T$ & 0.0092 & 0.0522 & 0.1064 && 0.0086 & 0.0486 & 0.0982 && 0.0092 & 0.0452 & 0.095 \\
			\hline
			\multirow{13}{*}{0.5} & \multirow{4}{*}{0.25} & $S_W$ & 0.0144 & 0.0612 & 0.111 && 0.0112 & 0.0492 & 0.0958 && 0.011 & 0.0554 & 0.1026 \\
			& & $S_{LR}$ & 0.015 & 0.0678 & 0.131 && 0.0164 & 0.0674 & 0.122 && 0.0112 & 0.0576 & 0.1062 \\
			& & $S_R$ & 0.0432 & 0.1368 & 0.208 && 0.0194 & 0.0806 & 0.1418 && 0.0164 & 0.0668 & 0.1248 \\
			& & $S_T$ & 0.0086 & 0.0502 & 0.1046 && 0.0068 & 0.0438 & 0.0916 && 0.008 & 0.0454 & 0.0922 \\
			\cline{2-14}
			& \multirow{4}{*}{0.5} & $S_W$ & 0.0188 & 0.0614 & 0.1116 && 0.0128 & 0.0544 & 0.097 && 0.0106 & 0.0518 & 0.0996 \\
			& & $S_{LR}$ & 0.0156 & 0.0694 & 0.1284 && 0.014 & 0.0586 & 0.1118 && 0.0108 & 0.0626 & 0.1174 \\
			& & $S_R$ & 0.041 & 0.1124 & 0.1846 && 0.021 & 0.0758 & 0.143 && 0.0124 & 0.0608 & 0.1186 \\
			& & $S_T$ & 0.006 & 0.0456 & 0.0988 && 0.0094 & 0.05 & 0.1024 && 0.0066 & 0.0452 & 0.0948 \\
			\cline{2-14}
			& \multirow{4}{*}{0.75} & $S_W$ & 0.0172 & 0.0548 & 0.1032 && 0.01 & 0.0566 & 0.1038 && 0.0108 & 0.0494 & 0.0972 \\
			& & $S_{LR}$ & 0.0144 & 0.072 & 0.1354 && 0.0142 & 0.059 & 0.1112 && 0.0126 & 0.0624 & 0.1218 \\
			& & $S_R$ & 0.0474 & 0.1302 & 0.2032 && 0.018 & 0.0736 & 0.1328 && 0.0138 & 0.0624 & 0.1186 \\
			& & $S_T$ & 0.0084 & 0.0496 & 0.1016 && 0.0086 & 0.0436 & 0.0912 && 0.0084 & 0.049 & 0.0946 \\
			\hline
			\multirow{13}{*}{1} & \multirow{4}{*}{0.25} & $S_W$ & 0.0118 & 0.0478 & 0.0898 && 0.0118 & 0.042 & 0.0864 && 0.0106 & 0.0446 & 0.0874 \\
			& & $S_{LR}$ & 0.0138 & 0.0604 & 0.1128 && 0.0134 & 0.058 & 0.1058 && 0.0138 & 0.055 & 0.1016 \\
			& & $S_R$ & 0.0544 & 0.1374 & 0.2124 && 0.0156 & 0.0708 & 0.1356 && 0.0144 & 0.0638 & 0.1252 \\
			& & $S_T$ & 0.0086 & 0.0514 & 0.105 && 0.0068 & 0.044 & 0.0916 &&  0.0098 & 0.0544 & 0.1016 \\
			\cline{2-14}
			& \multirow{4}{*}{0.5} & $S_W$ & 0.0102 & 0.043 & 0.0848 && 0.0102 & 0.048 & 0.094 && 0.01 & 0.0452 & 0.0936 \\
			& & $S_{LR}$ & 0.016 & 0.0664 & 0.1232 && 0.0136 & 0.0552 & 0.1074 && 0.0116 & 0.0498 & 0.097 \\
			& & $S_R$ & 0.0196 & 0.0844 & 0.1572 && 0.0122 & 0.0698 & 0.137 && 0.0134 & 0.0592 & 0.1218 \\
			& & $S_T$ & 0.008 & 0.0452 & 0.0998 && 0.0086 & 0.0508 & 0.0974 && 0.009 & 0.0472 & 0.096 \\
			\cline{2-14}
			& \multirow{4}{*}{0.75} & $S_W$ & 0.0144 & 0.0484 & 0.0882 && 0.0094 & 0.0464 & 0.094 && 0.011 & 0.0482 & 0.0974 \\
			& & $S_{LR}$ & 0.0178 & 0.0696 & 0.1236 && 0.0096 & 0.0526 & 0.1 && 0.0122 & 0.0528 & 0.1074 \\
			& & $S_R$ & 0.0416 & 0.1196 & 0.1964 && 0.0224 & 0.082 & 0.1416 && 0.0114 & 0.0592 & 0.1132 \\
			& & $S_T$ & 0.0072 & 0.049 & 0.0982 && 0.0082 & 0.0518 & 0.1028 && 0.011 & 0.051 & 0.099 \\
			\bottomrule
		\end{tabular}
	}
	\label{tab:null-3-tbs}
\end{table}

\end{document}